\catcode`\@=11

\newif\if@fewtab\@fewtabtrue

{\count255=\time\divide\count255 by 60
\xdef\hourmin{\number\count255}
\multiply\count255 by-60\advance\count255 by\time
\xdef\hourmin{\hourmin:\ifnum\count255<10 0\fi\the\count255}}
\def\ps@draft{\let\@mkboth\@gobbletwo
    \def\@oddhead{}
    \def\@oddfoot
       {\hbox to 7 cm{$\scriptstyle Draft\ version:\ \draftdate$
       \hfil}\hskip -7cm\hfil\rm\thepage \hfil}
    \def\@evenhead{}\let\@evenfoot\@oddfoot}

\def\ceqno{\global\@fewtabfalse
    \ifcase\@eqcnt \def\@tempa{& & &}\or \def\@tempa{& &}
      \or \def\@tempa{&}
      \or\def\@tempa{}\fi\@tempa
{\rm(\theequation)}}

\def\aeqno#1{\global\@fewtabfalse
    \ifcase\@eqcnt \def\@tempa{& & &}\or \def\@tempa{& &}
      \or \def\@tempa{&}
      \or\def\@tempa{}\fi\@tempa
{\rm(\theequation,#1)}}

\def\label#1{\ifnum\draftcontrol=1
 \global\def\draftnote{$\scriptstyle #1$}\fi
 \@bsphack\if@filesw {\let\thepage\relax
   \def\protect{\noexpand\noexpand\noexpand}%
\xdef\@gtempa{\write\@auxout{\string
      \newlabel{#1}{{\@currentlabel}{\thepage}}}}}\@gtempa
   \if@nobreak \ifvmode\nobreak\fi\fi\fi
  \@esphack}

\def\alabel#1#2{\label{#1}\global\@fewtabfalse
    \ifcase\@eqcnt \def\@tempa{& & &}\or \def\@tempa{& &}
      \or \def\@tempa{&}
      \or\def\@tempa{}\fi\@tempa
{\hbox to 3cm{\phantom{\rm(\theequation,#2)}
\draftnote \hfil}\hskip -3cm {\rm(\theequation,#2)}}}

\def\clabel#1{\label{#1}\global\@fewtabfalse
    \ifcase\@eqcnt \def\@tempa{& & &}\or \def\@tempa{& &}
      \or \def\@tempa{&}
      \or\def\@tempa{}\fi\@tempa
{\hbox to 3cm{\phantom{\rm(\theequation)}
\draftnote \hfil}\hskip -3cm{\rm(\theequation)}}}

\def\eqnarray{\def\draftnote{{}}\global\@fewtabtrue
\stepcounter{equation}\let\@currentlabel=\theequation
\global\@eqnswtrue
\global\@eqcnt\z@\tabskip\@centering\let\\=\@eqncr
$$\halign to \displaywidth\bgroup\@eqnsel\hskip\@centering\@eqcnt\z@
  $\displaystyle\tabskip\z@{##}$&\global\@eqcnt\@ne
  \hskip 1\arraycolsep \hfil${##}$\hfil
  &\global\@eqcnt\tw@ \hskip 1\arraycolsep
$\displaystyle\tabskip\z@{##}$
\hfil  \tabskip\@centering&\global\@eqcnt\thr@@\llap{##}\tabskip\z@
\cr}

\def\endeqnarray{\@@eqncr\egroup
      \global\advance\c@equation\m@ne$$\global\@ignoretrue}

\def\@eqnnum{\hbox to 3cm{\phantom{\rm(\theequation)} \draftnote
                         \hfil}\hskip -3cm {\rm(\theequation)}}

\def\@@eqncr{\let\@tempa\relax
    \ifcase\@eqcnt \def\@tempa{& & &}\or \def\@tempa{& &}
      \or \def\@tempa{&}
      \or\def\@tempa{}
\fi\@tempa
\if@eqnsw
\if@fewtab\@eqnnum\fi
\stepcounter{equation}\fi\global
\@eqnswtrue\global\@eqcnt\z@\global\@fewtabtrue\cr}

\def\draftcite#1{\ifnum\draftcontrol=1#1\else{}\fi}

\def\@lbibitem[#1]#2{\item{}\hskip -3cm \hbox to 2cm
{\hfil$\scriptstyle\draftcite{#2}$}\hskip
1cm[\@biblabel{#1}]\if@filesw
     {\def\protect##1{\string ##1\space}\immediate
      \write\@auxout{\string\bibcite{#2}{#1}}}\fi\ignorespaces}

\def\@bibitem#1{\item\hskip -3cm \hbox to 2cm
{\hfil $\scriptstyle\draftcite{#1}$}\hskip 1cm
\if@filesw \immediate\write\@auxout
       {\string\bibcite{#1}{\the\value{\@listctr}}}\fi\ignorespaces}

\def\nsection#1{\section{#1}\setcounter{equation}{0}}

\def\nappendix#1{\vskip 1cm\no{\bf Appendix #1}\def\thesection{#1}
\setcounter{equation}{0}}

\font\tendl=msbm10  scaled \magstep1
\font\sevendl=msbm7 scaled \magstep1
\font\fivedl=msbm5 scaled \magstep1
\font\tengl=eufm10  scaled \magstep1
\font\sevengl=eufm7 scaled \magstep1
\font\fivegl=eufm5 scaled \magstep1

\newfam\dlfam \def\dl{\fam\dlfam\tendl} 
\textfont\dlfam=\tendl \scriptfont\dlfam=\sevendl
\scriptscriptfont\dlfam=\fivedl
\newfam\glfam  
\textfont\glfam=\tengl \scriptfont\glfam=\sevengl
\scriptscriptfont\glfam=\fivegl

\def\draftdate{\number\month/\number\day/\number\year\ \ \ \hourmin }

\global\def\draftcontrol{0}
\catcode`\@=12
\def\tilde{\widetilde}

\documentstyle[11pt]{article}

\def\theequation{{\thesection.\arabic{equation}}}

\newcommand{\be}{\begin{eqnarray}}
\newcommand{\en}{\end{eqnarray}\vs 0.5 cm}

\newcommand{\no}{\noindent}
\newcommand{\vs}{\vskip}
\newcommand{\hs}{\hspace}

\newcommand{\NR}{{{\dl R}}}
\newcommand{\NP}{{{\dl P}}}
\newcommand{\NC}{{{\dl C}}}
\newcommand{\qq}{\begin{eqnarray}}
\newcommand{\de}{\bar\partial}
\newcommand{\da}{\partial}
\newcommand{\ee}{{\rm e}}

\newcommand{\qqq}{\end{eqnarray}}

\newcommand{\tr}{\hbox{tr}}

\newcommand{\CA}{{\cal A}}

\newcommand{\CD}{{\cal D}}

\newcommand{\CF}{{\cal F}}
\newcommand{\CG}{{\cal G}}
\newcommand{\CH}{{\cal H}}

\newcommand{\CL}{{\cal L}}
\newcommand{\CM}{{\cal M}}
\newcommand{\CN}{{\cal N}}
\newcommand{\CO}{{\cal O}}
\newcommand{\CP}{{\cal P}}

\newcommand{\CS}{{\cal S}}

\newcommand{\CU}{{\cal U}}

\newcommand{\CW}{{\cal W}}

\newcommand{\s}{\hspace{0.05cm}}
\newcommand{\m}{\hspace{0.025cm}}

\newcommand{\hf}{{_1\over^2}}

\newcommand{\bx}{{\bf x}}
\begin{document}
\title{\bf  \s$\bf{SU(2)}$ WZW theory
at higher genera}
\author{\\ \\Krzysztof Gaw\c{e}dzki \\C.N.R.S., I.H.E.S.,
Bures-sur-Yvette, 91440, France}
\date{ }
\maketitle

\vskip 1 cm

\begin{abstract}
We compute, by free field techniques, the scalar product of
the \s$SU(2)\s$ Chern-Simons states on genus $>\m1$ surfaces.
The result is a finite-dimensional integral over
positions of ``screening charges'' and one complex
modular parameter. It uses an effective description
of the CS states closely related to
the one worked out by Bertram \cite{Bertram}. The scalar
product formula allows to express the
higher genus partition functions of the WZW conformal field
theory by finite-dimensional integrals. It should provide the
hermitian metric preserved by the Knizhnik-Zamolodchikov-Bernard
connection describing the variations of the CS states under
the change of the complex structure of the surface.

\end{abstract}

\vskip 1cm

\nsection{\hspace{-.6cm}.\ \ Introduction}
\vs 0.5cm

As noted in \cite{WittenJones}, there exists a close
relation between the Chern-Simons (CS) topological
gauge theory in 3D and the Wess-Zumino-Witten (WZW)
model of conformal field theory in 2D. The (fixed time)
quantum states of the CS theory on a Riemann surface
\s$\Sigma\s$ without boundary are solutions of
the current algebra Ward identities of the WZW theory.
The states of the CS theory are holomorphic functionals
\s$\Psi\s$ on the space \s$\CA^{01}\s$ of (smooth) \s$0,1$-forms
\s$A^{01}\s$ with values in the complexified Lie algebra
\s${\bf g}^\NC\s$ of a compact Lie group \s$G\s$.
\s The functionals \s$\Psi\s$ are required to be
invariant under the complex (chiral)
gauge transformations \s\s$\Psi\s\mapsto\s
{}^h\hs{-0.04cm}\Psi\s\s$, where
\qq
{}^h\hs{-0.04cm}\Psi(A^{01})\s=\s\ee^{-kS(h,\m A^{01})}\s\s
\Psi({}^{h^{-1}}\hs{-0.17cm}A^{01})\
\label{Trans}
\qqq
for \s$h:\Sigma\m\rightarrow\m G^\NC\s$. \s In the above
formula, \s$S(h,\m A^{01})\s$ denotes the action of the
WZW model \cite{WitBos}
in the external gauge field \s$A^{01}\s$.
\s For a general gauge field, it takes the form
\qq
S(h,A^{10}+A^{01})\s=\s-{_{i}\over^{4\pi}}
\int_{_\Sigma}
\tr\s\m(h^{-1}\da h)\hs{-0.09cm}\wedge\m\hs{-0.1cm}
(h^{-1}\de h)-{_{i}\over^{12\pi}}
\int_{_\Sigma}d^{-1}\s\tr\s\m(h^{-1}dh)^{\wedge 3}\ \s\cr
+{_{i}\over^{2\pi}}\int_{_\Sigma}\tr\s\m
[\m(h\da h^{-1})\hs{-0.1cm}\wedge\hs{-0.1cm}\m
A^{01}+A^{10}\hs{-0.15cm}\wedge\m\hs{-0.1cm}\m(h^{-1}\de h)
+hA^{10}h^{-1}\hs{-0.15cm}\wedge\hs{-0.1cm}\m A^{01}
-A^{10}\hs{-0.1cm}\wedge\hs{-0.09cm}\m A^{01}]\ .
\label{dEf}
\qqq
The non-negative integer \s$k\s$ is called the level
of the model. The invariance
\s\s$\Psi={}^h\hs{-0.04cm}\Psi\s\s$ is exactly the
chiral gauge symmetry
Ward identity for the WZW partition function. Moreover,
adding static Wilson lines in the CS theory,
one obtains the chiral Ward identities for the Green functions
of the primary fields of the WZW theory. For the sake
of simplicity, we shall concentrate here on the case
of the WZW partition function and we shall take \s$G=SU(2)\s$.
\vs 0.5cm

The space of CS states has finite dimension.
The CS states \s$\Psi\s$ may be viewed as sections
of a complex line bundle over the orbit space \s$\CA^{01}/\CG^\NC\s$
of the group \s$\CG^\NC\s$ of complex gauge transformations.
The orbits \s$\CG^\NC\hs{-0.1cm}A^{01}\s$ are in
one to one correspondence
with the isomorphism classes of holomorphic
vector bundles (h.v.b.) \s$E\s$ of rank 2,
with trivial determinant,
given by the \s$SL(2,\NC)$-valued holomorphic 1-cocycles
\s$(g_{\alpha\beta})\s$,
\qq
g_{\alpha\beta}\s g_{\beta\gamma}=g_{\alpha\gamma}\ ,
\qqq
 where \s$A^{01}=g_\alpha^{-1}\de
g_\alpha\s$ locally and \s$g_{\alpha\beta}=g_\alpha g_\beta^{-1}\s$.
\s If, for the genus \s$g\s$ of \s$\Sigma\s$
\s$>\m 1\s$, \s one limits oneself to the open dense
(in the $C^\infty$ topology) subset \s$\CA^{01}_{s}
\subset\CA^{01}\s$ corresponding to the stable bundles,
then the orbit space \s$\CA^{01}_{s}/\CG^\NC\s$
becomes a complex variety \s$\CN_s\s$ of dimension
\s$3g-3\equiv N\s$. \s Besides,
\s$\CN_s\s$ possesses a natural compactification \s$\CN_{ss}\s$,
the Seshadri moduli space of semi-stable bundles
\cite{Sesh}\cite{NaraRama}. The CS states coincide with
the holomorphic sections of the \s$k^{\m th}\s$ power of the
natural determinant
bundle \s$\CD\s$ over \s$\CN_{ss}\s$. \s The spaces
\s$H^0(\CD^k)\s$ of such sections have dimensions given
by the Verlinde formula \cite{Verl}. They form a holomorphic
vector bundle \s$\CW_k\s$ over the moduli space \s$\CM\s$ of
complex curves.
This bundle may be equipped with a projectively flat
``heat kernel'' connection first described by Bernard
\cite{Bernard}, see also
\cite{WittenJones}\cite{Hitchin}\cite{Karpacz}\cite{Axel},
which generalizes the Knizhnik-Zamolodchikov connection
\cite{KZ} to the higher genus situation.
\vs 0.5cm

The partition function of the WZW model is formally given
by the functional integral
\qq
Z(A^{10}+A^{01})\s=\s\int\ee^{kS(g,\m A^{10}+A^{01})}\s\s Dg
\label{FI}
\qqq
where \s$Dg\s$ stands for the formal product \s$\prod
\limits_{{x}\in\Sigma}dg({x})\s$ of the
Haar measures on \s$G\s$. It has been argued in
\cite{Quadr} that the solution of (\ref{FI}) is given by
\qq
Z(A^{10}+A^{01})\s=\s\sum\limits_{r}{\overline{
\Psi_r(-(A^{10})^\dagger)}}
\s\s\Psi_r(A^{10})\ \s\ee^{\s{{ik}\over{2\pi}}
\smallint_{_\Sigma}{\rm tr}
\s A^{10}\wedge\m A^{01}}\ ,
\label{PartF}
\qqq
for an arbitrary basis \s$(\Psi_r)\s$ of the
CS states orthonormal with respect to the
scalar product corresponding to the norm
\qq
\|\Psi\|^2\s=\s\int|\Psi(A^{01})|^2\s\s\ee^{\s{ik\over^{2\pi}}
\smallint_{_\Sigma}{\rm tr}\s A^{10}\wedge A^{01}}\s\m DA\ .
\label{last}
\qqq
The functional integral in (\ref{last})
is over the unitary gauge fields
\s$A\equiv A^{10}+A^{01}\s$ with \s$A^{10}=-(A^{01})^\dagger\s$.
This way the calculation of the partition functions
in the WZW theory is reduced to that of the scalar
product of the CS states.
\vs 0.5cm

Formal arguments show that the scalar
product (\ref{last}) should supply the bundle \s$\CW_k\s$ of
CS states with a hermitian structure preserved by the
Knizhnik-Zamolodchikov-Bernard (KZB) connection,
see \cite{Karpacz}.
Proving rigorously the metricity of the KZB connection
is an important mathematical problem in conformal field
theory still left open. The purpose of this work is to
provide a candidate for its (constructive) solution
by computing exactly the functional integral
Eq.\s\s(\ref{last}). The result will have a form
of an explicit finite-dimensional integral of a positive
measure. Similar work has been done in the case of genus zero with
insertions in \cite{1}\cite{Quadr} for \s$G=SU(2)\s$ and in
\cite{FalGaw0}
for a general simple \s$G\s$. The integral formulae for
the scalar product are the dual versions of the
expressions for conformal blocks of the WZW theory in terms
of the generalized hypergeometric integrals
\cite{Flume}\cite{FadZam}\cite{VarSchecht}, at the core
of the relations between the WZW models and the quantum groups
and of the recently
discovered relation between the WZW model and the Bethe
ansatz \cite{Resh}\cite{Varch}\cite{Babu}\cite{FeiFrResh}.
The extension to higher genera has required a
nontrivial generalization of the
low genus arguments and has taken
some time.
\vskip 0.4cm

The paper is organized as follows. In Sect.\s\s2 we describe
a slice in the space of gauge fields, transversal to generic
\s$\CG^\NC$-orbits. It corresponds to realizing generic
rank $2$ determinant $0$ holomorphic vector bundles as extensions
of a one-parameter family of degree \s$g-1\s$ line bundles.
In Sect.\s\s3, we examine the restrictions
of CS states to the slice. They become sections of the
\s$k^{\m th}\s$ power of the determinant bundle of the family
of extensions. This picture of the higher genus CS states
is closely related to the one worked out in \cite{Bertram},
see also \cite{Thad}, based on considering the extensions of
fixed degree \s$g\s$ line bundle\footnote{We thank
B. Feigin and S. Ramanan for attracting our attention to
\cite{Bertram} and \cite{Thad}.}. The relation between the two
presentations is the subject of Sect.\s\s4.
Sect.\s\s5 describes a projective
version of the scalar product formula, from which
the surface dependent constants were omitted.
It has been extracted from the full-fledged
formula discussed later for the sake of a
moderately interested reader who would not
like to dwell into the details of the
functional integration which occupies most
of the rest of the paper. And so,
in Sect.\s\s6, using the slice of the space
\s$\CA^{01}\s$, we decompose the functional integral
(\ref{last}) to the one
over \s$\CG^\NC\s$ and over the orbit space. The Jacobian
of the relevant change of variables is computed in
Sect.\s\s6 by free field functional integration. The crucial
Sect.\s\s7 performs the integration over \s$\CG^\NC\s$
by reducing it to an iterative Gaussian integral. Finally,
Sect.\s\s8 assembles the complete formula for the scalar
product. In Appendices, besides treating
a number of technical points, we work out the details
of the relation of our description of CS states to
that of \cite{Bertram} (Appendix C) and submit the scalar
product formula to simple consistency checks (Appendix F).
What remains to be proven, however, is that the
finite-dimensional integrals appearing in the formula
actually always converge resulting in a hermitian
metric on the bundle \s$\CW_k\s$ of state spaces
which is preserved by the KZB connection.
What is also missing is an interpretation of the formula
in terms of modular geometry, providing
a counterpart of the analysis
of the KZB connection carried out in \cite{Hitchin}.
As the first step in this direction one could try
to simplify the formal arguments given below.
Numerous cancellations occurring in intermediate steps
of the calculation suggest that such simplification
should be possible.
\vs 0.4cm

This is a non-rigorous work in its manipulation of formal
functional integrals which lead, in the end, to a chain
of Gaussian integrations. Handling these integrals
required, nevertheless, careful treatment. As a result,
the paper employs relatively sophisticated mathematical
tools. It may be viewed as a piece of ``theoretical
mathematics'' in the sense of \cite{JaffeQ}: it uses formal
functional integral to extract an interesting mathematical
structure which should be submitted now to rigorous analysis.
Care was taken to clearly mark the non rigorous steps
in the discussion. The main result of the
calculation performed here was announced in the note
\cite{Ja}. The case with insertion points will be
treated in a separate publication.
\vskip 1cm

\nsection{\hspace{-.6cm}.\ \ Space of orbits}
\vs 0.5cm

We shall need below an effective description of generic orbits
\s$\CG^\NC\hs{-0.1cm}A^{01}\s$. It will be based on the fact that
every h.v.b. $E\s$ has a line subbundle \s$L^{-1}
\subset E\s$ or,
equivalently, that the cocycle \s$(g_{\alpha\beta})\s$ of \s$E\s$
may be always chosen in the triangular form
\qq
g_{\alpha\beta}\s=\left(\matrix{a_{\alpha\beta}^{-1}&
b_{\alpha\beta}\cr 0&a_{\alpha\beta}}\right)\ ,
\qqq
where $(a_{\alpha\beta})\s$ is a \s$1$-cocycle of a holomorphic
line bundle (h.l.b.) $L\s$ s.\s t. the dual bundle \s$L^{-1}
\subset E\s$. \s$(b_{\alpha\beta})\s$
satisfy the twisted cocycle condition
\qq
a_{\alpha\beta}^{-1} b_{\beta\gamma}+b_{\alpha\beta}\m
a_{\beta\gamma}\s=\s b_{\alpha\gamma}
\label{cocycle2}
\qqq
which means that they define a holomorphic
1-cocycle with values in the \m h.l.b. \m$L^{-2}\s$
(this may be better seen by rewriting Eq.\s\s(\ref{cocycle2})
\s as \s\s$a_{\alpha\beta}^{-2}
\m b_{\beta\gamma}'+b_{\alpha\beta}'
=b_{\alpha\gamma}'\s\s$, \s where \s$b_{\alpha\beta}'\equiv
a_{\alpha\beta}^{-1}\m b_{\alpha\beta}\m$)\m.
\s The corresponding
cohomology class \s$[(b_{\alpha\beta})]\s$ in
\s$H^1(L^{-2})\s$ describes
\s$E\s$ as an extension
\qq
0\s\longrightarrow\s L^{-1}\s\longrightarrow\s E
\s\longrightarrow\s L\s\longrightarrow\s0
\qqq
of the line bundle \s$L\s$ by \s$L^{-1}\s$. Proportional
\s$[(b_{\alpha\beta})]\s$ give rise to isomorphic bundles
(but the converse may be not true).
By the Riemann-Roch Theorem,
\qq
{\rm dim}\m(H^1(L^{-2}))\s=\s
g-1+2\s{\rm deg}\m(L)\hs{1cm}{\rm for\hs{1cm}deg}\m(L)>0\ .
\qqq
\vs 0.3cm

Let \s$L(\pm x)\s$ denote the h.l.b. \s$L\CO(\pm x)\s$ (omitting
the sign of the tensor product between the bundles),
\s where \s$\CO({\pm x})\s$ is the degree \s$\pm1\s$ h.l.b.
with divisor \s${\pm x}\s$. We shall fix for the rest of the paper
a h.l.b. \s$L\s$
of degree \s$g\s$. \s For later convenience, we shall assume
that this is done so that \s$L(-x)^2\s$ never
coincides\footnote{as opposed to the choice of \s$L\s$
employed in \cite{Ja}} with the
canonical line bundle \s$K\s$ of \s$\Sigma\s$. For any rank 2 h.v.b.
bundle \s$E\s$ with trivial determinant, there exists
\s${x}\in\Sigma\s$ and a non-trivial homomorphism
\qq
\phi:\s L(-x)^{-1}\s\longrightarrow\s E\ ,
\qqq
see \cite{NaraRama}, Lemma 5.4. If \s$\phi\s$ has zeros
(counted with multiplicities) at \s${x}_1\m,\s\dots\m,\s{x}_r\s$
then \s$\phi\s$ induces an embedding of
\s\s$L(-x-x_1\dots-x_r)^{-1}\s\s$ into \s$E\s$ or,
\m in other words, \s$E\s$ is an extension of
\s$L(-x-x_1\dots-x_r)\s$. Notice that
\s${\rm deg}\m(L(-x-x_1\dots-x_r))=g-1-r\s$.
If \s$E\s$ is
stable then in can have only negative degree subbundles so
that, necessarily, \s$r<g-1\s$
and, moreover, the extension has to be nontrivial.
\vs 0.5cm

The above discussion gives rise to the following description
of the orbits \s$\CG^\NC\hs{-0.1cm}A^{01}\m$, related
to the picture of the moduli space \s$\CN\s$ discussed in
the papers \cite{Bertram}\cite{Thad}.
For \s$0\leq r<g-1\s$, consider
a holomorphic family \s$(L(-x-x_1\dots-x_r))\s$
of line bundles. By definition, it is a holomorphic
line bundle \s$\CL_r\s$
over\footnote{We denote by \s$\Sigma^n\s$ the symmetrized
\s$n$-fold Cartesian product of \s$\Sigma\s$.}
\s$\Sigma^{r+1}\times\Sigma\s$ whose restriction
to the fiber \s$pr_1^{-1}(\{x,x_1,\dots,x_r\})\s$
of the projection on the first factor gives
\s$L(-x-x_1\dots-x_r)\equiv L(-X_r)\s$.
\s$\CL_r\s$ is not unique:
for each h.l.b. $M\s$ over \s$\Sigma^{r+1}\s$,
\s we may take \s${pr_1}^*(M)\m\CL_r\s$ instead
of \s$\CL_r\s$. \s Let \s$W_r\s$
denote the first direct image \s$R^1{pr_1}_*(\CL_r^{-2})\s$
of \s$\CL_r^{-2}\s$ by \s$pr_1\s$ (\s$W_r\s$ is a h.v.b.
of dimension \s$N-2r\s$ (\s$N\equiv3g-3\s$)
\s over \s$\Sigma^{r+1}\s$ with fibers
\s$H^1(L(-X_r)^{-2})\s$)\m. \s
Let \s$\NP W_r\s$ denote the corresponding holomorphic
bundle of projective spaces
\s$\NP H^1(L(-X_r)^{-2})\s$.
The total dimension of the compact complex manifold
\s$\NP W_r\s$ is \s$N-r\s$. Now, each element
\s$w\in \NP W_r\s$, with the base point
\s$\{{x},{x}_1,\dots,{x}_r\}\equiv X_r\s$,
\s defines (uniquely up to isomorphisms) a
holomorphic bundle \s$E\s$ of rank 2 and
trivial determinant which is an
extension of \s$L(-X_r)\s$.
(Many \s$w\s$'s may define the same \s$E\s$.\m) \s
The dimensions imply that, generically, \s$r=0\s$
(recall that \s${\rm dim}(\CN)=N\s$)\m. \s
{}From this and the analysis of \cite{Bertram} and
\cite{Thad}, the following picture of the orbit space
emerges:
\vs 0.3cm

\no\hspace*{0.6cm}\parbox{14.25cm}{an open dense subset
of \s$\NP W_0\s$ corresponding
to stable bundles is a ramified (\m$2g$-fold) cover of a dense
subset of the stable moduli space \s$\CN_s\s$. The rest
of \s$\CN_s\s$ is in the image of
subsets of \s$\NP W_r\s$. In particular,
the union of the \s$\CG^\NC$-orbits corresponding
to h.v.b.'s \s$E\s$ obtained from \s$W_0\s$ is dense
in \s$\CA^{01}\s$.}
\vs 0.3cm

\no Other details of
that geometry may be found in \cite{Bertram}\cite{Thad}.
\vs 0.5cm

We shall construct gauge fields corresponding to points
in \s$\NP W_0\s$ (\m a slice
\s$s:\NP W_0\rightarrow\CA^{01}\s$)\m.
Let us start by an explicit construction
of a family \s$(L(-x))\s$ of line
bundles on \s$\Sigma\s$ and of the corresponding bundle \s$W_0\s$.
\s A more natural but less explicit construction
will be discussed in the next section and, a somewhat pedantic,
distinction between different realizations of the family
\s$(L(-x))\s$ will later play an important role.
In order to describe the first construction, fix a point
\s${x}_0\in\Sigma\s$ and denote \s$L_0\equiv L(-x_0)\s$. \s
The family \s$(L(-x))\s$ will be obtained by twisting
the \s$\de$-operator in \s$L_0\s$.
Let \s$(\omega^i)_{i=1}^g\s$ be the basis of holomorphic
\s$0,1$-forms on \s$\Sigma\s$ adapted to a marking
of \s$\Sigma\s$ i.\s e. to a choice of a standard homology basis
\s$(a_i,b_j)\s$. \s$\int\limits_{a_i}\omega^j=\delta^{ij}\s$
and \s$\int\limits_{b_i}\omega^j=\tau^{ij}\s$ were \s$\tau\s$
is the period matrix. Define a \s$0,1$-form
\qq
a\s=\s \pi \sum\limits_{i,j=1}^g(\smallint_{_{{x}_0}}^{^{{x}}}
\omega^i)\s
({_1\over^{{\rm Im}\s\tau}})_{_{ij}}\s
\m\bar\omega^j\s\equiv\s\pi\s(
\smallint_{_{{x}_0}}^{^{{x}}}\omega)\s
({\rm Im}\s\tau)^{-1}\m\bar\omega\ .
\label{atilde}
\qqq
Notice, that \s$a\equiv a_{{\bf x}}\s$ depends on the lift
\s${\bf x}\s$ of the point \s${x}\s$ to the covering
space \s$\tilde\Sigma\s$ of \s$\Sigma\s$ (with the base point
\s${x}_0\s$)\s. Denote by \s$L_{{\bf x}}\s$
the line bundle \s$L_0\s$ with \s$\de\s$
replaced by \s$\de_{L_\bx}\equiv\de+a_{{\bf x}}\s$. \s It
is a standard fact that all
\s$L_{{\bf x}}\s$ corresponding to the same \s${x}\s$
are isomorphic to \s$L(-x)\s$.
Consider the holomorphic bundle
\s$\tilde\Sigma\times L_0\s$ over \s$\tilde\Sigma
\times\Sigma\s$ with the antiholomorphic derivative
\s$\bar\delta+\de\s$ where \s$\bar\delta\s$ differentiates
in the trivial direction of \s$\tilde\Sigma\s$.
We shall twist \s$\tilde\Sigma\times L_0\s$
by replacing its antiholomorphic derivative by
\s$\bar\delta+\de+a\s$. Denote the resulting h.l.b.
over \s$\tilde\Sigma\times\Sigma\s$ by \s$\tilde \CL_0\s$.
It gives a specific realization of a holomorphic
family \s$(L_{{\bf x}})\s$.
The action of the fundamental group \s$\Pi_1(\Sigma,{x}_0)
\equiv\Pi_1\s$ on \s$\tilde\Sigma\s$ lifts to
an action on \s$\tilde\CL_0\s$ preserving the structure
of the h.l.b.
The lifted action of \s$p\in\Pi_1\s$ is
\qq
({\bf x}\m,\m l_y)&\mapsto&(p{\bf x}\m,\m
c_p({y})^{-1}\m l_y)
\qqq
for \s$l_y\s$ in the fiber of \s$L_{\bf x}\s$ over
\s$y\in\Sigma\s$, \s where
\qq
c_p\m({y})&=&\ee^{\s2\pi\m i\m\s{\rm Im}\m[\m(\int_{_p}\omega)\s
({\rm Im}\s\tau)^{-1}\m(\int_{{x}_0}^{{y}}\bar\omega)\m]}\ .
\label{cp}
\qqq
Note that \s$c_{p}\s$ is a function
on \s$\Sigma\s$ (it does not
depend on the integration path from \s$x_0\s$ to \s$y\s$)\m. \s
Dividing \s$\tilde\CL_0\s$ by the action of \s$\Pi_1\s$,
\s we obtain a h.l.b. \s$\CL_0\s$ over
\s$\Sigma\times\Sigma\s$, \s the first explicit
realization of the holomorphic family \s$(L(-x))\s$.
\vs 0.5cm

For a line bundle \s$M\s$, we shall denote by \s$\Gamma(M)\s$
the space of smooth sections of \s$L\s$ and by
\s$\wedge^{01}(M)\s$ the space of smooth \s$0,1$-forms with
values in \s$M\s$. \s
The bundle \s$\tilde W_0\s=$ $R^1\m{pr_1}_*\m(\tilde
\CL_0^{-2})\s$
(the first direct image
of \s$\tilde\CL_0^{-2}\s$ under \s$pr_1\s$) \s may be viewed
as the quotient of the infinite-dimensional
trivial bundle \s$\tilde\Sigma\times
\wedge^{01}(L_0^{-2})\s$ by the subbundle
whose fiber over \s${\bf x}\s$ is the image by
\s\s$\de_{L^{-2}_{\bf x}}\equiv\de-2a_{{\bf x}}\s\s$
of \s$\Gamma(L_0^{-2})\s$.
Indeed,
\qq
\wedge{}^{01}(L_0^{-2})\m/\m\de_{L^{-2}_{\bf x}}
(\Gamma(L_0^{-2}))\s=\s\wedge{}^{01}(L_{{\bf x}}^{-2})\m/\m
\de_{L^{-2}_{\bf x}}
(\Gamma(L_{{\bf x}}^{-2}))\s\cong\s H^1(L_{{\bf x}}^{-2})\ ,
\qqq
which is the Dolbeault realization of \s$H^1(L_{{\bf x}}^{-2})\s$.
\s Division by the action of \s$\Pi_1\s$ \s gives
an explicit construction of the fiber bundle \s$W_0\s$.
\vs 0.5cm

Let us construct now a gauge field \s$A^{01}\s$
whose \s$\CG^\NC$-orbit corresponds to a
given point \s$w\in W_0\s$. To this end, we
shall choose a smooth isomorphism \s$U\s$
of rank 2 vector bundles over \s$\Sigma\s$
with trivial determinants,
\qq
U:\s L_0^{-1}\oplus L_0\s\longrightarrow
\s\Sigma\times\NC^2\ .
\label{U}
\qqq
Let us twist the holomorphic structure
of the vector bundle \s$E_0\equiv
L_0^{-1}\oplus L_0\s$
by replacing its \s$\de$-operator by
\qq
\de\s+\s\left(\matrix{_{-a_{\bf x}}&_b\cr^0&^{a_{\bf x}}}
\right)\s\equiv\s
\de+B^{01}_{{\bf x},b}\ ,
\qqq
where
\s$b\in\wedge{}^{01}(L_0^{-2})\s$. \s
We shall denote the twisted bundle by \s$E\s$.
Note that \s$E\s$ is an extension of the
line bundle \s$L_{\bf x}\s$ by \s$L_{\bf x}^{-1}\s$.
We may use the smooth isomorphism \s$U\s$ of (\ref{U})
to transport the holomorphic structure from
\s$E\s$ to the trivial bundle where we get
the \s$\de$-operator
\qq
\de\s+U\m B_{{\bf x},b}^{01}\m U^{-1}+U\de U^{-1}
\s\equiv\s\s\de+A^{01}_{{\bf x},b}\ .
\label{TRAN}
\qqq
Let \s$c\s$ be a constant \s$\not=0\s$
or \s$c=c_p\s$, see Eq.\s\s(\ref{cp}), and let
\s$v\in\Gamma(L_0^{-2})\s$.
\qq
g_{c,v}\s\equiv\s\left(\matrix{_{c^{-1}}&_{cv}
\cr ^0&^c}\right)
\label{2.13}
\qqq
is a smooth section of the bundle \s${\rm Aut}(L_0^{-1}
\oplus L_0)\s$ of automorphisms of \s$L_0^{-1}
\oplus L_0\s$. The gauge transformation
\s\s\s$B^{01}_{{\bf x},b}\s\s\mapsto\s\s {}^{g_{c,v}^{-1}}
\hs{-0.1cm}B^{01}_{{\bf x},b}\m
\m\equiv\m g_{c,v}^{-1}\m B^{01}_{{\bf x},b}\s g_{c,v}+
g_{c,v}^{-1}\de g_{c,v}\s\s\s$
preserves the form of the gauge field \s$B^{01}_{\bx,b}\s$
shifting
\qq
a_{\bf x}\s\mapsto\s a_{\bf x}+c^{-1}\de c
\hs{1cm}{\rm and}\hs{1cm}
b\s\mapsto c^2(b+(\de-2a_\bx)\m v)\ .\label{shift}
\qqq
In particular, for \s$c=c_p\s$,
\m\s$a_\bx\m\mapsto\m a_{\bf x}+c_p^{-1}\de
c_p=a_{p{\bf x}}\s$.
The corresponding fields \s$A^{01}_{\bx,b}\s$
are gauge-related by
\qq
h_{c,v}\equiv U\m g_{c,v}\m U^{-1}\in\CG^\NC\ ,\label{h}
\qqq
so that they lie in the same \s$\CG^\NC$-orbit.
Taking constant \s$c\not=0\s$, \s
we see that \s$b\s$ leading to the same class in the
projective space
\qq
\NP(\s\wedge{}^{01}(L_0^{-2})
\m/\m\de_{L_{\bf x}^{-2}}
(\Gamma(L_0^{-2}))\s)\s\s\cong\s\s \NP H^1(L_{\bf x}^{-2})
\qqq
give gauge fields \s$A^{01}_{\bx,b}\s$ in the same \s$\CG^\NC$-orbit.
\s The class of \s$b\s$ in \s$\NP H^1(L_{\bf x}^{-2})\s$
is exactly the one describing the rank 2 h.v.b. \s$E\s$,
\s an extension of \s$L_{\bf x}\s$, \s
associated to the orbit \s$\CG^\NC\hs{-0.1cm}A^{01}_{{\bf x},b}\s$.
\s Choosing \s$\bf x\s$ in a fundamental
domain in \s$\tilde\Sigma\s$ and one \s$b\s$ in each class of
\s$\NP H^1(L^{-2}_{\bf x})\s$, \s we obtain a slice \s$s:\NP W_0
\rightarrow\CA^{01}\s$ which cuts a generic orbit
a finite number (\s$=2g\s$) of times.\s One may take \s$s\s$
to be piecewise holomorphic.
\vskip 1cm

\nsection{\hspace{-.6cm}.\ \ Determinant bundle}
\vs 0.5cm

Let us fix a hermitian structure on the h.l.b. $L_0\s$.
\s It induces a metric connection on \s$L_0\s$ whose
covariant derivative in the antiholomorphic direction
coincides with the \s$\de\s$ operator. Let \s$F_0\s$ denote
the curvature form of this connection (normalized so that
\s$\int_{_\Sigma}{i\over 2\pi}F_0={\rm deg}(L_0)=g-1\s$)\m.
\s The hermitian metric
on \s$L_0\s$ induces a hermitian structure and a
connection on \s$L_0^{-1}\s$ and, putting
both together, a hermitian structure and a connection
on \s$L_0^{-1}\oplus L_0\s$.
Let us denote by \s$\nabla\s$ the holomorphic covariant derivative
in \s$L_0^{-1}\oplus L_0\s$. The
complete covariant derivative is \s$\nabla+\de\s$.
Clearly, its curvature form
\qq
{\rm curv}(\nabla+\de)\s=\left(\matrix{_{-F_0}&_0\cr^0&^{F_0}}
\right)\ .
\qqq
Let us assume that the smooth isomorphism \s$U:L_0^{-1}\oplus
L_0\rightarrow\Sigma\times\NC^2\s$ maps the hermitian
metric of \s$L_0^{-1}\oplus L_0\s$ into the one
coming from the standard scalar product of \s$\NC^2\s$.
\s Using \s$U\s$, \s we may transport the connection
on \s$L_0^{-1}\oplus L_0\s$
to the trivial bundle \s$\Sigma\times\NC^2\s$:
\qq
U(\nabla+\de)U^{-1}=d+U\nabla U^{-1}+U\de U^{-1}\s\equiv\s
d+A^{10}_0+A^{01}_0\equiv d+A_0\ .
\qqq
The right hand side gives a unitary connection, so that
\s$A^{10}_0=-(A_0^{01})^\dagger\s$.
The curvature forms are related by
\qq
{\rm curv}(U(\nabla+\de)U^{-1})\s=\s
F(A_0)\equiv dA_0+A_0\hs{-0.1cm}\wedge\hs{-0.08cm}A_0
\s=\s U\left(\matrix{_{-F_0}&_0\cr^0&^{F_0}}\right)U^{-1}\ .
\qqq
\vs 0.5cm

We shall represent the CS states by holomorphic sections of a
line bundle over \s$\NP W_0\s$.
To this end, let us define, for a CS state \s$\Psi\s$, a
holomorphic function
\qq
\psi({\bf x},b)\s\ \equiv\m\s\ \exp[\s{_{ik}\over^{2\pi}}
\smallint_{_\Sigma}\tr\s\s A^{10}_0\hs{-0.1cm}\wedge
\hs{-0.08cm}A^{01}_{{\bf x},b}]\s\ \s
\Psi(A^{01}_{{\bf x},b})
\label{newCS}
\qqq
of
\s${\bf x}\in\tilde\Sigma\s$
and \s$b\in\wedge^{01}(L_0^{-2})\s$. As is shown
in Appendix F, only the normalization of \s$\psi\s$ depends
on the choice of the hermitian structure on \s$L_0\s$
and of the isomorphism \s$U\s$. \s
Since the \s$\CG^\NC$-orbits of the chiral gauge
fields \s$A^{01}_{{\bf x},b}\s$ form a dense
subset of \s$\CA^{01}\s$, \s$\Psi\s$ is uniquely
determined by the function \s$\psi\s$.
The gauge relations between the forms \s$A^{01}_{{\bf x},b}\s$
induce constraints for functions \s$\psi\s$,
\s due to the gauge invariance of CS states \s$\Psi\s$.
\s In particular,
\qq
\hbox to 3.2 cm{$\psi({\bf x}\m,\m b+
(\de-2a_{{\bf x}})v)$\hfill}
&=&\s\s\exp[\s kS(h_v,\m A^{10}_0+A^{01}_{{\bf x},
b})\s]\s\hs{0.1cm}\
\s\psi({\bf x},b)\ ,\label{1}\\
\hbox to 3.2 cm{$\psi({\bf x}\m,\m c^{2}b)$\hfill}&=&\s\s
\exp[\s kS(h_c,\m A^{10}_0+A^{01}_{{\bf x},b})\s]\s\
\hs{0.13cm}\s\psi({\bf x},b)\ ,\label{2}\\
\hbox to 3.2 cm{$\psi(p{\bf x}\m,\m c_p^{2}b)$\hfill}&=&\s\s
\exp[\s kS(h_{c_p},\m A^{10}_0+A^{01}_{{\bf x},b})\s]\s\
\s\psi({\bf x},b)\ ,\label{3}
\qqq
where \s$h_v\s$ is given by Eq.\s\s(\ref{h}) with
\s$c=1\s$, \s$h_c\s$
by the same formula with \s$v=0\s$ and
\s$c\in\NC^\times\equiv\NC\setminus\{0\}\s$
and \s$h_{c_p}\s$ by setting \s$v=0\s$
and \s$c=c_p\s$, \s see (\ref{cp}).
\vs 0.5cm

Let us study these transformation properties in greater detail.
The first two equations  become much more transparent
if we rewrite them in the infinitesimal form. Since
\s\s\s${_\delta\over^{\delta h}}|_{_{h\equiv 1}}\s\s
S(h,A)\s=\s{_{i}\over^{2\pi}}
\s F(A)\ ,\s\s$ we obtain
\qq
&{_\delta\over^{\delta v}}|_{_{v\equiv 0}}\s\m
S(h_v,\m A_0^{10}\hs{-0.05cm}+\hs{-0.05cm}A^{01}_{{\bf x},b})
={_i\over^{2\pi}}\s\s\tr\s\s\sigma^+U^{-1}F(\hs{-0.03cm}
A^{10}_0\hs{-0.05cm}+\hs{-0.05cm}A^{01}_{{\bf x},b})\s U
\hs{-0.1cm}&\cr
&={_i\over^{2\pi}}\s\s\tr\s\s\sigma^+
({\rm curv}(\nabla\hs{-0.05cm}+\de)\hs{-0.02cm}+\hs{-0.05cm}
\nabla(B^{01}_{{\bf x},b}))
=\s{_i\over^{2\pi}}\s\s\tr\left(\hs{-0.08cm}
\matrix{_0&\hs{-0.1cm}_1\cr^0&\hs{-0.1cm}^0}\hs{-0.13cm}\right)
\hs{-0.1cm}\left(\hs{-0.05cm}\left(\hs{-0.1cm}
\matrix{_{-F_0}&\hs{-0.13cm}_0\cr^0&\hs{-0.13cm}^{F_0}}
\hs{-0.16cm}\right)\hs{-0.05cm}+
\hs{-0.05cm}\left(\hs{-0.1cm}
\matrix{_{-\da a_{{\bf x}}}&\hs{-0.14cm}_{\nabla(b)}\cr^0&
\hs{-0.14cm}^{\da a_{{\bf x}}}}\hs{-0.14cm}\right)
\hs{-0.05cm}\right)
\s=\s\s 0\s\m,\hs{0.76cm}&\label{1p}\\
\nonumber
\qqq
where we have used the identity\m: \s\s${\rm curv}
(\nabla+\de+B^{01}_{{\bf x},b})\s=\s
{\rm curv}(\nabla+\de)+\nabla(B^{01}_{{\bf x},b})\s$.
\s Above, \s$\nabla(b)\s$ stands for the holomorphic
covariant derivative of the \s$L_0^{-2}$-valued
\s$0,1$-form \s$b\s$. \s Similarly,
\qq
{_\delta\over^{\delta c}}|_{_{c=1}}\s\s
S(h_c,\s A_0^{10}+A^{01}_{{\bf x},b})
\s=\s{_1\over^{2\pi i}}\int_{_\Sigma}
\s\tr\left(\hs{-0.08cm}
\matrix{_1&\hs{-0.1cm}_0\cr^0&\hs{-0.1cm}^{-1}}\hs{-0.15cm}\right)
\hs{-0.1cm}\left(\hs{-0.05cm}\left(\hs{-0.1cm}
\matrix{_{-F_0}&\hs{-0.13cm}_0\cr^0&\hs{-0.13cm}^{F_0}}
\hs{-0.16cm}\right)\hs{-0.05cm}+
\hs{-0.05cm}\left(\hs{-0.1cm}
\matrix{_{-\da a_{{\bf x}}}&\hs{-0.14cm}_{\nabla(b)}\cr^0&
\hs{-0.14cm}^{\da a_{{\bf x}}}}\hs{-0.14cm}\right)
\hs{-0.05cm}\right)\ \m
\cr\cr
\hs{7cm}\s=\s{_i\over^{\pi}}
\int_{_\Sigma}\hs{-0.05cm} F_0\s=\s 2(g-1)\ .
\label{2p}
\qqq
As for the relation (\ref{3}),
first notice that we may reduce the calculation
to the case \s$b=0\s$ since for \s$h=h_{c_p}\s$,
\qq
S(h,\m A^{10}_0\hs{-0.09cm}+\hs{-0.07cm}
A^{01}_{{\bf x},b})=
S(h,\m A^{10}_0\hs{-0.09cm}+\hs{-0.07cm}
A^{01}_{{\bf x},0})
+\hs{-0.02cm}{_{i}\over^{2\pi}}
\hs{-0.06cm}\int_{_\Sigma}
\hs{-0.05cm}\tr\m\s[(h\da h^{-1}\hs{-0.05cm}
+\hs{-0.03cm}h\hs{-0.09cm}A_0^{10} h^{-1}\hs{-0.05cm}-
\hs{-0.09cm}A^{10}_0)
\hs{-0.04cm}\wedge\hs{-0.04cm}
U\sigma^+b\m\hs{-0.02cm} U^{-1}]\cr
=\m S(h,\m A^{10}_0\hs{-0.09cm}+
\hs{-0.07cm}A^{01}_{{\bf x},0})
+{_{i}\over^{2\pi}}\hs{-0.05cm}
\int_{_\Sigma}\hs{-0.03cm}\tr\m\s[\sigma^3\sigma^+
c_{p}^{-1}\da c_{p}\hs{-0.08cm}\wedge\hs{-0.05cm}b\m]
\m=\m  S(h,\m A^{10}_0
\hs{-0.09cm}+\hs{-0.07cm}A^{01}_{{\bf x},0})\ .
\hs{1cm}\nonumber
\qqq
We shall show in Appendix A that for \s$c\s$
a non-vanishing function on \s$\Sigma\s$, \s for
\s$g_c\s=\left(\matrix{_{c^{-1}}&_0\cr^0&^c}\right)\s$
and for \s$h_c=U\m g_c U^{-1}\s$,
\qq
\exp[\s S(h_{c},\s A^{10}_0+A^{01}_{{\bf x},0})]\s=
\s\ee^{{_1\over^{2\pi i}}\int_{_\Sigma} c^{-1}\da c\wedge
\m(c^{-1}\de c+2a_{{\bf x}})}\s\ \nu\m(c)\ .
\qqq
Note that, except for the last factor, the expression
on the right hand side is
\s$\ee^{\s S(g_c,-\sigma^3 a_{{\bf x}})}\s$
where \s$g_c\s$ is viewed as a standard \s$SL(2,\NC)$-valued
field. The correction term is
\qq
\nu\m(c)\s=\s
\ee^{\s{i\over\pi}\smallint_{_\Sigma}F_0\s{\rm ln}\m c}\s\s\prod
\limits_{j=1}^g\bigg(W_{a_j}^{-{_i\over^\pi}
\smallint_{_{b_j}}c^{-1}dc}\s W_{b_j}^{{_i\over^\pi}
\smallint_{_{a_j}}c^{-1}dc}\bigg)\ ,
\label {corrt}
\qqq
where \s$W_{a_j}\s$ (\s$W_{b_j}\s$)
stand for the holonomy of the metric connection on
\s$L_0\s$ along the \s$a_j\s$ ($\m b_j\m$) cycle.
\s${\rm ln}\m c\m(x)=\int_{_{x_0}}^{^x}c^{-1}dc\s$,
\s where the integration
path is taken inside a fundamental domain of
\s$\Sigma\s$ obtained by cutting the surface along
the cycles \s$a_j,\ b_j\s$ starting at \s$x_0\s$.
Altogether,
\qq
\exp[\s S(h_{c_p},\m A^{10}_0+A^{01}_{{\bf x},b})\s]\s
\s=\s\s\ee^{\s\pi\s(\int_{_p}\bar\omega)\s
({\rm Im}\s\tau)^{-1}\m(\int_{_p}\omega)\s\m+\s\m 2\pi
\s(\int_{_p}\bar\omega)\s({\rm Im}\s\tau)^{-1}\m
(\int_{_{x_0}}^{^{{\bf x}}}\omega)}\ \s\nu\m(c_p)\ \s\cr
\equiv\ \mu\m(p,{\bf x})\ \nu\m(c_p)\ .\hs{1cm}
\label{3p}
\qqq
Gathering Eqs. (\ref{1}) and (\ref{1p}), \s(\ref{2}) and
(\ref{2p}), \s(\ref{3}) and (\ref{3p}), we obtain
\vs 1cm

\no\hspace*{0.6cm}\parbox{13.8cm}{{\bf PROPOSITION}. \s
{\it Holomorphic
functions \s$\psi\s$ possess the following transformation
properties}:}
\qq
\hbox to 3.5cm{$\psi({\bf x}\m,\s \lambda b+
(\de-2a_{{\bf x}})v\m)$
\hfill}&=&\lambda^{k(g-1)}\s\s\s\m\psi({\bf x}\m,\s b)\ ,
\label{5}\\
\hbox to 3.5cm{$\psi(p\m{\bf x},\s c_{p}^{2}b)$\hfill}
&=&\mu(p,\bx)\ \s\nu\m(c_{p})^k
\ \s\psi({\bf x}\m,\s b)\ .
\label{6}
\qqq
\vskip 0.8cm

In particular, for fixed \s$\bx\s$, \s$\psi(\bx,\m\cdot\m)\s$
is a homogeneous polynomial of degree \s$k(g-1)\s$ on
\s$H^1(L_\bx^{-2})\s$. \s Note that the
factor \s$\mu(p,{\bf x})\s$ appears in the
transformation property for the square of Riemann's theta
function:
\qq
\ee^{\m\pi k\m(\smallint_{_{x_0}}^{^{p{\bf x}}}\hs{-0.08cm}\omega)
\s{1\over{\rm Im}\s\tau}\s(\smallint_{_{x_0}}^{^{p{\bf x}}}
\hs{-0.08cm}\omega)
}\ \s\vartheta(\smallint_{_{x_0}}^{^{p{\bf x}}}\hs{-0.08cm}
\omega\s|\s\tau\m)^2
\s=\s\s\mu(p,{\bf x})\s\s\s\m\ee^{\m\pi k\m
(\smallint_{_{x_0}}^{^{{\bf x}}}\hs{-0.08cm}\omega)
\s{1\over{\rm Im}\s\tau}\s(\smallint_{_{x_0}}^{^{{\bf x}}}
\hs{-0.08cm}\omega)
}\ \s\vartheta(\smallint_{_{x_0}}^{^{{\bf x}}}\hs{-0.08cm}
\omega\s|\s\tau\m)^2\m.\ \
\qqq
The map \s\m${\bf x}\m\mapsto\m\vartheta
(\smallint_{_{x_0}}^{^{{\bf x}}}\hs{-0.08cm}
\omega\s|\s\tau\m)^2\s\m$ defines
a holomorphic section of the bundle
\s$K(2x_0)\s$. The map \s$\Pi_1\ni p\s\mapsto\s\nu\m(c_p)\s$
is a character of the fundamental group \s$\Pi_1\s$.
We show in Appendix A that it defines the flat bundle
\s$L(-g x_0)^2\s$. Hence a holomorphic function
\s$\phi({\bf x})\s$ s.\s t.
\qq
\phi(p{\bf x})=\mu(p,{\bf x})\ \nu\m(c_p)\ \phi({\bf x})
\label{PhI}
\qqq
defines \s\s(\s upon multiplication by
\s\s\s$\ee^{-\pi k\m
(\smallint_{_{x_0}}^{^{{\bf x}}}\hs{-0.08cm}\omega)
\s(\m{\rm Im}\s\tau\m)^{-1}(\smallint_{_{x_0}}^{^{{\bf x}}}
\hs{-0.08cm}\omega)}\s$) \s\s a section of the line bundle
\s$L^2 K((2-2g)x_0)\s$.
\s The transformation properties (\ref{5}) and (\ref{6})
may be rephrased by saying that \s$\psi\s$
is a holomorphic section of the \s$k^{\rm th}\s$ power of
a h.l.b., which we shall
denote \s${\rm DET}\s$, \s over the total space
of the fiber bundle \s$\NP W_0\s$. Explicitly,
\qq
{\rm DET}\s=\s\varpi^*
(L^2 K((2-2g)x_0))\s\s\s{\rm Hf}\m(W_0)^{(1-g)}
\label{DeT}
\qqq
where \s$\varpi\s$ is the bundle projection of \s$\NP W_0\s$
and \s${\rm Hf}\m(W_0)\s$ is the
tautological bundle over \s$\NP W_0\s$. \s
In particular, the restriction
of the h.l.b. ${\rm DET}\s$ to the fiber \s$\NP H^1(L(-x)^{-2})\s$
of \s$\NP W_0\s$ over \s$x\in\Sigma\s$ is the \s$(1-g)^{\rm th}\s$
power of the tautological bundle over \s$\NP H^1(L(-x)^{-2})\s$.
\s As we shall see in the next section, the h.l.b.
$L^2 K((2-2g)x_0)\s$ is isomorphic to
(the \s$\Pi_1\s$ quotient of) the determinant bundle of the
family \s$(\de-\sigma^3a_{\bf x})\s$ of \s$\de$-operators
in \s$L_0^{-1}\oplus L_0\s$. \s
In turn, the line bundle \s${\rm DET}\s$ is isomorphic to
(the \s$\Pi_1$ quotient of) the determinant bundle for
the family \s$(\de+B^{01}_{{\bf x},b})\s$ of
\s$\de$-operators in \s$L_0^{-1}\oplus L_0\s$.
Note the simple way by which the addition of the
upper-diagonal gauge field \s$b\s$ in the \s$\de$-operator
manifests itself in the determinant bundle.
The map \s$\Psi\s\mapsto\s\psi\s$ embeds
the space of CS states onto a subspace
\s$\CH\subset H^0({\rm DET})\s$.
The space \s$H^0({\rm DET})\s$ of the holomorphic sections
of a line bundle over the compact space \s$W_0\s$ is
finite dimensional so that the finite-dimensionality
of the space of CS states follows. Given
\s$\psi\in H^0({\rm DET})\s$,
the question whether it comes from a CS state should be
determined by its behavior at the codimension one
subvariety defined by \s$\NP W_1\s$ in the moduli space
\s$\CN_s\s$ of stable bundles. We shall return below to this
question which requires some refinement of the analysis of
\cite{Bertram}\cite{Thad}.
\vs 0.5cm

For the later use, it will be convenient to write
the homogeneous polynomial \s$\psi(\bx,\m\cdot\m)\s$
on \s$H^1(L_\bx^{-2})\s$ in an integral form
following from the Serre duality:
\qq
\psi(\bx,b)\s\s\s=\s\int\limits_{\Sigma^{k(g-1)}}
\chi(\bx;\m x_1,\m\dots\m,\m x_{k(g-1)})\s\s\s b(x_1)\s\cdots
\s b(x_{k(g-1)})
\label{hoMO}
\qqq
for \s$b\in\wedge^{01}(L_0)\s$ and
\s$\chi(\bx;\m\cdot\m)\in H^0(S^{k(g-1)}(L_\bx^{2}K))\s$,
\s where, for a h.l.b. $M\s$ over \s$\Sigma\s$, \s$S^nM\s$
stands for the \s$n$-fold (symmetrized) tensor product
with the bases space \s$\Sigma^n\s$, \s the \s$n$-fold
symmetric Cartesian product of \s$\Sigma\s$.
The \s$\bx$-dependence of \s$\chi\s$ again gives rise to
a section of \s$L^2K((2-2g)x_0)\s$ so that we may view
\s$\chi(\m\cdot\s;\m\cdot\m)\s$ as a holomorphic,
\s${pr_2}$-horizontal \s$k(g-1),0$-form on
\s$\Sigma\times\Sigma^{k(g-1)}\s$ with values in
the h.l.b.
\qq
{pr_1}^*(L^2 K((2-2g)x_0))^k\s\s S^{k(g-1)}({\CL_0}^2)
\s\s\s\equiv\s\s\s B_k\ .
\label{B}
\qqq
\vskip 1cm

\nsection{\hspace{-.6cm}.\ \ Relation to Bertram's picture}
\vs 0.5cm

Paper \cite{Bertram}, see also \cite{Thad}, describes a somewhat
different, simpler, construction of the space of CS states.
It is based on the realization of a generic rank $2$ determinant
$0$ bundle as an extension of a fixed degree \s$g\s$
line bundle \s$L\s$. Taking
\qq
{B'}^{01}_{b'}\equiv\left(
\matrix{_0&_{{b'}}\cr^0&^0}\right)\ ,
\qqq
where \s$b'\in\wedge^{01}(L^{-2})\s$, we shall put
\qq
{A'}^{01}_{b'}=\m U'\m {A'}^{01}_{b'}
{U'}^{-1}\m+\m U'\de{U'}^{-1}
\qqq
where \s\s\s$U'\m:\s L^{-1}\oplus L\s
\rightarrow\s\Sigma\times{\NC^2}\s\s\s$
is a fixed smooth isometric isomorphism. Defining
\qq
\psi'(b')\m\s=\s\exp[\s{_{ik}\over^{2\pi}}
\smallint_{_\Sigma}\tr\s\s {A'}^{10}_0\hs{-0.1cm}\wedge
\hs{-0.08cm}{A'}^{01}_{b'}]\s\ \s
\Psi({A'}^{01}_{b'})\ ,
\label{vnewCS}
\qqq
we infer in the same way as above that
\qq
\psi'(\lambda(b'+\de v'))=\lambda^{kg}\s
\psi'(b')\ ,
\label{6prime}
\qqq
compare (\ref{5}). Hence each \s$\psi'\s$ is
a degree \s$kg\s$ homogeneous polynomial on \s$H^1(L^{-2})\s$.
The latter space has dimension \s$3g-1\s$ and the number of
independent homogeneous polynomials of degree \s$kg\s$ on it
is \s$({kg+3g-2\atop 3g-2})\s$. By the Serre duality,
we may write
\qq
\psi'(b')\s=\s\int\limits_{\Sigma^{kg}}
\chi'(x_1,\dots,x_{kg})\s\s\m b'(x_1)\s\cdots\s
b'({x_{kg}})
\label{present}
\qqq
where \s$\chi'\in H^0(S^{kg}(L^2K))\s$.
\vs 0.8cm

\no\hspace*{0.6cm}{\bf THEOREM 2a}
of \cite{Bertram}\footnote{We have learned this reformulation
of the result of \cite{Bertram} from B. Feigin who discussed in
\cite{Feigin} its generalization to the case with
insertions and with arbitrary simple groups.}.
\m{\it The CS
states correspond
exactly to the polynomials \s$\psi'\s$ s\s. t.\hfill}\break
\no\hspace*{0.6cm}$\chi'\in H^0(S^{kg}(L^2K))\s$
{\it vanish whenever \s$k+1\s$ of their arguments
\s$x_n\s$ coincide}.
\vs 0.8cm

\no The dimension of the space
of such polynomials was shown in \cite{Thad} to be
\qq
({_{k+2}\over^2})^{^{g-1}}\sum
\limits_{j=0,{1\over2},\dots,{k\over 2}}
(\m\sin\s{_{\pi(2j+1)}\over^{k+2}}\m)^{^{2-2g}}\ ,
\qqq
in agreement with the Verlinde formula \cite{Verl}
for the dimensions of the spaces of CS states.
Let us set
\qq
\chi(x;x_1,\dots,x_{k(g-1)})\s\equiv\s
\chi'(\smash{\mathop{x,\dots,x}\limits_{^{k\
{\rm times}}_{}}},x_1,
\dots,x_{k(g-1)})\ .
\label{chi}
\qqq
Note that
\qq
\chi(x;x,x_2,\dots,x_{k(g-1)})\s=\s0\hs{0.6cm}{\rm and}
\hs{0.6cm}{_{\da}\over^{\da x_1}}|_{_{{x_1=x}}}\s
\chi(x;x_1,\dots,x_{k(g-1)})\s=\s0
\qqq
where the second equality is obtained by differentiating
the first one over \s$x\s$. \s In particular, for fixed
\s$x\s$, \s$\chi(x;\m\cdot\m) \in (L^2K)^k|_x\m\otimes\m
H^0(\m S^{k(g-1)}(L^2K(-2x))\m)\s$. \s We may also
interpret \s$\chi(x;\m\cdot\m)\s$ as a holomorphic
\s$k(g-1),0\m$-form on \s$\Sigma^{k(g-1)}\s$
with values in the line bundle \s$(L^2 K)^k|_x\otimes
S^{k(g-1)}(L^2(-2x))\s$.
\vs 0.5cm

We shall show that giving \s$\chi\s$ is equivalent to giving
\s$\psi\s$ in the description of the CS states
of the previous section. For this purpose,
let us consider a line bundle \s$\CO(-\Delta)\s$
over \s$\Sigma\times\Sigma\s$
where \s$\Delta\s$ denotes the diagonal. \s$\CL_0'\equiv pr_2^*(L)
\m\CO(-\Delta)\equiv{pr_2}^*(L)(-\Delta)\s$ is another
realization of a family
\s$(L(-x))\s$ of the h.l.b.'s, different one
from \s$\CL_0\s$ described before. It is not difficult
to see that
\qq
\CL_0'\s\cong\s pr_1^*(\CO(-x_0))\s\CL_0\ .
\label{notdiff}
\qqq
An explicit isomorphism is given in Appendix B.
\s If \s$W'_0\equiv R^1{pr_1}_*({\CL_0'}^{-2})\s$ then
\qq
W'_0\s\cong\s \CO(2x_0)\m W_0\ ,\hs{0.6cm}\NP W'_0\s
\cong\s\NP W_0\hs{0.6cm}{\rm and}\hs{0.6cm}{\rm Hf}
\m(W_0')\s\cong\s\varpi^*(\CO(2x_0))\s\s
{\rm Hf}\m(W_0)\ .\hs{0.5cm}
\label{IsOm}
\qqq
Now \s$\chi\s$\s, \s with its \s$x$-dependence taken
into account, may be viewed as a holomorphic
\s(\nobreak\s\nobreak$pr_2$-horizontal)
\s$k(g-1),0\m$-form on
\s$\Sigma\times\Sigma^{k(g-1)}\s$ with the values
in the h.l.b.
\qq
B'_k\equiv{pr_1}^*(L^2 K)^k\s\s
S^{k(g-1)}({\CL'_0}^2)\ .
\label{BB}
\qqq
Since the h.l.b.'s \s$B'_k\s$ of Eq.\s\s(\ref{BB}) and
\s$B_k\s$ of (\ref{B}) are isomorphic due to
(\ref{notdiff}), \s$\chi\s$ introduced by formula
(\ref{chi}) is the same type of object as \s$\chi\s$
considered in Sect.\s\s3, see Eq.\s\s(\ref{hoMO}).
Indeed, in Appendix C we show that the two \s$\chi$'s
coincide completely under an explicit isomorphism.
This will establish the precise relation between
the two descriptions of the CS states: by
functions \s$\psi\s$ which we shall employ in this
work and by polynomials \s$\psi'\s$.
Let us briefly sketch here the geometric
picture due to \cite{Bertram}\cite{Thad} which is
at the core of the detailed analysis of
Appendix C. One may embed the curve \s$\Sigma\s$ into
\s$\NP H^1(L^{-2})\s$ by
\qq
\Sigma\ni x\s\mapsto\s[b_x']\in\NP H^1(L^{-2})\ ,
\label{Emb}
\qqq
where, for \s$\eta'\in
H^0(L^2K)\s$, \s \s$\smallint_{_\Sigma}\eta'\m b_x'
=\eta'(x)\s$ in some trivialization of \s$L^2 K\s$
around \s$x\s$. \s We shall see in Appendix C
that \s$[b_x]\s$ corresponds to an an extension
of \s$L\s$ which, as a rank two bundle, is
isomorphic to the split bundle \s$L(-x)^{-1}
\oplus L(-x)\s$. \s The condition of Theorem 2a of
\cite{Bertram} means\footnote{this was the original
formulation of Theorem 2a of \cite{Bertram}}
that \s$\psi'\s$ vanishes
to order \s$k(g-1)-1\s$ on (the image of) \s$\Sigma\s$.
\s In this description, the CS states are thus realized
as homogeneous polynomials \s$\psi'\s$ on
\s$(3g-1)$-dimensional vector space which vanish to some
order on the explicit embedding of \s$\Sigma\s$ into
the space. \s$\psi$'s are given by the
\s$k(g-1)^{\m{\rm th}}\s$ Taylor coefficients
(the first non-trivial order)
of \s$\psi'\s$ on \s$\Sigma\s$.
\vs 0.5cm

Similarly, one may embed \s$\Sigma\s$ into
\s$\NP H^1(L(-x)^{-2})\s$
by mapping \s$y\s$ to \s$[b_y]\s$ s.\s t.
\s$\smallint_{_\Sigma}\eta\m b_y=\eta(y)\s$ for each
\s$\eta\in H^0(L(-x)^2 K)\s$. \s$[b_y]\s$ defines
an extension of \s$L(-x)\s$ which, as a h.v.b.,
is isomorphic to \s$L(-x-y)^{-1}\oplus L(-x-y)\s$.
Changing also \s$x\s$, we get an embedding of
\s$\Sigma^2\s$ into the bundle \s$\NP W_0\s$ considered
in Sect.\s\s2. Replacement of the image of
\s$\Sigma^2\s$ in \s$\NP W_0\s$ by \s$\NP W_1\s$
is the second blow-up step of \cite{Bertram}\cite{Thad}.
Their analysis shows that, again, \s$\chi(x;\m y_1,
\dots,y_{k(g-1)})\s$ vanishes
whenever \s$k+1\s$ of \s$y_n$'s \s coincide.
We may then equate \s$k\s$ of \s$y_n$'s \s and
continue the process including higher and higher
\s$\NP W_r$'s \s into the game. It is interesting to know
whether the vanishing
of \s$\chi(\bx;\m x_1,\dots,x_{k(g-1)})\s$
at \s$k+1\s$ coincident points characterizes completely
the sections \s$\psi\in H^0({\rm DET})\s$ coming from
the CS states.
\vs 0.5cm

Let us remark, \s in the end, \s
that the h.l.b.'s isomorphisms
\s\s${\rm det}\m\s R\m\s{pr_1}_*
(\CL'_0)\cong L^{-1}\s\s$ and
\s\s${\rm det}\s\m R\m\s{pr_1}_*({\CL'_0}^{-1})\s\cong\s
(L\m K)^{-1}\s\s$ following from the exact sequence
$$0\s\longrightarrow\s{pr_2}^*(L)(-\Delta)\s\longrightarrow
\s{pr_2}^*(L)\s\longrightarrow\s{pr_2}^*(L)|_{_\Delta}\s
\longrightarrow\s0$$
and the relation (\ref{notdiff}) imply that
the determinant bundle \s\s${\rm det}^{-1}\m R\s\m\s
{pr_1}_*(\CL_0^{-1}\oplus\CL_0)\s\s$
of the family \s$(\de-\sigma^3 a_{\bf x})\s$
of \s$\de$-operators
in \s$L_0^{-1}\oplus L_0\s$ is
isomorphic to the h.l.b. \s\s$L^2 K((2-2g)x_0)\s$.
\s\s This provides an interpretation of the first
factor on the right hand side of Eq.\s\s(\ref{DeT}).
\vskip 1cm

\nsection{\hspace{-.6cm}.\ \ Projective formula}
\vs 0.5cm

The outcome of the calculation performed
in this paper is much simpler than the calculation
itself. We shall start by describing a softened
version of its result. It will give
the scalar product of CS
states up to a \s$\Sigma$-dependent constant. Such
data are enough to generate a projective connection
on the bundle \s$\CW_k\s$ of state spaces.
This should coincide with the projective class of
the KZB connection and hence be flat.
Sect.\s\s9 contains a more detailed scalar
product formula with the normalization
fixed, up to an overall constant depending only
on the level \s$k\s$ and genus \s$g\s$
(which could be traced
through the calculation). The detailed formula
gives also the dependence of the scalar product
on the metric of the surface
(i.\s e., in particular, on its complex structure).
\vs 0.4cm

In the simplified formula, we shall use,
as the geometric input, the representation
of the CS states by the holomorphic
\s$k(g-1),0$-forms \s$\chi\s$
with values in the line bundle \s$B'_k\s$, \s see (\ref{BB}),
as well as hermitian structures on these bundles.
It will be convenient to choose the latter in a specific
way. Following \cite{AGBMNV}, we
shall call a hermitian metric on a h.l.b. $M\s$
on \s$\Sigma\s$ admissible, if the curvature form of the
induced connection is proportional to the \s$2$-form
\s$\alpha\equiv{i\over 2g}\s
\omega\s({\rm Im}\s\tau)^{-1}\hs{-0.1cm}
\wedge\bar\omega\s$. \s Admissible hermitian structures
exist and are unique up to normalization.
We shall call a Riemannian metric
on \s$\Sigma\s$ admissible
if it induces an admissible hermitian structure on the
holomorphic tangent bundle of \s$\Sigma\s$. \s Let
\s$G(x,y)\s$ denote the Green function of the scalar
Laplacian \s$\Delta\s$ on \s$\Sigma\s$
chosen so that \s$\smallint_{_\Sigma}\hs{-0.05cm}G(\m\cdot\m,
\m y)\m\alpha(y)=0\s$. \s$G(x,y)\s$ has a logarithmic
singularity at coinciding points. We shall define
\qq
:G(x,x):\s=\s{\displaystyle{\lim_{\epsilon\rightarrow 0}}}
\s\m(\m G(x,x')-{1\over 2\pi}\s\ln\m\epsilon\m)\ ,
\label{Wick}
\qqq
with the distance \s$d(x,x')=\epsilon\s$. An admissible
Riemannian metric on \s$\Sigma\s$ normalized so that
\s$:G(x,x):\s\equiv 0\s$ is called the Arakelov metric
\cite{Arakel}.
It will be also convenient to fix the dependence of
the hermitian structures on the parameter for families of
h.l.b.'s on \s$\Sigma\s$. In particular, the line
bundle \s$\CO(\Delta)\s$ over \s$\Sigma\times\Sigma\s$
may be provided with a hermitian metric by
setting
\qq
|1(x,y)|^2\s=\s\ee^{4\pi\s G(x,y)}
\label{G1}
\qqq
for its canonical section \s$1\s$. Fixing also
an admissible hermitian structure on the h.l.b. $L\s$,
we obtain this way a hermitian metric on
\s$\CL'={pr_2}^*(L)(-\Delta)\s$ which may be viewed
as a family of admissible hermitian structures
on the family \s$(L(-x))\s$ of h.l.b.'s realized
as \s$\CL_0'\s$. \s The above choices determine
a hermitian metric on the h.l.b. \s$B_k'\s$
of (\ref{BB}) in which forms \s$\chi\s$
of Eq.\s\s(\ref{chi}) take values.
\vs 0.5cm

Another geometric input in the scalar product formula
comes from the linear map
\qq
H^0(K)\ni\s\nu\s\s\ {\smash{\mathop{\longmapsto}
\limits^{l\m(x,b'')}}}\ \s\s\nu\s\s[b'']\m\in H^1(L(-x)^{-2}K)
\cong H^0(L(-x)^2)^*
\label{RANK}
\qqq
defined for each \s$b''\in\wedge^{01}(L(-x)^2)\s$ and
depending only on the class of \s$b''\s\s$
in \s\s$H^1(L(-x)^{-2})\s$. \s
As we shall see in Sect.\s\s6.3, the rank of this map
controls the (local) regularity of the projection
from \s$\CA^{01}\s$ into the orbit space
\s$\CA^{01}/\CG^\NC\m$.
\s$l\m(x,b'')\s$ may be viewed as an element of the
vector space \s$H^0(K)^*\otimes H^0(L(-x)^2)^*\m$. \s Since
\qq
\wedge\hs{-0.05cm}^{^{g-1}}\hs{-0.05cm}H^0(K)^*
\s\s\cong\s\s H^0(K)\s\otimes\s\m
{\rm det}^{-1}H^0(K)\ ,
\qqq
one may consider \s\s$\wedge\hs{-0.05cm}^{^{g-1}}\hs{-0.05cm}
l\m(\m\cdot\m,b'')\s\s$ as a
holomorphic \s$1,0$-form on \s$\Sigma\s$ with values
in the bundle \s\s${\rm det}^{-1}H^0(K)\m\otimes\m
{\rm det}^{-1}R^0{pr_1}_*{\CL'_0}^2\s$.
Since it depends homogeneously
on \s$[b'']\in H^1(L(-x)^{-2})\s$, we may write
\qq
\wedge\hs{-0.05cm}^{^{g-1}}\hs{-0.05cm}
l\m(x,b'')\s\s=\s\s\int\limits_{\Sigma^{g-1}}
\phi(x;\m x_1,\dots,x_{g-1})\s\s\s
b''(x_1)\s\cdots\s b''(x_{g-1})
\label{jesz}
\qqq
where \s$\phi\s$ is a holomorphic \s$g,0$-form
on \s$\Sigma\times\Sigma^{g-1}\s$ with values in
\qq
{\rm det}^{-1}H^0(K)\s\otimes\s{pr_1}_*(
{\rm det}^{-1}R^0{pr_1}_*{\CL_0'}^2)\s\s S^{g-1}
({\CL_0'}^2)\ .
\label{L}
\qqq
Note that the choices of the metric on \s$\Sigma\s$
and of the hermitian structure on \s$\CL_0'\s$ described
above induce a hermitian metric on the h.l.b.
(\ref{L}).
\vs 0.5cm

The functional integral calculation which we describe
in this paper implies the following scalar
product formula for the CS states:
\qq
\|\Psi\|^2\ \s\s
=\ \s\s{\rm const}.\ \s i^{^{-1-M}}\m
\s\int\s\s{\rm det}'\m
(\m\de_{L(-x)^{2}}^\dagger\m
\de_{L(-x)^2})\hs{5.3cm}\cr
\cdot\ |\m\CS_M(\m\phi(x;\m x_1,\dots,x_{g-1})\s\s\s
\chi(x;\m x_g,\dots,x_M)\m)\m|^{^{\wedge 2}}\
\prod\limits_{m_1\not=m_2}\hs{-0.2cm}
\ee^{-{4\pi\over k+2}\m G(x_{m_1},\m x_{m_2})}\ \
\label{final0}
\qqq
Above, \s$M\equiv(k+1)(g-1)\s$, \s$\CS_M\s$ stands for
the symmetrizer of \s$(x_m)_{_{m=1}}^{^M}\s$,
\s$|\s\cdot\s|^{^{\wedge 2}}\s$
denotes the \s$(1+M),(1+M)$-form obtained by
pairing a \s$(1+M),0$-form
with values in a h.l.b. with itself using the hermitian
structures described above. The prefactor \s$i^{^{-1-M}}\s$
assures the positivity of the integrated form.
The determinant of the operator
\s$\de_{L(-x)^{2}}^\dagger\m
\de_{L(-x)^2}\s$ restricted to the subspace
orthogonal to the zero modes should be
zeta-function regularized. Notice that the product
on the right hand side of Eq.\s\s(\ref{final0})
has a form of the Boltzmann factor
for a gas of two-dimensional particles interacting
with attractive Coulomb forces. Appearance of such
``Coulomb (or, more properly, Newton) gas representation''
was a characteristic
feature of the genus zero scalar product formulae,
see \cite{Quadr}\cite{FalGaw0}. As mentioned above,
we have not proven that the above equation defines
the scalar product of CS states which induces
(projectivized) KZB connection. The first
thing which remains to be shown is
that the integral on the right hand side of (\ref{final0})
(over the modular parameter \s$x\s$ and over
\s$M\s$ positions of ``screening charges'' at points
\s$x_m\s$) converges for \s$\chi\s$ corresponding to the CS
states. Although not proven in general, the convergence
seems very plausible in view of the
analysis of Sect.\s\s9 below. In particular,
it is evident for genus \s$2\s$.
\s Appendix F discusses
other consistency checks of the complete scalar
product formula worked out in Sect.\s\s9.
\vskip 1cm

\nsection{\hspace{-.6cm}.\ \ Change of variables}
\vs 0.5cm

As we have mentioned, the main idea of this work
is a brute-force calculation of the functional
integral (\ref{last}) giving the formal
scalar product of the CS states. This will be
a long process in which the first step
is the change of variables
\qq
A^{01}\s=\s{}^{h^{-1}}\hs{-0.16cm}A^{01}(n)
\label{CHV}
\qqq
where \s$n\mapsto A^{01}(n)\s$ parametrizes holomorphically
a \s$(3g-3)\equiv N$-dimensional slice
of \s$\CA^{01}\s$ (generically) transversal to the
chiral gauge orbits.
The reparametrization (\ref{CHV}) permits
to transform the formal scalar product formula into
\qq
\|\Psi\|^2=\int|\Psi(\m{}^{h^{-1}}\hs{-0.18cm}A^{01}(n)\m)|^2
\s\s\ee^{-{{ik}\over{2\pi}}\smallint_{_\Sigma}{\rm tr}\s\s
({}^{h\hs{-0.05cm}^{-1}}
\hs{-0.14cm}A^{01}(n))^\dagger\wedge\s{}^{h\hs{-0.05cm}^{-1}}
\hs{-0.14cm}A^{01}(n)}\s
\left|{\da\s(\s{}^{h^{-1}}\hs{-0.16cm}A^{01}(n)\s)\over
\da\s(\s h\m,\s n\s)}\right|^2 Dh \prod\limits_\alpha
d^2n_\alpha\ \ \ \label{scpr}
\qqq
where \s$Dh=\prod_{_x}\hs{-0.05cm}dh(x)\s$ is a
formal local product
of the Haar measures on \s$SL(2,\NC)\s$. \s We have to
compute the Jacobian \s\s$\left|{\da\s(\s{}^{h\hs{-0.05cm}^{-1}}
\hs{-0.16cm}A^{01}(n)\s)\over
\da\s(\s h\m,\s n\s)}\right|^2\s\s$ of the change of
variables. Notice that under an infinitesimal variation \s
\s of \s$h\s$ and \s$n\s$,
\qq
\delta\s(\s{}^{h^{-1}}\hs{-0.16cm}A^{01}(n)\s)\s=\s
{}^{h^{-1}}\hs{-0.1cm}\bar D_n
(h^{-1}\delta h)\s+\s
h^{-1}{_{\da\s A^{01}(n)}\over^{\da\s n_\alpha}}
\s\delta n_\alpha\s h
\qqq
where \s\s$\bar D_n\s\equiv\s\de\s+\s[\s A^{01}(n)\s,
\s\s\cdot\s\s\s]\s$ and \s\s${}^{h^{-1}}
\hs{-0.1cm}\bar D_n\s\equiv
\s{\rm Ad}_{h^{-1}}\bar D_n\s{\rm Ad}_{h}\s$.
\s Assume that \s$A^{01}(n_0)\s$ defines a stable vector bundle.
Then, for \s$n\s$ close to \s$n_0\s$ one may choose
a basis \s$(\omega^\alpha(n))_{_{\alpha=1}}^{^{N}}\s$
of \s$1,0$-forms with values in \s$sl(2,\NC)\s$
such that \s$\bar D_n\s\omega^\alpha(n)=0\s$
and \s$\omega^\alpha(n)\s$
depends holomorphically on \s$n\s$. Notice that the
relation \s$\bar D_n\s\omega^\alpha(n)=0\s$ holds if and only if
\qq
\int_{_\Sigma}\tr\s\s\s \omega^\alpha(n)
\wedge\bar D_n\Lambda\s=\s0
\qqq
for all (smooth) \s$sl(2,\NC)$-valued function \s$\Lambda\s$
so that \s$\omega^\alpha(n)\s$ may be viewed as covectors
tangent to the orbit space \s$\CA^{01}/\CG^\NC\s$ since
they vanish on the variations \s$\bar D_n\Lambda\s$ tangent
to the orbit of \s$\CG^\NC\s$ through \s$A^{01}(n)\s$.
\s The space \s$\CA^{01}\s$ has a natural scalar product
corresponding to the norm
\qq
\|A^{01}\|^2\s=\s{{i}}\s\int_{_\Sigma}\tr\s\m(A^{01})^\dagger
\wedge A^{01}\ .
\qqq
Using this scalar product, we may decompose
\qq
\CA^{01}\s=\s{\rm im}({}^{h^{-1}}\hs{-0.1cm}\bar D_n)
\m\oplus\m({\rm im}
({}^{h^{-1}}\hs{-0.1cm}\bar D_n))^\perp
\qqq
Notice that the subspace \s$({\rm im}({}^{h^{-1}}\hs{-0.1cm}
\bar D_n))^\perp\s$
orthogonal to the image of \s${}^{h^{-1}}\hs{-0.1cm}
\bar D_n\s$ is spanned
by the forms \s$(h^{-1}\omega^\alpha(n)\s h)^\dagger\s$.
The (holomorphic) derivative of the change of variables
(\ref{CHV}) is
\qq
{\delta\s(\s{}^{h^{-1}}\hs{-0.16cm}A^{01}(n)\s)\over
\delta\s(\s h\m,\s n\s)}\s=\s\left(\matrix{{}^{h^{-1}}\hs{-0.1cm}
\bar D_n&
{{}_{{}_{h^{-1}}{\underline{\da A^{01}(n)}}_{\s h}}}\cr
0&{}^{\da{n_\alpha}}}\right)\s
=\s\left(\matrix{{}^{h^{-1}}\hs{-0.1cm}\bar D_n&
\cdot\ \cdot\ \cdot\cr
0&(h^{-1}{\da A^{01}(n)\over\da n_\alpha}\s h)^\perp}\right)\ ,
\qqq
where
\qq
(h^{-1}{_{\da A^{01}(n)}\over^{\da n_\alpha}}\s h)^\perp\s=\s {i}
\s\s(h^{-1}\omega^\gamma(n)\s h)^\dagger\s\s\
(\s\Omega(hh^\dagger,n)^{-1})_{\gamma\beta}
\s\int_{_\Sigma}\tr\s\s\s\omega^\beta(n)\wedge
{_{\da A^{01}(n)}\over^{
\da n_\alpha}}
\qqq
with \s\s$\Omega(hh^\dagger,n)^{\beta\gamma}\s
=\s{{i}}\s\int_{_\Sigma}
\tr\s\s\s h^{-1}\omega^\beta(n)\s h\wedge
(h^{-1}\omega^\gamma(n)\s h)^\dagger\s$.
It follows that the Jacobian of the change of variables (\ref{CHV})
is
\qq
\left|{\da\s(\s{}^{h^{-1}}\hs{-0.16cm}A^{01}(n)\s)\over
\da\s(\s h\m,\s n\s)}\right|^2\s=\s
{\rm det}\s(\s({}^{h^{-1}}\hs{-0.1cm}
\bar D_n)^\dagger\s\s {}^{h^{-1}}\hs{-0.1cm}\bar D_n\s)
\ \s{\rm det}\m(\Omega(hh^\dagger,n))^{-1}\cr
\cdot\ \s\left|\s{\rm det}\left(
\int_{_\Sigma}\s\tr\s\s\s\omega^\beta(n)\wedge{_{\da A^{01}(n)}
\over^{\da n_\alpha}}\right)\right|^2\ .
\label{ins1}
\qqq
Of course, \s\s${\rm det}\s(\m({}^{h^{-1}}\hs{-0.1cm}
\bar D_n)^\dagger
\s\s {}^{h^{-1}}\hs{-0.1cm}
\bar D_n\m)\s\s$ has to be regularized, e.\s g.
by the zeta-function prescription.
The chiral anomaly permits to compute the \s$h$-dependence
of the regularized Jacobian\m:
\qq
&{\rm det}\s(\s({}^{h^{-1}}\hs{-0.1cm}
\bar D_n)^\dagger\s\s {}^{h^{-1}}\hs{-0.1cm}
\bar D_n\m)
\ \s{\rm det}\m(\m\Omega(hh^\dagger,n))^{-1}&
\ \cr\cr
&=\s\ee^{\s4\s S(hh^\dagger,\m A(n))}\ \s
{\rm det}\s(\bar D_n^\dagger\s\bar D_n\m)\
\s{\rm det}\m(\m\Omega(1,n))^{-1}\s ,&
\label{ins2}
\qqq
\vskip 0.01cm
\no where \s$A(n)\equiv-(A^{01}(n))^\dagger+A^{01}(n)\s$.
Also a short calculation using the transformation
properties (\ref{Trans}) shows that
\qq
|\Psi(\m{}^{h^{-1}}\hs{-0.16cm}A^{01}(n)\m)|^2
\s\s\ee^{-{{ik}\over{2\pi}}\smallint_{_\Sigma}{\rm tr}
\s\s({}^{h\hs{-0.05cm}^{-1}}
\hs{-0.16cm}A^{01}(n))^\dagger\wedge\s{}^{h\hs{-0.05cm}^{-1}}
\hs{-0.16cm}A^{01}(n)}\s=\s
|\Psi(A^{01}(n)|^2\s\s\ee^{-{{ik}\over{2\pi}}\smallint_{_\Sigma}
{\rm tr}\s\s (A^{01}(n))^\dagger\wedge A^{01}(n)}\ \cr
\cdot\ \ee^{\s k\m S(hh^\dagger,\s A(n))}\ .\hs{0.6cm}
\label{ins3}
\qqq
Inserting Eqs.\s\s(\ref{ins1})\m,\s(\ref{ins2}) and (\ref{ins3})
into the functional integral (\ref{scpr}), we obtain
\qq
\|\Psi\|^2\s=\s\int|\Psi(A^{01}(n)|^2\ \s\ee^{-{{ik}\over{2\pi}}
\smallint_{_\Sigma}
{\rm tr}\s\s (A^{01}(n))^\dagger\wedge A^{01}(n)}
\ \s\ee^{\m(k+4)\m S(hh^\dagger,\s A(n))}
\ \s {\rm det}\s(\bar D_n^\dagger\s\bar D_n\m)
\ \cr
\cdot\ \s{\rm det}\m(\Omega(1,n))^{-1}\s\s
\left|\s{\rm det}\left(
\int_{_\Sigma}\m\tr\m\s\s\omega^\beta(n)\wedge{_{\da A^{01}(n)}
\over^{\da n_\alpha}}\right)\right|^2
D(hh^\dagger)\prod\limits_\alpha
d^2n_\alpha\ ,
\label{newI}
\qqq
where we have used the fact that the integrand depends on
\s$h\s$ only through \s$hh^\dagger\s$, \s related to the gauge
invariance of the original integral (\ref{last}),
to reduce the \s$h$-integration to that over the
\s$hh^\dagger\s$ fields effectively taking values
in the hyperbolic space \s$SL(2,\NC)/SU(2)\s$. \s$D(hh^\dagger)\s$
should then be interpreted as the local formal product \s$\prod_{_x}
\hs{-0.05cm}d(hh^\dagger)(x)\s$ of
\s$SL(2,\NC)$-invariant measures on \s$SL(2,\NC)/SU(2)\s$.
\vs 0.5cm

Formula (\ref{newI}) decomposes the original functional
integral (\ref{last}) over \s$\CA^{01}\s$ into the one along
the orbits of the chiral gauge
transformations, which has the form of the partition function
of an \s$SL(2,\NC)/SU(2)$-valued WZW model \cite{Haba}\cite{GK}
\qq
\int\ee^{\m(k+4)\m S(hh^\dagger,\s A(n))}\ D(hh^\dagger)\ ,
\qqq
and the integral along a slice \s$n\mapsto A^{01}(n)\s$ of
\s$\CA^{01}\s$ which we shall parametrize by the complex bundle
\s$\NP W_0\s$. More exactly, as discussed in Sect.\s\s2,
we shall consider the map
\qq
({\bf x},b)\s\ {\smash{\mathop{\longmapsto}\limits^{s}}}
\  A^{01}_{{\bf x},b}
\label{slice}
\qqq
with \s$\bf x\s$ running through a fundamental domain of
\s$\Pi_1\s$ in
\s$\tilde\Sigma\s$ and one \s$b\s$
in each class of \s$\NP H^1(L_{\bf x}^{-2})\s$.
Such a map gives a multiply parametrized slice of
\s$\CA^{01}\s$ (as we have
seen in Section 2, the induced map
from \s$\NP W_0\s$ is essentially a multiple covering
of \s$\CA^{01}/\CG^\NC\s$; \s one may show that multiplicity
is equal to \s$2g\s$)\s. \s
Let \s$(\eta^\alpha_{\bf x})_{_{\alpha=1}}^{^{N}}\s$
be a basis of \s$H^0(L_{\bf x}^{2}K)\s$.
\s$\eta^\alpha_{\bf x}\s$ may be chosen locally as
depending holomorphically on \s$\bf x\s$. The integrals
\qq
z^\alpha_{\bf x}=\smallint_{_\Sigma}\hs{-0.04cm}
\eta^\alpha_{\bf x}\wedge b
\qqq
provide coordinates on \s$H^1(L^{-2}_{\bf x})\s$
(homogeneous coordinates on \s$\NP H^1(L_{\bf x}^{-2})\s$)
and a local (holomorphic) trivialization of \s$W_0\s$.
We shall have to find explicit expressions for
various terms under the integral (\ref{newI}).
\vskip 0.8cm

\subsection{\hspace{-.5cm}.\ \ Term \s\s\s$|\Psi(A^{01}(n)|^2\
\ee^{-{{ik}\over{2\pi}}\smallint_{_\Sigma}
{\rm tr}\s\s (A^{01}(n))^\dagger\wedge A^{01}(n)}\s$}
\vs 0.4cm

Using Eq.\s\s(\ref{newCS}), we obtain
\qq
|\Psi(A^{01}_{{\bf x},b})|^2\ \s
\ee^{-{{ik}\over{2\pi}}\smallint_{_\Sigma}
{\rm tr}\s\s (A^{01}_{{\bf x},b})^\dagger
\wedge A^{01}_{{\bf x},b}}
\s\s=\s\s|\psi({\bf x},\m b)|^2\
\hs{4.5cm}\cr
\cdot\ \ee^{-{{ik}\over{2\pi}}\int_{_\Sigma}{\rm tr}\s\s
(A_0^{10}+\s(A^{01}_{{\bf x},b})^\dagger)\m
\wedge\m(\m-A_0^{01}
+A^{01}_{{\bf x},b})\s-\s{{ik}\over{2\pi}}
\int_{_\Sigma}{\rm tr}
\s\s A^{10}_0\wedge\m A^{01}_0}\ .
\qqq
Recall that \s\s\s$A^{01}_{\bx,b}=U B^{01}_{\bx,b}\m U^{-1}
+\s U\de U^{-1}\ ,\s\s$ \s\s$\hs{0.6cm}A^{10}_0=
U\nabla U^{-1}=-(U\de U^{-1})^\dagger=-(A_0^{01})^\dagger\s\s\s$
with \s$B^{01}_{\bx,b}=(\matrix{_{-a_{\bx}}&_b\cr^0&^{a_\bx}}
)\s$. \s Hence \s\s$-A^{01}_0\m+\s A^{01}_{\bx,b}\s
=\s U B^{01}_{\bx,b}\s U^{-1}\s\s$ and \s\s$A^{10}_0\s
+\s(A^{01}_{\bx,b})^\dagger\s=\s U\m
(B^{01}_{\bx,b})^\dagger\m U^{-1}\s,$ \s
where \s$(B^{01}_{\bx,b})^\dagger=(\matrix{_{\overline{a_\bx}}&
_0\cr^{b^\dagger}&^{-{\overline{a_\bx}}}})\s$ with
the \s$L^{-2}_0$-valued \s$1,0$-form \s$b^\dagger\s$
being the conjugate
of \s$b\s$ with respect to the hermitian metric \s$\langle
\m\cdot\s,\m\cdot\m\rangle\s$ of
\s$L_0^{-2}\s$.
\s It follows that \s\s$\int_{_\Sigma}{\rm tr}\s\s
(A_0^{10}+\s(A^{01}_{\bx,b})^\dagger)\m\wedge\m(\m-A_0^{01}
+A^{01}_{\bx,b})\s=\s2\int_{_\Sigma}{\overline{a_{\bx}}}\wedge a_\bx
\s+\s\int_{_\Sigma}\langle b\m,\wedge\m b\rangle\s$.
\s\s Consequently,
\qq
|\Psi(A^{01}_{{\bf x},b})|^2\
\ee^{-{{ik}\over{2\pi}}\smallint_{_\Sigma}\hs{-0.06cm}
{\rm tr}\s\s (\hs{-0.04cm}A^{01}_{{\bf x},b})^\dagger
\wedge A^{01}_{{\bf x},b}}
\s=\s\ee^{-{{ik}\over{2\pi}}\int_{_\Sigma}\hs{-0.08cm}{\rm tr}
\s\m A^{10}_0\wedge\hs{-0.04cm}A^{01}_0
\s-\s2\pi k\s(\smallint_{_{x_0}}^{^{{\bf x}}}\hs{-0.06cm}
\bar\omega)\s{1\over{\rm Im}\s\tau}\s
(\smallint_{_{x_0}}^{^{{\bf x}}}\hs{-0.06cm}
\omega)}\ \s|\psi({\bf x},\m b)|^2\m.\ \ \ \ \
\label{5.20}
\qqq
\vskip 0.8cm

\subsection{\hspace{-.5cm}.\ \ Term \s\s
\s$\ee^{\m(k+4)\s S(hh^\dagger,\s A(n))}\s$}
\vs 0.4cm

Consider the field \s$U^{-1}hh^\dagger\m U\s$. It is a smooth
section of the bundle \s${\rm End}(L_0^{-1}\oplus L_0)\s$
taking values in the positive endomorphisms.
It is easy to see that, necessarily,
\qq
U^{-1}hh^\dagger\m U=\left(\matrix{
\ee^\varphi+\ee^{-\varphi}\langle v,\m v\rangle&
\ee^{-\varphi}\m v\cr\ee^{-\varphi}\m
v^\dagger&\ee^{-\varphi}}\right)=
\left(\matrix{\ee^{\varphi/2}&
\ee^{-\varphi/2}\m v\cr0&\ee^{-\varphi/2}}\right)
\hs{-0.16cm}\left(
\matrix{\ee^{\varphi/2}&\ee^{-\varphi/2}\m v\cr
0&\ee^{-\varphi/2}}
\right)^{\hs{-0.07cm}\dagger}\equiv\s gg^\dagger\hs{0.6cm}
\label{param}
\qqq
for unique real function
\s$\varphi\s$ on \s$\Sigma\s$ and \s$v\in
\Gamma(L_0^{-2})\s$.
We shall prove that
\qq
S(hh^\dagger,\s A_{\bx,b})\s=
\s{_i\over 2\pi}\int_{_\Sigma}\hs{-0.1cm}
\varphi\s(\m\da\de\varphi-2\m F_0\m)\s\
\hs{5cm}\cr
-\s{_i\over^{2\pi}}\int_{_\Sigma}
\ee^{-2\varphi}\s\langle b+(\de-2a_\bx)v)\m\s,\s\wedge\s(\m b+
(\de-2a_\bx)\m v)\s\rangle\s
+\s{_i\over 2\pi}\int_{_\Sigma}
\langle b\s,\s\wedge\m b\rangle\ .
\label{toprove}
\qqq
Using the formula (\ref{var})
of Appendix A, we see that
\qq
\delta\m S(hh^\dagger,\s A_{\bx,b})\s=\s
{_i\over^{2\pi}}\int_{_\Sigma}\tr\s\s\s(hh^\dagger)^{-1}
\delta(hh^\dagger)\s\ \s F(-(A^{01}_{\bx,b})^\dagger\s+\s
{}^{(hh^\dagger)^{-1}}\hs{-0.18cm}
A^{01}_{\bx,b}\s)\ \s\cr
=\s{_i\over^{2\pi}}\int_{_\Sigma}\tr\s\s (gg^\dagger)^{-1}
\delta (gg^\dagger)
\ \s\s{\rm curv}\m(\s\nabla-(B^{01}_{\bx,b})^\dagger+
(gg^\dagger)^{-1}(\de+B^{01})\m (gg^\dagger)\s)
\qqq
By a straightforward computation, under holomorphic
variations of \s$v\s$
\qq
\delta\m S(hh^\dagger,\m A^{01}_{\bx,b})\s=\s
-{{_i}\over^{2\pi}}\int_{_\Sigma}\delta v\ \m(\s\de+2a_{\bx})
\left(\ee^{-2\varphi}(b+(\de-2a_\bx)\m v\m)\right)^\dagger
\qqq
and under antiholomorphic ones
\qq
\delta\m S(hh^\dagger,\m A^{01}_{\bx,b})\s=\s
{{_i}\over^{2\pi}}\int_{_\Sigma}\delta v^\dagger\ \m(\nabla+2
\m\overline{a_{\bx}})
\left(\ee^{-2\varphi}(b+(\de-2a_\bx)\s v\m)\right)
\qqq
which coincides with the \s$v$-variations of the right hand side
of (\ref{toprove}). Thus we may assume that \s$v=0\s$.
Then the variation of \s$S\s$ with respect
to \s$\varphi\s$ becomes
\qq
\delta\m S(hh^\dagger,\m A^{01}_{\bx,b})\s=\s
{{_i}\over^{\pi}}\int_{_\Sigma}
\delta\varphi\m\left(\s\da\de\varphi
-F_0+\s\ee^{-2\varphi}\s b^\dagger
\wedge\m b\m\right)
\qqq
(recall that \s$\de\m a_\bx=0\s$) \s which is also
the variation of the right hand side of
Eq.\s\s(\ref{toprove}) for \s$v=0\s$. \s This ends the proof
of (\ref{toprove}) since \s$S(hh^\dagger,
\s A^{01}_{\bx,b})=0\s$ for \s$hh^\dagger=1\s$.
\vskip 0.8cm

\subsection{\hspace{-.5cm}.\ \ Term \s\s
\s${\rm det}\m(\Omega(1,n))^{-1}\s
\left|\s{\rm det}\left(
\int_{_\Sigma}\s{\rm tr}\s\s\s\omega^\beta(n)
\wedge{_{\da A^{01}(n)}
\over^{\da u_\alpha}}\right)\right|^2\s\prod\limits_\alpha
d^2n_\alpha\s$}
\vs 0.4cm

We have to look for \s${sl(2,\NC)}$-valued
\s$1,0$-forms \s$\omega(\bx,b)\equiv\omega\s$ such that
\s$\de\omega+A^{01}_{\bx,b}\m\omega
+\omega A^{01}_{\bx,b}
=0\s$. \s Such forms represent vectors cotangent to the orbit
space \s$\CA^{01}/\CG^\NC\s$. \s Writing
\qq
\omega\s=U\left(\matrix{_{-\mu}&_\lambda\cr
^\eta&^\mu}\right)U^{-1}\equiv
U\rho\s U^{-1}\ ,
\qqq
where \s$\eta\in\wedge^{10}(L_0^{2})\s$,
\s$\mu\in\wedge^{10}\s$ and \s$\lambda\in\wedge^{10}
(L_0^{-2})\s$, the condition for \s$\omega\s$
becomes \s$\de\m\rho+B^{01}_{\bx,b}\m
\rho+\rho\m
B^{01}_{\bx,b}=0\m\s$  or, \s in components,
\qq
(\de+2a_{\bx})\s\eta=0\ ,\hs{0.6cm}
\de\mu=-\eta\wedge\m b\ ,\hs{0.6cm}
(\de-2a_{\bx})\m\lambda=2\m\mu\m\wedge\m b\ .
\label{three}
\qqq
The first of these equations requires that \s$\eta\in H^0
(L_{\bf x}^{2} K)\s$ which has dimension \s$N\s$.
\s The second one has a solution
if and only if \s$\int_{_\Sigma}\eta\m\wedge\m b=0\s$ which,
for \s$b\s$ corresponding to a non-zero element
in \s$H^1(L_\bx^{-2})\s$, \s defines a \s$N-1\s$
dimensional subspace in \s$H^0(L_{\bf x}^{2}K)\s$.
For \s$\eta\s$ in this subspace,
\qq
\mu=2i\s\da\int G
(\m\cdot\s,\m y)\s(\eta\m\wedge\m b)\m(y)
\s+\s\nu\s\equiv\s\mu^0(\eta)\s+\s\nu\ ,
\label{mu0}
\qqq
where \s$G(x,y)\s$ is a Green function of the Laplacian
on \s$\Sigma\s$ and \s$\nu\s$ is an arbitrary holomorphic
\s$1,0$-form on \s$\Sigma\s$. \s Finally, let
\s$(\kappa^r)_{_{r=1}}^{^{g-1}}\s$ be a basis
of \s$H^0(L^{2}_{\bx})\s$.
\s$\kappa^r\s$ may be chosen locally
depending holomorphically
on \s$\bx\s$. \s The third of the equations (\ref{three})
has a solution for \s$\lambda\s$ if and only if
\qq
\int_{_\Sigma}\kappa^r\s\mu\wedge\m b
\s=\s\int_{_\Sigma}\kappa^r\s\mu^0(\eta)\wedge\m b
\s+\s\int_{_\Sigma}\kappa^r\s\nu\m\wedge\m b\s=\s0
\label{CoN}
\qqq
for each \s$r\s$. Then the solution for \s$\lambda\s$
is unique since \s$H^0(L_\bx^{-2}K)=\{0\}\s$
since we have chosen \s$L\s$ so that \s$L^{2}_\bx\s$ is
never isomorphic to \s$K\s$.
Let us consider more carefully the condition (\ref{CoN}).
Notice that the exterior multiplication by \s$b\s$ induces
a linear map
\qq
{l(\bx,b)}:\s H^0(K)\s\longrightarrow\s
H^1(L_{\bf x}^{-2} K)\ .
\label{Map}
\qqq
${l(\bx,b)}\s$
depends only on the class
\s\m$[b]\m\s$ of \m\s$b\m\s$ in \m\s$H^1(L_{\bx}^{-2})\s$.
\s The dimensions of the spaces are \s dim($H^0(K))=g\s$
and \s dim($H^1(L_{\bx}^{-2}K))$ $
={\rm dim}(H^0(L_\bx^2))=g-1\s$.
\s If \s${l(\bx,b)}\s$ maps onto then, for fixed \s$\mu^0(\eta)\s$,
there exists \s$\nu\s$ solving (\ref{CoN}) and it
is unique up to the addition of \s$\nu\s$
from the one-dimensional
kernel of \s${l(\bx,b)}\s$.
Altogether, the space of solutions of Eqs.\s\s(\ref{three})
is then \s$N$-dimensional: \s$N-1\s$
dimensions of the freedom to choose \s$\eta\s$
and $1$ dimension in the choice of \s$\nu\s$.
\s Let us examine the condition of surjectivity of \s${l(\bx,b)}\s$
which guarantees that the the space tangent to the
\s$\CG^\NC$-orbit through \s$A_{\bx,b}\s$ is
of maximal codimension (\s$=N\s$). \s
Taking the standard basis \s$(\omega^i)_{_{i=1}}^{^g}\s$
of \s$H^0(K)\s$, \s this condition means that the matrix
\qq
\left(\hs{0.02cm}\smallint_{_\Sigma}\hs{-0.02cm}
\kappa^r\s\omega^i\wedge\m b\right)
\label{matr}
\qqq
has rank $=\s g-1\s$.  The \s$[b]$'s \s in \s$H^1(L_\bx^{-2})\s$
for which this fails are common zeros of \s$g\s$
homogeneous polynomials giving the \s$(g-1)\times(g-1)\s$
minors of the matrix (\ref{matr}). If these equations are
non-trivial, it follows that \s${l(\bx,b)}\s$ is surjective
except for a subvariety of positive codimension. To see their
non-triviality notice\footnote{We thank J.-B. Bost for this
argument.} that
\s${l(\bx,b)}\s$ fails to be surjective if and only if for
some \s$0\not=\kappa\in
H^0(L_{\bx}^{2})\s$
$$[b]\s\in\s B_{\kappa}\s\equiv\s\{\ [b]\ \s\s|\s\s
\smallint_{_\Sigma}\kappa\s\nu\m\wedge\m b=0\ \ {\rm for\ all}
\ \ \nu\in H^0(K)\ \}\ .$$
But \s$\cup_{_\kappa}B_\kappa\s$
is at most \s$3g-5\s$ dimensional \s (\s${\rm dim}(B_\kappa)
=2g-3\s$
and \s$B_\kappa\s$ depends only on the class of \s$\kappa\s$
in the \s$(g-2)$-dimensional projective space
\s$\NP H^0(L_\bx^{2})\s$).
\vs 0.5cm

For given \s$b\in\wedge^{01}(L^{-2}_\bx)\s$
corresponding to a non-trivial element in
\s$H^1(L_\bx^{-2})\s$, \s we may choose
the basis \s$(\eta^\alpha)_{_{n=1}}^{^{N}}\s$ of
\s$H^0(L_{\bx}^{2} K)\s$ so that \s$z^1\s(\equiv
\int_{_\Sigma}\eta^1\wedge\m b\s)\not=0\s$ and \s$z^\alpha=0\s$
for \s$\alpha>1\s$. Suppose also that the non-zero
\s$(g-1)\times(g-1)\s$ minor
of the matrix (\ref{matr}) corresponds to \s$i<g\s$.
Then, we may take
\qq
\eta=0\ ,\ \ \ \mu^1\s\equiv\s\omega^g\s-\s\omega^i\s
\s M_{ir}\m\s\s(\smallint_{_\Sigma}
\kappa^r\s\omega^g\wedge\m b\s)\ ,
\ \ \ \lambda^1\s\equiv\s2\s\de^{-1}(\mu^1\wedge\s b\s)\ ,
\label{5.33}
\qqq
where \s$(M_{ir})\s$ is the matrix inverse to
\s$(\smallint_{_\Sigma}\kappa^r\m\omega^i\wedge\m
b\s)_{_{i<g}}\s$, \s and, \s for \s$\alpha>1\s$,
\qq
\eta^\alpha\ ,\ \ \ \mu^\alpha
\s\equiv\s\mu^0(\eta^\alpha)
\s-\s\omega^i\s\s M_{ir}\m\s\s(\smallint_{_\Sigma}\kappa^r(
\m\mu^0(\eta^\alpha))\m\wedge\m b\s)
\ ,\ \ \ \lambda^\alpha\s\equiv\s 2\s\de^{-1}(\mu^\alpha
\wedge\s b\s)
\label{5.34}
\qqq
as giving a basis of solutions of Eqs.\s\s(\ref{three})
and, consequently, a basis \s$(\omega^\alpha(\bx,b))\s$
of the \s$sl(2,\NC)$-valued \s$1,0$-forms representing covectors
tangent to the orbit space \m$\CA^{01}/\CG^\NC\m$
at the orbit passing through \s$A^{01}_{\bx,b}\s$.
\s Above \s$\de^{-1}\equiv(\de_{L^{-2}_{\bx}K})^{-1}\s$. \s
With this choice,
\qq
{\rm det}\s(\Omega(1,n))\s\equiv\s{\rm det}\s(\Omega(1,\bx,b))
\s=\s{\rm det}\left(
\int_{_\Sigma}{_1\over^i}\s(\s2\m
\overline{\mu^\alpha}\wedge\mu^\beta
+\langle\eta^\alpha,\wedge\eta^\beta
\rangle+\langle\lambda^\alpha,\wedge\lambda^\beta\rangle)\s\right)
\ \ \
\label{5.35}
\qqq
(\s$\eta^1\s$ should be replaced by zero).
Moreover, since \s\s$\int_{_\Sigma}\tr\s\s\m\omega^\beta(\bx,b)
\s\m\delta\hs{-0.04cm} A^{01}_{\bx,b}
\s=\s\int_{_\Sigma}(\s 2\m\mu^\beta\m\delta
a_{\bx}\s+\s\eta^\beta\m\delta b\s)\s,$
\qq
\left|\s{\rm det}\hs{-0.1cm}\left(
\int_{_\Sigma}\hs{-0.12cm}{\rm tr}\s\s
\omega^\beta(\hs{-0.03cm}n\hs{-0.03cm})
\hs{-0.04cm}\wedge\hs{-0.09cm}{_{\da A^{01}(n)}
\over^{\da n_\alpha}}\hs{-0.06cm}\right)
\hs{-0.06cm}\right|^2\hs{-0.09cm}\prod\limits_\alpha
\hs{-0.08cm}d^2n_\alpha=
{_{8\pi^2}\over^i}\m
|\omega^g(x)\hs{-0.03cm}-\hs{-0.03cm}
\omega^i(x)\s\m M_{ir}\m\s(\hs{-0.04cm}\smallint_{_\Sigma}
\hs{-0.05cm}\kappa^r\omega^g\hs{-0.05cm}
\wedge\hs{-0.05cm} b)\m|^{^{\wedge 2}}\hs{-0.04cm}
\prod\limits_{\alpha=2}^{{N}}
\hs{-0.15cm}d^2\hs{-0.02cm}z^\alpha\hs{-0.04cm},\hs{0.55cm}
\label{5.36}
\qqq
where, for a form \s$\omega\s$, \s$|\omega|^{^{\wedge 2}}\s$
denotes the form \s$\bar\omega\wedge\omega\s$.
A simple algebra shows that
\qq
\omega^g(\hs{-0.02cm}x\hs{-0.02cm})\hs{-0.03cm}-\hs{-0.06cm}
\omega^i(\hs{-0.02cm}x\hs{-0.02cm})\s M_{ir}\s(\hs{-0.02cm}
\smallint_{_\Sigma}\hs{-0.03cm}
\kappa^r\omega^g\hs{-0.07cm}\wedge\hs{-0.02cm} b)=
{\rm det}\m(\hs{-0.02cm}\smallint_{_\Sigma}\hs{-0.03cm}
\kappa^r\omega^i\hs{-0.07cm}
\wedge\hs{-0.02cm} b)_{\hs{-0.01cm}_{i<g}}^{-1}\m
\sum\limits_{j=1}^g(-1)^{^{g-j}}\m{\rm det}\m(\hs{-0.02cm}
\smallint_{_\Sigma}\hs{-0.03cm}\kappa^r\omega^i\hs{-0.07cm}
\wedge\hs{-0.02cm} b)_{\hs{-0.01cm}_{i\not=j}}
\m\s\omega^j(x).\ \ \ \ \
\label{5.37}
\qqq
It will be convenient to represent \s\s${\rm det}\m
(\Omega(1,\bx,b))\s\s$ as a finite dimensional integral.
To this end, consider a linear map
\s\s$B:\s H^0(L^{2}_\bx K)\oplus
H^0(K)\s\rightarrow\s\NC^{g}\s$
given by
\qq
B(\eta\m,\s\nu)\s=\s(\m\smallint_{_\Sigma}
\eta\m\wedge\m b\s,\s\s
\smallint_{_\Sigma}\kappa^r\m(\mu^0(\eta)+\nu)\m\wedge\m b\s)
\qqq
and another linear map \s\s$C:\s {\rm ker}(B)\s\rightarrow
\s\wedge^{10}(\m{\rm End}(L_0^{-1}\oplus
L_0)\m)\s\s$ s.\s t.
\qq
C(\eta\m,\s\nu)\s=\s\left(\matrix{-\mu^0(\eta)-\nu &
2\m\de^{-1}\m(\m(\mu^0(\eta)+\nu)\wedge\m b\m)\cr
\eta&\mu^0(\eta)+\nu}\right),
\qqq
(with \s$\de^{-1}\equiv(\de_{L^{-2}_\bx K})^{-1}\s$)\s.
\s A straightforward calculation shows that with
\s$V\equiv(\eta,\m\nu)\s$ and \s$DV\s$ standing for the
volume element of \s$H^0(L^{2}_\bx K)\oplus
H^0(K)\s$ coming from the scalar product induced by
the hermitian metric of \s$L_0\s$ and the metric of
\s$\Sigma\s$,
\qq
\int\delta(BV)\ \s\ee^{-\|CV\|^2}\ DV
\s=\s|z^1|^{-2}\ |{\rm det}\s(\smallint_{_\Sigma}\kappa^r
\omega^i\wedge\m b\s)_{\hs{-0.01cm}_{i<g}}|^{-2}\
{\rm det}\m(\s H_0\m)\ \ \ \cr
\cdot\ {\rm det}\m(\m{\rm Im}\s\tau)
\ \s
{\rm det}\s(\Omega(1,\bx,b))^{-1}\ ,
\label{llast}
\qqq
where
\qq
(H_0)^{\alpha\beta}\s\equiv\s\s {_1\over^i}\smallint_{_\Sigma}\m
\langle\eta^\alpha,\wedge\eta^\beta\rangle\ .
\label{h0}
\qqq
Putting together Eqs.\s\s(\ref{5.36}), (\ref{5.37})
and (\ref{llast}), we obtain
\qq
&{\rm det}\s(\Omega(1,n))^{-1}\s
\left|\s{\rm det}\left(
\int_{_\Sigma}\s{\rm tr}\s\s\s\omega^\beta(n)
\wedge{_{\da A^{01}(n)}
\over^{\da n_\alpha}}\right)\right|^2\s\m\prod\limits_\alpha
d^2n_\alpha\hs{2.3cm}&\cr\cr
&=\s\s {\rm const.}\ i^{^{-N}}
\ \s{\rm det}\m(\s H_0\m)^{-1}\ {\rm det}
\m(\m{\rm Im}\s\tau)^{-1}
\m\left(\int\delta(BV)\ \s\ee^{-\|CV\|^2}
\ DV\right)\ \hs{1.9cm}&\cr
&\cdot\ \s|\s\epsilon_{\alpha_1,\dots,\alpha_{{N}}}
\s\m z^{\alpha_1}\m dz^{\alpha_2}\hs{-0.03cm}\wedge\m
\dots\m\wedge dz^{\alpha_{{N}}}\s|^{^{\wedge 2}}
\ |\sum\limits_{j=1}^g
\s(-1)^j\s\s{\rm det}\m(
\smallint_{_\Sigma}\kappa^r\omega^i
\wedge\m b\s)_{\hs{-0.01cm}_{i\not=j}}\
\omega^j(x)\s|^{^{\wedge 2}}\ \ \ \ \ &
\label{omega}
\qqq
(with a numerical, easy to trace,
\s$g$-dependent positive constant in front; the power of
\s$i\s$ makes the right hand side a positive measure).
We have given the term \s$|\m z^1dz^2\hs{-0.03cm}\wedge\dots
\wedge dz^{{N}}\s|^2\s$ a form independent of the assumed
relations \s$z^1\not=0\s$,\ \s$z^\alpha=0\s$ for \s$\alpha>1\s$.
\vskip 1cm

\nsection{\hspace{-.6cm}.\ \ Calculation of \s\s\s${\rm det}
\s(\bar D_n^\dagger\s\bar D_n\m)\s$}
\vs 0.5cm

Let \s$\Lambda\s$ be an \s$sl(2,\NC)\s$ valued function on
\s$\Sigma\s$. Writing
\qq
\Lambda\s=\s U\left(\matrix{-X&Y\cr Z&X}\right)U^{-1}\ ,
\qqq
where \s$X\s$ is a function, \s$Y\in\Gamma(L_0^{-2})\s$
and \s$Z\in\Gamma(L^{2}_0)\s$, we obtain
\qq
\bar D_n\m\Lambda\s=\s U\left(
\matrix{-\de X+Zb&(\de-2a_\bx)Y+2Xb\cr
(\de+2a_\bx)Z&\de X-Zb}\right)U^{-1}
\qqq
It follows that
\qq
(\Lambda\m,\s\bar D_n^\dagger\m\bar D_n\Lambda)\s\equiv\s
i\int_{_\Sigma}\tr\s\s\m(\bar D_n\Lambda)^\dagger\wedge\s\bar D_n
\Lambda\s=\s i\int_{_\Sigma}(\s 2\m(\overline{\de X-Zb})\m\wedge
\m(\de X-Zb)\ \ \ \cr
+\s\langle\de Y+2Xb\m,\s\wedge\s(\de Y+2Xb)\rangle
\s+\s\langle\de Z\m,\s\wedge\s\de Z\rangle\s)\ ,
\qqq
where in the last line \s$\de\equiv\de_{L_\bx^{-2}}\s
\s(\s\de_{L_\bx^{2}}\s)\s$ when
acting on \s$Y\s$ (\s$Z\s$)\s.
Formally,
\qq
{\rm det}\m(\bar D_n^\dagger\m\bar D_n)^{-1}
\s=\s\int\ee^{-(\Lambda\m,\s\bar D_n^\dagger\m\bar D_n\Lambda)}
\ D\Lambda\hs{7cm}\cr
=\s\int\ee^{-i\int_{_\Sigma}
(\s 2\m(\overline{\de X-Zb})\m\wedge
\m(\de X-Zb)\s
+\s\langle\de Y+2Xb\m,\s\wedge\s(\de Y+2Xb)\rangle
\s+\s\langle\de Z\m,\s\wedge\s\de Z\rangle\s)}\ DY\ DX\ DZ
\qqq
and we shall compute the latter Gaussian integral iteratively,
first integrating over \s$Y\s$ then over \s$X\s$ and in the end
over \s$Z\s$. \s As we shall see, this procedure requires
a correction if we want to assure that the final result
gives the zeta-function regularized determinant of
\s$\bar D_n^\dagger\m\bar D_n\s$. \s It would be more natural
to consider \s$\Lambda\s$ as anticommuting ghost field
rather than the commuting one. Indeed, this is
\s${\rm det}\m(\bar D_n^\dagger\m\bar D_n)\s$ and not its
inverse which appears in the expression for the scalar
product of the CS states. The choice of commuting fields
\s$\Lambda\s$ in this calculation is purely a matter
of convenience.
\vskip 0.8cm

\subsection{\hspace{-.5cm}.\ \ Integral over \s\s$Y\s$}
\vs 0.4cm

\qq
I_Y\s\equiv\s\int\ee^{-i\int_{_\Sigma}\langle\de Y+2Xb\m,
\s\wedge\s(\de Y+2Xb)\rangle}\ DY
\s=\s{\rm det}\s(\m\de_{L^{-2}_\bx}^\dagger
\de_{L^{-2}_\bx})^{-1}
\ \ee^{-4\m i\int_{_\Sigma}\langle (Xb)^\perp,\s(Xb)^\perp
\rangle}\ ,
\label{Y1}
\qqq
where \s$(Xb)^\perp\s$ is the component of \s$Xb\s$
orthogonal to \s$\de(\Gamma(L_\bx^{-2}))\subset
\wedge^{01}(L_{\bf x}^{-2})\s$. Recall that the scalar product
in the spaces of sections is induced by the fixed hermitian
structure of \s$L_0\s$ and the metric on \s$\Sigma\s$.
Explicitly,
\qq
(Xb)^\perp\s=\s i\m(\eta^\alpha)^\dagger\m\s\
(H_0^{-1})_{_{\beta\alpha}}\m\s
\s\smallint_{_\Sigma}
\eta^\beta\wedge\m Xb\ ,
\qqq
where \s$(H_0)^{\alpha\beta}\s$ is given by Eq. (\ref{h0}). \s
We have
\qq
i\smallint_{_\Sigma}\m\langle (Xb)^\perp,\s(Xb)^\perp
\rangle\s=\s\smallint_{_\Sigma}\overline{\eta^\alpha\wedge\m Xb}
\ \m\s(H_0^{-1})_{\beta\alpha}\s\s\m\smallint_{_\Sigma}
\eta^\beta\wedge\m Xb\ .
\qqq
It will be convenient to express
\s\s$\ee^{-4\m i\int_{_\Sigma}\langle (Xb)^\perp,\s(Xb)^\perp
\rangle}\s\s$ as a \s$6(g-1)$-dimensional Gaussian integral.
Namely
\qq
\ee^{-4\m i\int_{_\Sigma}\langle (Xb)^\perp,\s(Xb)^\perp
\rangle}\s=\s\pi^{-{N}}\s\s{\rm det}\m(H_0)\
\int\ee^{\s2i\m\s c_\alpha\int_{_\Sigma}\hs{-0.05cm}
\eta^\alpha\wedge\m Xb\s\s+\s\s2i\m\s\bar c_\alpha\s
\int_{_\Sigma}\hs{-0.05cm}\overline{\eta^\alpha\wedge\m Xb}\s
\s-\s\s\bar c_\alpha\s(H_0)^{\alpha\beta}\m c_\beta}\cr
\cdot\ \prod\limits_\alpha d^2c_\alpha\ .\hs{0.6cm}
\label{Y2}
\qqq
\vskip 0.8cm

\subsection{\hspace{-.5cm}.\ \ Integral over \s\s$X\s$}
\vs 0.4cm

We have to calculate
\qq
I_X\s\equiv\s\int\ee^{-i\int_{_\Sigma}\hs{-0.05cm}
2\m(\overline{\de X-Zb})\m
\wedge\m(\de X-Zb)\s\s
+\s\s2i\m\s c_\alpha\int_{_\Sigma}\hs{-0.05cm}
\eta^\alpha\wedge\m Xb\s\s+\s\s2i\m\s\bar c_\alpha
\int_{_\Sigma}\hs{-0.05cm}\overline{\eta^\alpha\wedge\m Xb}}\ DX\ .
\label{X0}
\qqq
We have to fix the constant mode \s$X_0\equiv(\int X\m{\rm vol}\m)/
({\rm area})^{1/2}\s$ which corresponds
to the flat direction in the above Gaussian integral.
\s${\rm vol}\s$ is the Riemannian volume
form of \s$\Sigma\s$ and \s${\rm area}\equiv
\int_{_\Sigma}\hs{-0.03cm}{\rm vol}\s$. \s
Let us multiply \s$I_X\s$ by \s$1\s
={\rm area}\s\cdot\int\delta\s(X_0\hs{-0.05cm}-
\hs{-0.05cm}a\s({\rm area})^{1/2})\s\s d^2a\s$.
Changing the order
of integration and shifting \s$X\s$ to \s$X+a$, \s we obtain
\qq
I_X\s=\s{\rm area}\s\cdot\smallint
\delta\s(X_0\hs{-0.05cm}-\hs{-0.05cm}
a\s({\rm area})^{1/2})\s\s d^2a
\ \s\s I_X\s=\s{\rm area}\s\cdot
\int\ee^{-i\int_{_\Sigma}\hs{-0.05cm}2\m(\overline{\de X-Zb})\m
\wedge\m(\de X-Zb)\s\s}\ \ \ \cr
\cdot\ \ee^{\s\s2i\m\s c_\alpha\int_{_\Sigma}\hs{-0.05cm}
\eta^n\wedge\m Xb\s\s+\s\s2i\m\s\bar c_\alpha
\int_{_\Sigma}\hs{-0.05cm}\overline{\eta^n\wedge\m Xb}
\s\s+\s\s2i\m\s a\m\s c_\alpha\int_{_\Sigma}\hs{-0.05cm}
\eta^n\wedge\m b\s\s+\s\s2i\m\s\bar a\m\s\bar c_\alpha
\int_{_\Sigma}\hs{-0.05cm}\overline{\eta^\alpha\wedge\m b}}
\ \s\m\delta(X_0)\s\ DX\ \s d^2a\ .\ \ \
\qqq
Performing the \s$a$-integral first, we obtain
\qq
I_X&=&{_{\pi^2}\over^4}\s\cdot\s\m{\rm area}\m\s
\cdot\s\s\delta\s(\s c_\alpha\smallint_{_\Sigma}\hs{-0.05cm}
\eta^\alpha\wedge\m b\s)\hs{7cm}\cr
&&\cdot\s\int
\ee^{-i\int_{_\Sigma}\hs{-0.05cm}2\m(\overline{\de X-Zb})\m
\wedge\m(\de X-Zb)\s\s
+\s\s2i\m\s c_\alpha\int_{_\Sigma}\hs{-0.05cm}
\eta^\alpha\wedge\m Xb\s\s+\s\s2i\m\s\bar c_\alpha
\int_{_\Sigma}\hs{-0.05cm}\overline{\eta^\alpha\wedge\m Xb}}
\ \s\m\delta(X_0)\s\ DX\cr\cr
&=&{_{\pi^2}\over^4}\s\cdot\s\m{\rm area}\m\s
\cdot\s\s\delta\s(\s c_\alpha\smallint_{_\Sigma}\hs{-0.05cm}
\eta^\alpha\wedge\m b\s)\ \s\ee^{-2i\int_{_\Sigma}
\hs{-0.05cm}\overline{Zb}\m
\m\wedge\m Zb}\hs{5cm}\cr
&&\cdot\s\int\ee^{-\int_{_\Sigma}\hs{-0.05cm}\bar X
\m(-\Delta X)\s{\rm vol}\s\s+\s\s2i\int_{_\Sigma}\hs{-0.05cm}
X\m(c_\alpha\eta^\alpha
\wedge\m b\m+\m\de(\overline{Zb})\m)
\s\s+\s\s2i\int_{_\Sigma}\hs{-0.05cm}
\bar X\m(\overline{c_\alpha\eta^\alpha
\wedge\m b}\m-\m\da(Zb)\m)}
\ \s\delta(X_0)\s\ DX\ \s d^2a\cr\cr
&=&{\rm const}.\s\left({_{{\rm det}'\m(-\Delta)}\over^{\rm area}}
\right)^{\hs{-0.05cm}-1}\ \s
\ee^{-2i\int_{_\Sigma}\hs{-0.05cm}\overline{Zb}\m
\m\wedge\m Zb}\s\ \s\ee^{\s
4\int_{_\Sigma}\hs{-0.05cm}\int_{_\Sigma}
\hs{-0.05cm}(\s\overline{c_\alpha\eta^\alpha\wedge\m b}\m
-\m\da(\hs{-0.03cm}Zb)\s)(x)
\s\s\m G(x,y)\m\s\s\s(\s c_\alpha\eta^\alpha\wedge\m b\m+
\m\de(\hs{-0.03cm}\overline{Zb})\s)(y)}\cr
&&\hspace{9cm}\cdot\ \s\delta\s(\s c_\alpha
\smallint_{_\Sigma}
\hs{-0.05cm}\eta^\alpha\wedge\m b\s)\cr\cr
&=&{\rm const}.\s\left({_{{\rm det}'\m(-\Delta)}\over^{\rm area}}
\right)^{\hs{-0.05cm}-1}\ \s
\ee^{-2i\int_{_\Sigma}\hs{-0.05cm}\overline{Zb}\m
\m\wedge\m Zb\s\s-\s\s 4\int_{_\Sigma}\hs{-0.05cm}\int_{_\Sigma}
\hs{-0.05cm}(\m\da(\hs{-0.03cm}Zb)\m)(x)
\s\s G(x,y)\s\s\s(\m\de(\hs{-0.03cm}\overline{Zb})\m)(y)}\cr
&&\cdot\ \s\ee^{-2i\int_{_\Sigma}
\hs{-0.05cm}\mu^0(c_\alpha\eta^\alpha)\m
\wedge\m Zb\s\s-\s\s2i\m\int_{_\Sigma}\hs{-0.05cm}\overline{
\mu^0(c_\alpha\eta^\alpha)
\m\wedge\m Zb}\s\s+\s\s
2i\s\int_{_\Sigma}\hs{-0.05cm}\overline{\mu^0(c_\alpha
\eta^\alpha)}
\m\wedge\m\mu^0(c_\alpha\eta^\alpha)}\ \m
\delta\s(\m c_\alpha\smallint_{_\Sigma}
\hs{-0.05cm}\eta^\alpha\wedge\m b\m)\s ,\ \label{X1}
\qqq
where \s${\rm det}'\s$ denotes the determinant of the operator
restricted to the subspace orthogonal to its kernel,
\s$G(x,y)\s$ is a Green function of the Laplacian \s$\Delta\s$
on \s$\Sigma\s$ and \s$\mu^0(\eta)\s\equiv\s{2i}\s\da\int G
(\m\cdot\s,\m y)\s(\eta\m\wedge\m b)\m(y)\s$, \s as in Eq.\s\s
(\ref{mu0}). \s It is easy to see that
\qq
\ee^{-2i\int_{_\Sigma}\hs{-0.05cm}\overline{Zb}\m
\m\wedge\m Zb\s\s-\s\s 4\int_{_\Sigma}\hs{-0.05cm}\int_{_\Sigma}
\hs{-0.05cm}(\da(Zb))(x)
\s\s G(x,y)\s\s(\de(\overline{Zb}))(y)}\s=\s
\ee^{-2i\int_{_\Sigma}\hs{-0.05cm}\overline{(Zb)^\perp}
\m\wedge\m(Zb)^\perp}\ ,
\label{X2}
\qqq
where \s$(Zb)^\perp\s$ is the component of
\s$Zb\s$ orthogonal to the image of \s$\de\s$ acting on
functions on \s$\Sigma\s$.
\s Explicitly,
\qq
(Zb)^\perp\ &=&\ {_i\over^2}\s\m\bar\omega^i\s\s
({_1\over^{{\rm Im}\s\tau}})_{_{ij}}
\s\s\smallint_{_\Sigma}\omega^j\wedge\m Zb\ ,\cr
2i\smallint_{_\Sigma}\hs{-0.03cm}\overline{(Zb)^\perp}
\wedge\m(Zb)^\perp&=&\ (\smallint_{_\Sigma}\hs{-0.05cm}
\overline{\omega^i\wedge\m Zb}\m)\s\s\s({_1\over^{
{\rm Im}\s\tau}})_{_{ij}}\s
\s\s(\smallint_{_\Sigma}\hs{-0.05cm}\omega^j\wedge\m Zb\m)\ .
\qqq
We shall rewrite the exponential of the latter expression
as a \s$2g$-dimensional Gaussian integral:
\qq
\ee^{-2i\smallint_{_\Sigma}\hs{-0.05cm}\overline{(Zb)^\perp}
\m\wedge\m(Zb)^\perp}&=&\pi^{-g}\
{\rm det}\s(\m{\rm Im}\s\tau)\hs{5cm}\cr
&\cdot& \int\ee^{-\m 4\s\m\bar e_i\s\m({\rm Im}\s\tau)^{ij}\s e_j
\s\s-\s\s2i\s\m e_i\int_{_\Sigma}
\hs{-0.05cm}\omega^i\wedge\m Zb\s\s-\s\s2i\s\m\bar e_i
\int_{_\Sigma}\hs{-0.05cm}\overline{\omega^i\wedge\m Zb}}
\ \prod\limits_i d^2e_i\ .
\label{X3}
\qqq
Gathering Eqs.\s\s(\ref{X1}),\ (\ref{X2}) and (\ref{X3}),
we obtain
\qq
I_X\s\s=\s\s{\rm const}.\ \s{\rm det}\s(\m{\rm Im}\s\tau)\
\left({_{{\rm det}'\m(-\Delta)}\over^{\rm area}}
\right)^{\hs{-0.05cm}-1}\ \s\ee^{- 2i\int_{_\Sigma}
\hs{-0.05cm}\mu^0(c_\alpha\eta^\alpha)\m
\wedge\m Zb\s\s-\s\s2i\m\int_{_\Sigma}\hs{-0.05cm}\overline{
\mu^0(c_\alpha\eta^\alpha)\m\wedge\m Zb}}\hs{1.4cm}\cr
\cdot\ \s\ee^{\s2i\s\int_{_\Sigma}\hs{-0.05cm}
\overline{\mu^0(c_\alpha\eta^\alpha)}
\m\wedge\m\mu^0(c_\alpha\eta^\alpha)}\
\int\ee^{-\m 4\s\m\bar e_i\s\m({\rm Im}\s\tau)^{ij}\s e_j
\s\s-\s\s2i\s\m e_i\int_{_\Sigma}
\hs{-0.05cm}\omega^i\wedge\m Zb\s\s-\s\s2i\s\m\bar e_i
\int_{_\Sigma}\hs{-0.05cm}\overline{\omega^i\wedge\m Zb}}
\ \prod\limits_i d^2e_i\ \ \ \ \ \cr
\hs*{9cm}\cdot\ \s\delta\s(\s c_\alpha\smallint_{_\Sigma}
\hs{-0.03cm}\eta^\alpha\wedge\m b\s)\ .\ \ \ \
\label{X4}
\qqq
\vskip 0.5cm

\subsection{\hspace{-.5cm}.\ \ Integral over \s\s$Z\s$}
\vs 0.4cm
The integral to calculate is
\qq
I_Z\s\s\equiv\s\int\ee^{-i\int_{_\Sigma}\hs{-0.05cm}\langle
\de Z,\m\wedge\de Z\rangle\s\s-\s\s2i\int_{_\Sigma}
\hs{-0.05cm}Z\s(\m\mu^0(c_\alpha\eta^\alpha)\m
+\m e_i\omega^i\m)\m\wedge\m b
\s\s-\s\s2i\int_{_\Sigma}\hs{-0.05cm}\bar Z\s\overline{
(\m\mu^0(c_\alpha\eta^\alpha)\m
+\m e_i\omega^i\m)\m\wedge\m b}}\ \s DZ\ .
\label{Z0}
\qqq
Let us decompose \s$Z\s$ into the part \s$Z_0\s$ in
the kernel of \s$\de\s$ (i.\s e. in \s$H^0(L_\bx^{2})\s$)
\s and the part \s$Z'\s$ orthogonal to \s$H^0(L_\bx^{2})\s$.
Writing \s$Z_0=f_r\m\kappa^r\s$, \s where
\s$(\kappa^r)_{_{r=1}}^{^{g-1}}\s$ is a basis
of \s$H^0(L_\bx^{2})\s$, we obtain
\qq
I_Z\s\s=\s\int\ee^{-i\int_{_\Sigma}\hs{-0.05cm}\langle
\de Z',\m\wedge\de Z'\rangle\s\s-\s\s2i\int_{_\Sigma}
\hs{-0.05cm}Z'\s(\m\mu^0(c_\alpha\eta^\alpha)\m
+\m e_i\omega^i\m)\m\wedge\m b
\s\s-\s\s2i\int_{_\Sigma}\hs{-0.05cm}\bar Z'\s\overline{
(\m\mu^0(c_\alpha\eta^\alpha)\m+
\m e_i\omega^i\m)\m\wedge\m b}}\ \s DZ'
\hs{0.8cm}\cr
\cdot\ {\rm const}.\ {\rm det}\s(\s K_0\m)\s\int
\ee^{\m-2i\m\s f_r\int_{_\Sigma}
\hs{-0.05cm}\kappa^r\s(\m\mu^0(c_\alpha\eta^\alpha)
\m+\m e_i\omega^i\m)\m\wedge\m b
\s\s-\s\s2i\s\m\bar f_r\int_{_\Sigma}\hs{-0.05cm}
\bar\kappa^r\s\overline{
(\m\mu^0(c_\alpha\eta^\alpha)\m+\m
e_i\omega^i\m)\m\wedge\m b}}\
\prod\limits_r d^2f_r\ \cr
\s\s=\s\s{\rm const}.\s\ {\rm det}\s(\s K_0\m)\s
\ \s{\det}'\m(\de_{L_\bx^{2}}^\dagger\m\de_{L_\bx^{2}})^{-1}
\ \prod\limits_r\delta\s(\s
\smallint_{_\Sigma}\hs{-0.02cm}\kappa^r\s
(\m\mu^0(c_\alpha\eta^\alpha)\m+\m e_i
\omega^i\m)\wedge b\s)\ \cr
\cdot\ \s\ee^{\s 4i\int_{_\Sigma}\hs{-0.05cm}\langle
\de^{-1}\m(\m(\mu^0(c_\alpha\eta^\alpha)\m+
\m e_i\omega^i)\m\wedge\m b\m)\s\m,
\m\s\de^{-1}\m(\m(\mu^0(c_\alpha\eta^\alpha)\m
+\m e_i\omega^i)\m\wedge\m b\m)
\m\rangle}\ ,\ \ \ \ \ \ \
\label{Z1}
\qqq
where
\qq
(K_0)^{rs}\s\equiv\s
\smallint_{_\Sigma}\m\langle\kappa^r,
\m\kappa^s\rangle\s{\rm vol}
\qqq
and, in the last line of (\ref{Z1}), \s$\de^{-1}\s$
stands for the inverse
of \s\s$\de_{L^{-2}_\bx K}\s$.
\s\s We may finally collect Eqs.\s\s(\ref{Y1}), (\ref{Y2}),
(\ref{X0}), (\ref{X4}) , (\ref{Z0}) and (\ref{Z1})\m:
\vs 0.02cm
\qq
I_{XYZ}\s\s\equiv\s\s\ee^{-i\int_{_\Sigma}
(\s 2\m(\overline{\de X-Zb})\m\wedge
\m(\de X-Zb)\s
+\s\langle\de Y+2Xb\m,\s\wedge\s(\de Y+2Xb)\rangle
\s+\s\langle\de Z\m,\s\wedge\s\de Z\rangle\s)}\ DY\ DX\ DZ
\hs{0.8cm}\cr\cr
=\s{\rm const}.\ \s{\rm det}\s(\s H_0\m)\ \s
{\rm det}(\m{\rm Im}\s\tau)\ \s{\rm det}\s(\s K_0\m)
\ \s{\det}\s(\de_{L_\bx^{-2}}^\dagger\m\de_{L_\bx^{-2}})^{-1}
\left({_{{\rm det}'\m(-\Delta)}\over^{\rm area}}
\right)^{\hs{-0.05cm}-1} {\det}'\m
(\de_{L_\bx^{2}}^\dagger\m\de_{L_\bx^{2}})^{-1}\cr
\cdot\m \int\hs{-0.05cm}\delta\s(\s\smallint_{_\Sigma}
\hs{-0.04cm}c_\alpha\eta^\alpha
\hs{-0.05cm}\wedge b\m)\s \prod
\limits_r\hs{-0.04cm}\delta\s(\s
\smallint_{_\Sigma}\hs{-0.02cm}\kappa^r\s
(\m\mu^0(c_\alpha\eta^\alpha\hs{-0.025cm})+\hs{-0.02cm}
e_i\omega^i\m)\hs{-0.03cm}\wedge\hs{-0.03cm}b\m)
\ \s\s\ee^{-\bar c_\alpha\s(H_0)^{\alpha\beta}\m
c_\beta\s+\s\m
2i\m\int_{_\Sigma}\hs{-0.05cm}
\overline{\mu^0(c_\alpha\eta^\alpha)}
\m\wedge\m\mu^0(c_\alpha\eta^\alpha)}\cr
\cdot\ \s\ee^{\s-\m 4\s\m\bar e_i\s
\m({\rm Im}\s\tau)^{ij}\s e_j\s\s+\s\s
4i\int_{_\Sigma}\hs{-0.05cm}\langle
\de^{-1}\m(\m(\mu^0(c_\alpha\eta^\alpha)\m+\m e_i\omega^i)
\m\wedge\m b\m)\s\m,
\m\s\de^{-1}\m(\m(\mu^0(c_\alpha\eta^\alpha)\m+\m
e_i\omega^i)\m\wedge\m b\m)
\m\rangle}\ \prod\limits_\alpha
d^2c_\alpha\ \prod\limits_i d^2e_i\cr\cr
=\s{\rm const}.\ \s{\rm det}\s(\s K_0\m)
\ \s{\det}\s(\de_{L_\bx^{-2}}^\dagger\m\de_{L_\bx^{-2}})^{-1}
\left({_{{\rm det}'\m(-\Delta)}\over^{\rm area}}
\right)^{\hs{-0.05cm}-1} {\det}'\m
(\de_{L_\bx^{2}}^\dagger\m\de_{L_\bx^{2}})^{-1}\hs{3.3cm}\cr
\cdot\s\left(\int\delta(BV)\ \s\ee^{-\|CV\|^2}
\ DV\right)\ ,\hs{1cm}
\label{determ}
\qqq
where the last integral is the same as the one introduced
in Section 4.3. Notice that, with the use  of
Eq.\s\s(\ref{llast}), one obtains then
\qq
I_{XYZ}\s\s{\rm det}\m(\m\Omega(1,n))\s\s=\s\m
{\rm const}.\ |z^1|^{-2}\ |{\rm det}\s(\smallint_{_\Sigma}\kappa^r
\omega^i\wedge\m b\s)_{\hs{-0.01cm}_{i<g}}|^{-2}
\s\s{\rm det}\s(\s H_0\m)\s\m
{\rm det}\m(\m{\rm Im}\s\tau)\s\m{\rm det}\s(\s K_0\m)\cr
\ \s{\det}\s(\de_{L_\bx^{-2}}^\dagger\m\de_{L_\bx^{-2}})^{-1}
\left({_{{\rm det}'\m(-\Delta)}\over^{\rm area}}
\right)^{\hs{-0.05cm}-1} {\det}'\m
(\de_{L_\bx^{2}}^\dagger\m\de_{L_\bx^{2}})^{-1}\ ,\hs{1cm}
\label{jeszcz}
\qqq
i.\s e. the \s$V$-integrals cancel.
This ends the formal calculation of
\s\s${\rm det}\s(\bar D_n^\dagger\m\bar D_n)\s$.
\vs 0.5cm

Clearly, the determinants appearing on the right hand side of
the expression (\ref{determ}) need regularization.
If we use the zeta-function
procedure to give sense to them, it is not guaranteed that
the result will coincide with \s\s${\rm det}\s(\bar
D_n^\dagger\m\bar D_n)\s\s$ regularized by the zeta-function
prescription. Indeed, the latter should satisfy the chiral anomaly
relation (\ref{ins2}) but for
\s\s$h\s=\s U\s(\matrix{_{1}&_{v}
\cr^{0}&^{1}})\s U^{-1}\s$, \s\m we obtain
\qq
{\rm det}\s(\s({}^{h^{-1}}\hs{-0.1cm}\bar D_n)^\dagger
\s\s {}^{h^{-1}}\hs{-0.1cm}\bar D_n\s)
\ \s{\rm det}\m(\m\Omega(hh^\dagger,n))^{-1}
\s=\s{\rm det}\s(\bar D_n^\dagger\s\bar D_n\m)\
\s{\rm det}\m(\m\Omega(1,n))^{-1}\ ,\ \ \
\qqq
if we use for \s${\rm det}
\s(\bar D_n^\dagger\m\bar D_n)^{-1}\s$ the expression
on the right hand side of Eq.\s\s(\ref{determ}), instead of
\qq
{\rm det}\s(\s({}^{h^{-1}}\hs{-0.1cm}\bar D_n)^\dagger\s\s
{}^{h^{-1}}\hs{-0.1cm}\bar D_n\s)
\ \s{\rm det}\m(\m\Omega(hh^\dagger,n))^{-1}\hs{5cm}
\ \ \ \ \ \ \cr\cr
=\s\ee^{-{2i\over\pi}\int_{_\Sigma}\hs{-0.04cm}
\langle b+\de v\m,\s
\wedge\m(b+\de v)\rangle\s\s+\s\s{2i\over\pi}
\int_{_\Sigma}\hs{-0.04cm}
\langle b\m,\s\wedge\m b\rangle}\ \s
{\rm det}\s(\bar D_n^\dagger\s\bar D_n\m)\
\s{\rm det}\m(\m\Omega(1,n))^{-1}\ \ \
\label{dobry}
\qqq
given by (\ref{ins2}) (and
Eqs.\s\s(\ref{param}), (\ref{toprove})\m)\m.
\s It is easy to guess that we should correct the formal
result for \s${\rm det}\s(\bar D_n^\dagger\m\bar D_n)\s$
by taking
\qq
{\rm det}\s(\bar D_n^\dagger\m\bar D_n)\s=\s
{\rm const}.\ \s{\rm det}\s(\s K_0\m)^{-1}
\s\s{\det}\s(\de_{L_\bx^{-2}}^\dagger\m\de_{L_\bx^{-2}})\s
\left({_{{\rm det}'\m(-\Delta)}\over^{\rm area}}
\right)\s {\det}'\m
(\de_{L_\bx^{2}}^\dagger\m\de_{L_\bx^{2}})\ \ \cr
\cdot\ \s\ee^{-{2i\over\pi}\int_{_\Sigma}\hs{-0.04cm}
\langle b,\m\wedge b\rangle}\s
\left(\int\delta(BV)\ \s\ee^{-\|CV\|^2}
\ DV\right)^{\hs{-0.1cm}-1}.
\label{determi}
\qqq
Indeed, Eq.\s\s(\ref{jeszcz}) is replaced then by
the relation
\qq
{\rm det}\s(\bar D_n^\dagger\m\bar D_n)
\ \s{\rm det}\m(\m\Omega(1,n))^{-1}\s\s=\s\s
{\rm const}.\ |z^1|^{2}\ |{\rm det}
\s(\smallint_{_\Sigma}\kappa^r
\omega^i\wedge\m b\s)_{\hs{-0.01cm}_{i<g}}|^{2}\
\ee^{-{2i\over\pi}\int_{_\Sigma}\hs{-0.04cm}
\langle b,\m\wedge b\rangle}\hs{1cm}\cr
\cdot\ \s{\rm det}\s(\s H_0\m)^{-1}\ \s
{\rm det}(\m{\rm Im}\s\tau)^{-1}\ \s{\rm det}\s(\s K_0\m)^{-1}
\ \s{\det}\s(\de_{L_\bx^{-2}}^\dagger\m\de_{L_\bx^{-2}})
\left({_{{\rm det}'\m(-\Delta)}\over^{\rm area}}
\right) {\det}'\m
(\de_{L_\bx^{2}}^\dagger\m\de_{L_\bx^{2}})\ ,\hs{0.4cm}
\label{QuiL}
\qqq
and (\ref{dobry}) follows.
We shall show in Appendix D that formula (\ref{determi})
is, indeed, the right expression for the zeta-function
regularized determinant.
Putting it together with the Eq.\s\s(\ref{omega}) from
the end of Section 4.3, we obtain the following
explicit expression
\qq
{\rm det}\s(\bar D_n^\dagger\m\bar D_n)\
{\rm det}\m(\Omega(1,n))^{-1}\m
\left|\s{\rm det}\left(
\int_{_\Sigma}{\rm tr}\s\s\s\omega^\beta(n)
\wedge{_{\da A^{01}(n)}
\over^{\da n_\alpha}}\right)\right|^2\m\prod\limits_\alpha
d^2n_\alpha
\ =\
{\rm const.}\ i^{^{-N}}\ \ \ \ \ \s\s\cr
\cdot\ \s{\rm det}\m(\s H_0\m)^{-1}\
{\rm det}\m(\m{\rm Im}\s\tau)^{-1}
\ {\rm det}\s(\s K_0\m)^{-1}\
{\det}\s(\de_{L_\bx^{-2}}^\dagger\m\de_{L_\bx^{-2}})\s
\left({_{{\rm det}'\m(-\Delta)}\over^{\rm area}}
\right)\s {\det}'\m
(\de_{L_\bx^{2}}^\dagger\m\de_{L_\bx^{2}})\ \ \ \ \ \ \
\m\cr
\cdot\ \s\ee^{-{2i\over\pi}\int_{_\Sigma}\hs{-0.04cm}
\langle b,\m\wedge b\rangle}\ \s
|\s\epsilon_{\alpha_1,\dots,\alpha_{{N}}}
\s\m z^{\alpha_1}\m dz^{\alpha_2}\hs{-0.03cm}\wedge\m
\dots\m\wedge dz^{\alpha_{{N}}}\s|^{^{\wedge 2}}\s
\ \ \ \ \ \ \m\cr
\cdot\s\ {_1\over^i}\s
|\sum\limits_{j=1}^g\s(-1)^j\s\s{\rm det}\m(
\smallint_{_\Sigma}\kappa^r\omega^i
\wedge\m b\s)_{\hs{-0.01cm}_{i\not=j}}\
\omega^j(x)\s|^{^{\wedge 2}}.\ \s\ \
\s\ \
\label{DETer}
\qqq
\vskip 1cm

\nsection{\hspace{-.6cm}.\ \ Functional integral over
\s\s$hh^\dagger\s$}
\vs 0.5cm

We shall attempt now a direct calculation of the functional
integral (\ref{newI}).
Only the \s$\ee^{\m(k+4)\m S(hh^\dagger,\m A_{{\bf x},b})}\s$ term
under it depends on
\s$hh^\dagger\s$ and, as noticed before, its
\s$hh^\dagger$-integral should give the partition function of the
\s$SL(2,\NC)/SU(2)$-valued WZW model. Below, we shall find
its form somewhat surprising.
The action \s$S(hh^\dagger,\m A_{\bx,b})\s$
is explicitly given by Eq.\s\s(\ref{toprove}) in the parametrization
(\ref{param}) of \s$hh^\dagger\s$ by real functions \s$\varphi\s$
and \s$v\in\Gamma(L_0^{-2})\s$. It will be more convenient to use
\s$w\equiv\ee^{-\varphi}v\s$ instead of \s$v\s$. The formal measure
\s$D(hh^\dagger)\s$ becomes then the product
of the formal Lebesgue measures \s$Dw\s$ and \s$D\varphi\s$
determined by the \s$L^2\s$
scalar products \s\s$\smallint_{_\Sigma}\hs{-0.04cm}
\langle w,\m w\rangle\s{\rm vol}\s\s$ and
\s\s$\smallint_{_\Sigma}\hs{-0.04cm}
|\varphi|^2{\rm\s vol}\s$. \s\s Thus we have\m:
\qq
\int\ee^{\m(k+4)\m S(hh^\dagger,\m A_{\bx,b})}\ D(hh^\dagger)
\s\s=\s\s\ee^{-{k+4\over 2\pi i}\hs{-0.1cm}
\int_{_\Sigma}\hs{-0.1cm}
\langle b\s,\s\wedge\m b\rangle}\ \ \hs{6.5cm}\cr
\cdot\s\int\ee^{{k+4\over 2\pi i}\int_{_\Sigma}\hs{-0.02cm}[\s
-\varphi\s(\da\de\varphi-2F_0\m)\s\m+
\m\s\langle\s\ee^{-\varphi}b
+(\de+(\de\varphi))\m w\s,\m\s\wedge\m(
\ee^{-\varphi}b+
(\de+(\de\varphi))\m w)\s\rangle\m]}\ Dw\s\s D\varphi\s ,\ \ \ \
\qqq
where \s$\de\equiv\de_{L^{-2}_\bx}\s$ when acting on \s$w\s$
and \s$F_0\s$ is the curvature form of the holomorphic
connection of \s\m$L_0\s$
preserving the fixed hermitian metric.
Note that the field \s$w\s$ enters quadratically in
the action so that the \s$w$-integral is Gaussian and may
be easily performed:
\qq
\int\ee^{{k+4\over 2\pi i}\int_{_\Sigma}\hs{-0.02cm}
\langle\s\ee^{-\varphi}b
+(\de+(\de\varphi))\m w\s,\m\s\wedge\m(
\ee^{-\varphi}b+
(\de+\de(\varphi))\m w)\s\rangle}\ Dw\hs{2cm}\cr
=\s\ee^{{k+4\over 2\pi i}\int_{_\Sigma}\hs{-0.02cm}
\langle\s P_\varphi(\ee^{-\varphi}b)\s,
\s\m\wedge\m P_\varphi(\ee^{-\varphi}b)
\s\rangle}\ \s{\rm det}\left((\de+(\de\varphi))^\dagger\s
(\de+(\de\varphi))\right)^{-1}\ ,
\label{HH0}
\qqq
where \s$P_\varphi\s$ denotes the orthogonal projector on
the kernel of \s$(\de+(\de\varphi))^\dagger\s$. \s Explicitly,
\qq
P_\varphi(\ee^{-\varphi}b)\s=\s i\s\ee^{\varphi}\s
(\eta^\alpha)^\dagger\ \m
(H_\varphi^{-1})_{_{\beta\alpha}}\ \smallint_{\Sigma}\hs{-0.04cm}
\eta^\beta\wedge\m b\ ,\hs{3.5cm}\cr
i\smallint_{_\Sigma}
\langle\s P_\varphi(\ee^{-\varphi}b)\s,
\s\m\wedge\m P_\varphi(\ee^{-\varphi}b)
\s\rangle\s=\s(\smallint_{_\Sigma}\overline{\eta^\alpha
\wedge\m b}\m)
\ \m(H_\varphi^{-1})_{_{\beta\alpha}}\ \m(\smallint_{_\Sigma}
\eta^\beta\wedge\m b
\m)\s\equiv\s \bar z^\alpha\ (H_\varphi^{-1})_{_{\beta\alpha}}
\ z^{\beta}\ ,\hs{0.3cm}
\qqq
with the matrix of the modified scalar products of the vectors
of the basis \s$(\eta^\alpha)\s$ of \s$H^0(L^2_\bx K)\s$
\qq
(H_\varphi)^{\alpha\beta}\s\equiv\s{_1\over^i}
\smallint_{_\Sigma}\hs{-0.04cm}
\ee^{\m 2\varphi}\s\langle\s\eta^\alpha\m,
\m\wedge\eta^\beta\s\rangle\ ,
\qqq
compare Eq.\s\s(\ref{h0}). For convenience, we shall rewrite
the exponential on the right hand side of Eq.\s\s(\ref{HH0})
as a finite-dimensional Gaussian integral:
\qq
\ee^{{k+4\over 2\pi i}\int_{_\Sigma}\hs{-0.02cm}
\langle\s P_\varphi(\ee^{-\varphi}b)\s,
\s\m\wedge\m P_\varphi(\ee^{-\varphi}b)
\s\rangle}\s=\s\s\ee^{-{k+4\over 2\pi}\s
\bar z^\alpha(H^{-1}_\varphi)_{_{\beta\alpha}} z^\beta}\hs{4cm}\cr
=\s({_2\over^{k+4}})^{^{N}}\s{\rm det}\s(\m H_\varphi)\s
\int\ee^{-{2\pi\over k+4}\s\m\bar c_\alpha\s
(H_\varphi)^{\alpha\beta}
\s c_\beta\m\s+\s\m i\s c_\alpha z^{\alpha}\s\s
+\s\s i\s\bar c_\alpha\bar z^{\alpha}}
\s\m\prod\limits_\alpha d^2c_\alpha\ .
\label{HH1}
\qqq
The \s$\varphi$-dependence of the product of determinants
\s${\rm det}\m(H_\varphi)\s\s{\rm det}
\hs{-0.1cm}\left((\de+(\de\varphi))^\dagger\m
(\de+(\de\varphi))\right)^{\hs{-0.07cm}-1}\s$
with the second one
regularized by the zeta-function prescription (or any other
gauge invariant procedure) is given by the chiral anomaly\m:
\qq
\delta \s\ln\left({\rm det}\s(\m H_\varphi)\ \s{\rm det}
\hs{-0.08cm}\left((\de_{L^{-2}_\bx}+(\de\varphi))^\dagger\s
(\de_{L^{-2}_\bx}+(\de\varphi))\right)^{-1}\right)\s=\s
{_{2}\over^{\pi i}}\smallint_{_\Sigma}\m(\delta\varphi)
\s(\da\de\varphi-F_0)\ \ \cr
+\s{_1\over^{2\pi i}}\smallint_{_\Sigma}\m
(\delta\varphi)\m R\ ,
\label{anoma}
\qqq
where \s$R\s$ is the metric curvature form of (the holomorphic
tangent bundle of) \s$\Sigma\s$ normalized so that
\s$\smallint_{_\Sigma}\hs{-0.05cm}
R\s=\s4\pi i(g-1)\s$. \s The global form of the formula
(\ref{anoma}) is
\qq
{\rm det}\s(\m H_\varphi)\ \s{\rm det}
\left((\de_{L^{-2}_\bx}+(\de\varphi))^\dagger\s
(\de_{L^{-2}_\bx}+(\de\varphi))\right)^{-1}\hs{2.2cm}\cr\cr
=\s\s\ee^{\m{1\over\pi i}\int_{_\Sigma}\hs{-0.04cm}
\varphi\s(\da\de\varphi-2F_0)\s\s+
\s\s{1\over 2\pi i}\int_{_\Sigma}\hs{-0.05cm}\varphi\m R}
\ \s{\rm det}\s(\m H_0)\ \s\s
{\rm det}\s(\m\de_{L^{-2}_\bx}^\dagger\m
\de_{L^{-2}_\bx}\m)^{-1}\ .
\label{HH2}
\qqq
Gathering Eq.\s\s(\ref{HH0}), (\ref{HH1}) and (\ref{HH2}),
we obtain
\qq
\int\ee^{{k+4\over 2\pi i}\int_{_\Sigma}\hs{-0.02cm}
\langle\s\ee^{-\varphi}b
+(\de+(\de\varphi))\m w\s,\m\s\wedge\m(
\ee^{-\varphi}b+
(\de+\de(\varphi))\m w)\s\rangle}\ Dw\hs{2.8cm}\cr
=\s{\rm const}.\ \s\s{\rm det}\s(\m H_0)\ \s\s
{\rm det}\s(\m\de_{L^{-2}_\bx}^\dagger\m
\de_{L^{-2}_\bx}\m)^{-1}
\s\s\ee^{\m{1\over\pi i}\int_{_\Sigma}\hs{-0.04cm}
\varphi\s(\da\de\varphi-2F_0)\s\s+
\s\s{1\over 2\pi i}\int_{_\Sigma}
\hs{-0.05cm}\varphi\m R}\ \s\s\cr
\cdot\s \int\ee^{-{2\pi\over k+4}
\s\m\bar c_\alpha\s(H_\varphi)^{\alpha\beta}
\s c_\beta\m\s+\s\s i\s c_\alpha z^{\alpha}\s\s
+\s\m i\s\bar c_\alpha\bar z^{\alpha}}
\ \prod\limits_\alpha d^2c_\alpha\ .
\label{winteg}
\qqq
Here appears a new difficulty in the calculation
of the scalar product of CS states,
as compared to the genus zero and one cases
studied in \cite{Quadr}\cite{FalGaw0} and \cite{GK},
respectively. There, the \s$hh^\dagger\s$ integral
for the partition function of the \s$SL(2,\NC)/SU(2)$-valued
WZW theory led, after parametrization of \s$hh^\dagger\s$
by \s$\varphi\s$ and \s$w\s$, to an iterative Gaussian
integral\m: after calculation of the Gaussian \s$w\s$
integral, the remaining \s$\varphi\s$ integral was,
miraculously, also becoming Gaussian. This does not seem
to be the case here. The right hand side of Eq.\s\s(\ref{winteg})
includes the term
\qq
\ee^{-{2\pi\over k+4}
\s\m\bar c_\alpha\s(H_\varphi)^{\alpha\beta}
\s c_\beta}\s\equiv\s\s\ee^{{2\pi i\over k+4}\int_{_\Sigma}
\hs{-0.03cm}\ee^{2\varphi}\s\langle\m c_\alpha\eta^\alpha\m,
\m\wedge c_\beta\eta^\beta
\rangle}
\qqq
with the Liouville type terms containing \s$\ee^{2\varphi}\s$
in the exponential. So the \s$\varphi$-integral obtained
after integrating out \s$w\s$ seems to be of a non-Gaussian
type, in contrast to the low genera situation.
\vs 0.5cm

We shall show however, that this difficulty may be solved
by a trick used in the Liouville
theory \cite{GTW}\cite{GL}.
The functional integral over \s$\varphi\s$ which we are
left with has the form\m:
\qq
I_\varphi\s\s\equiv\s
\int\ee^{-{k+2\over 2\pi i}\int_{_\Sigma}\hs{-0.02cm}\s
\varphi\s(\da\de\varphi-2F_0\m)\s\s
+\s\s{1\over 2\pi i}\int_{_\Sigma}
\hs{-0.05cm}\varphi\m R\s\s-\s\s{2\pi\over k+4}
\s\m\bar c_\alpha\s(H_\varphi)^{\alpha\beta}\s c_m}
\ \s D\varphi\ .
\qqq
We shall integrate first over the zero
mode of \m\s$\varphi_0\equiv
(\smallint_{_\Sigma}\hs{-0.03cm}
\varphi\s{\rm vol}\m)/({\rm area})^{1/2}\s$.
For this purpose, let us multiply
\s$I_\varphi\s$ by \s\s$1\s=\s
({\rm area})^{1/2}\s\cdot\smallint\delta
(\varphi_0-a\s({\rm area})^{1/2})\s da\s$. \s\s Changing
the order of integration, shifting \s$\varphi\s$ to
\s$\varphi+a\s$ and setting \s$M\equiv(k+1)(g-1)\s$,
\s we obtain
\qq
({\rm area})^{-1/2}\s I_\varphi\s=\m
\int\hs{-0.03cm}\ee^{-2aM}\ \hs{-0.04cm}
\ee^{-{k+2\over 2\pi i}\hs{-0.05cm}\int_{_\Sigma}\hs{-0.1cm}
\varphi\m(\da\de\varphi-2F_0\m)\s
+\s{1\over 2\pi i}\int_{_\Sigma}
\hs{-0.08cm}\varphi\m R\s\m-\s\m{2\pi\over k+4}
\s\m\ee^{2a}\s\bar c_\alpha\s(H_\varphi)^{\alpha\beta}
\m c_\beta}\s\m
da\ \delta(\varphi_0)\ D\varphi\hs{-0.01cm}\cr
=\s\s{_1\over^2}\s\Gamma(-M)\s\s
({_{2\pi}\over^{k+4}})^{M}\hs{-0.08cm}\int\hs{-0.05cm}
\ee^{-{k+2\over 2\pi i}
\s\int_{_\Sigma}\hs{-0.1cm}
\varphi\s(\da\de\varphi-2F_0\m)\s
+\s{1\over 2\pi i}\int_{_\Sigma}
\hs{-0.08cm}\varphi\m R}
\ (\m\bar c_\alpha\s(H_\varphi)^{\alpha\beta}
\s c_\beta\m)^{M}
\ \delta(\varphi_0)\s\ D\varphi\ ,\hs{0.75cm}
\label{GL}
\qqq
where we have used the relations \s${_i\over^{2\pi}}
\smallint_{_\Sigma}\hs{-0.03cm}F_0={\rm deg}(L_0)=g-1\s\s$
and \s\s${_i\over^{2\pi}}
\smallint_{_\Sigma}\hs{-0.03cm}R={\rm deg}(K^{-1})=2(1-g)\s$.
As we see, the integration over the zero mode of \s$\varphi\s$
diverges but may be easily (multiplicatively) regularized
by removing the overall divergent factor
\s$(-1)^M\m\Gamma(-M)\s$.
Now, the \s$c$-integral is easy to perform\m:
\qq
\int(\m-\m\bar c_\alpha\s(H_\varphi\m)^{\alpha
\beta}\s c_\beta)^{M}
\s\s\ee^{\s i\s c_\alpha z^{\alpha}\m
+\s\m i\s\bar c_\alpha\bar z^{\alpha}}\m\prod\limits_\alpha
d^2c_\alpha\s
=\s(2\pi)^{2N)}\s\m
((H_\varphi)^{\alpha\beta}\m\da_{\bar z^{\alpha}}
\da_{z^\beta}\hs{-0.03cm})^{M}\prod\limits_\alpha
\hs{-0.05cm}
\delta(z^{\alpha})\s.\hs{0.6cm}
\qqq
Gathering the above results, we obtain the following
``Coulomb gas representation'' for the higher genus
partition function of the \s$SL(2,\NC)/SU(2)$-valued
WZW model\m:
\qq
\int\hs{-0.1cm}\ee^{\m(k+4)\m S(hh^\dagger\hs{-0.02cm},
\m A_{{\bf x},b})}\s\m D(hh^\dagger)
\s=\s{\rm const}.\ ({\rm area})^{1/2}\s\s {\rm det}\s(\m H_0)
\s\s{\rm det}\s(\m\de_{L^{-2}_\bx}^\dagger\m
\de_{L^{-2}_\bx})^{-1}\s
\ee^{-{k+4\over 2\pi i}\hs{-0.05cm}\int_{_\Sigma}\hs{-0.05cm}
\langle b\s,\s\wedge\m b\rangle}\ \ \cr
\cdot\s\left(
\int\ee^{-{k+2\over 2\pi i}\int_{_\Sigma}\hs{-0.1cm}\s
\varphi\s(\da\de\varphi-2F_0\m)\s\s
+\s\s{1\over 2\pi i}\int_{_\Sigma}
\hs{-0.05cm}\varphi\m R}\ ((H_\varphi)^{\alpha\beta}
\m\da_{\bar z^{\alpha}}
\da_{z^\beta}\hs{-0.03cm})^{M}\
\delta(\varphi_0)\s\ D\varphi\right)
\prod\limits_\alpha\hs{-0.03cm}
\delta(z^{\alpha})\ \ \hs{0.35cm}\cr
=\s{\rm const}.\ ({\rm area})^{1/2}\s\s{\rm det}\s(\m H_0)\
{\rm det}\s(\m\de_{L^{-2}_\bx}^\dagger\m
\de_{L^{-2}_\bx}\hs{-0.04cm})^{-1}\s\s
\ee^{-{k+4\over 2\pi i}\hs{-0.05cm}\int_{_\Sigma}\hs{-0.05cm}
\langle b\s,\s\wedge\m b\rangle}\m\bigg(\hs{-0.09cm}
\prod\limits_m
\m\da_{\bar z^{\alpha_m}}\da_{z^{\beta_m}}
\hs{-0.09cm}\prod\limits_\alpha
\delta(z^{\alpha})\hs{-0.03cm}\bigg)\hs{0.7cm}\cr
\cdot\s \int\hs{-0.16cm}\left(
\int\hs{-0.07cm}
\ee^{-{k+2\over 2\pi i}\int_{_\Sigma}\hs{-0.1cm}\s
\varphi\s(\da\de\varphi-2F_0\m)\s\s
+\s{1\over 2\pi i}\int_{_\Sigma}
\hs{-0.05cm}\varphi\m R\s\s+\s\s2\sum_{_m}\hs{-0.1cm}
\varphi(x_m)}\s\s\delta(\varphi_0)\s\s D\varphi\right)
\hs{-0.05cm}\prod\limits_m\hs{-0.06cm}
{_1\over^i}\langle\eta^{\alpha_m}\hs{-0.06cm},
\wedge\m\eta^{\beta_m}
\hs{-0.05cm}\rangle(x_m\hs{-0.04cm})\s,\hs{0.7cm}
\label{PFN}
\qqq
where \s$m\s$ runs from \s$1\s$ to \s$M\s$. \s
Observe, that the integrand in the functional integral
over \s$\varphi\s$ is now invariant under
constant shifts of \s$\varphi\s$, \s except for
the term \s$\delta(\varphi_0)\s$ (the neutrality
of the Coulomb gas). The \s$\varphi$-integral is of the
Gaussian form
\qq
\int\hs{-0.05cm}\ee^{{k+2\over 2\pi i}
\int_{_\Sigma}\hs{-0.1cm}\da\varphi\s\wedge
\de\varphi\s\s-\s\s i\int_{_\Sigma}\hs{-0.05cm}\varphi\s\sigma}
\s\s\delta(\varphi_0)\ D\varphi\s
=\s{\rm const}.\ {\rm det}'\m(-\Delta)^{-1/2}\
\ee^{\s{\pi\over k+2}\int_{_\Sigma}\hs{-0.05cm}\int_{_\Sigma}
\hs{-0.05cm}\sigma(x)\s G(x,y)\s\m\sigma(y)}\m,\hs{0.48cm}
\label{GI}
\qqq
where \s$G(\m\cdot\s,\m\cdot\m)\s$ is a Green function
of the Laplacian on \s$\Sigma\s$ and \s\s$\sigma
={_{k+2}\over^\pi} F_0+{_1\over^{2\pi}} R
+{2i}\sum_{_m}\hs{-0.08cm}\delta_{x_m}\s$.
Since \s$G(x,y)\sim{_1\over^{2\pi}}\s\ln\s d(x,y)\s$, \s
where \s$d(\m\cdot\s,\m\cdot\m)\s$ is the metric distance,
the terms \s\s$\ee^{-{4\pi\over k+2}\m G(x_m,\m x_m)}\s\s$
on the right hand side of (\ref{GI}) are divergent. They
may be easily multiplicatively renormalized by replacing them
by their normal ordered version \s\s$\ee^{-{4\pi\over k+2}
\m:G(x_m,\m x_m):}\s$, \s\m see Eq.\s\s(\ref{Wick})
for the definition.
This way, choosing for simplicity the Green function
satisfying \s$\smallint_{_\Sigma}\hs{-0.05cm}
G(\m\cdot\s,\m y)\s((k+2)F_0(y)+\hf R(y))=0\s$, \s we obtain
\qq
\int\hs{-0.05cm}\ee^{{k+2\over 2\pi i}
\int_{_\Sigma}\hs{-0.1cm}\da\varphi\s\wedge
\de\varphi\s\s-\s\s i\int_{_\Sigma}\hs{-0.05cm}\varphi\s\sigma}
\s\s\delta(\varphi_0)\ D\varphi\s\s
=\s\s{\rm const}.\s\ {\rm det}'\m(-\Delta)^{-1/2}
\s\bigg(\prod\limits_{m_1\not=m_2}
\ee^{-{4\pi\over k+2}\m G(x_{m_1},\m x_{m_2})}\bigg)\ \cr
\cdot\s\bigg(\prod\limits_m \ee^{-{4\pi\over k+2}
\m :G(x_m,\m x_m):}\bigg)\m.\hs{0.75cm}
\qqq
The substitution of this result into Eq.\s\s(\ref{PFN})
results in the relation
\qq
\int\hs{-0.05cm}\ee^{\m(k+4)\m S(hh^\dagger\hs{-0.02cm},
\m A_{\bx,b})}\ D(hh^\dagger)\s\s
=\s\s{\rm const}.\hs{7.7cm}\s\ \cr
\cdot\ \s{\rm det}\s(\m H_0)\s\
{\rm det}\s(\m\de_{L^{-2}_\bx}^\dagger\m
\de_{L^{-2}_\bx})^{-1}\s
\left({_{{\rm det}'\m(-\Delta)}\over^{{\rm area}}}
\right)^{\hs{-0.09cm}-1/2}\s
\ee^{-{k+4\over 2\pi i}\hs{-0.05cm}\int_{_\Sigma}\hs{-0.05cm}
\langle b\s,\s\wedge\m b\rangle}\s
\bigg(\hs{-0.09cm}
\prod\limits_m
\m\da_{\bar z^{\alpha_m}}\da_{z^{\beta_m}}
\hs{-0.09cm}\prod\limits_\alpha
\delta(z^{\alpha})\hs{-0.03cm}\bigg)\s\ \ \ \cr
\cdot\s\int\bigg(\prod\limits_{m_1\not=m_2}
\ee^{-{4\pi\over k+2}\m G(x_{m_1},\m x_{m_2})}\bigg)\s
\bigg(\prod\limits_m \ee^{-{4\pi\over k+2}
\m :G(x_m,\m x_m):}
\s\m{_1\over^i}\langle\eta^{\alpha_m},\wedge\eta^{\beta_m}
\hs{-0.05cm}\rangle(x_m\hs{-0.04cm})\bigg)
\hs{0.3cm}\ \hs{0.6cm}\label{HYPF}
\qqq
which is the final formula for the higher genus
partition function of the \s$SL(2,\NC)/SU(2)$-valued
WZW model.
\vs 0.5cm

Equation (\ref{HYPF}) reduces the functional integral
over \s$hh^\dagger\s$ to a finite dimensional integral
over \s$M\s$ copies of \s$\Sigma\s$. The integrand is a
smooth function except for
\s$\CO(d(x_{m_1},\m x_{m_2})^{-{4\over k+2}})\s$ singularities
at coinciding points. Power counting shows, that the
integral converges for \s$g=2\s$ but for higher
genera it diverge unless special combinations
\qq
\gamma_{(\alpha_m),\m(\beta_m)}\s\prod\limits_m
\langle\eta^{\alpha_m},\wedge\eta^{\beta_m}
\hs{-0.03cm}\rangle(x_m\hs{-0.02cm})
\qqq
of forms are integrated. We shall return to this issue below.
Another feature of the right hand side of Eq.\s\s(\ref{HYPF})
may look even more surprising in a candidate for the
partition function:
its dependence of the external field \s$A_{\bx,b}\s$
is not functional but distributional\m! The entire
dependence on \s$b\in\wedge^{01}(L_0)\s$
resides in the term \s\s$\prod\limits_m
\da_{\bar z^{\alpha_m}}\da_{z^{\beta_m}}
\hs{-0.09cm}\prod\limits_\alpha
\delta(z^{\alpha})\s\s$ (recall that
\s$z^{\alpha}\equiv\smallint_{_\Sigma}
\hs{-0.06cm}\eta^\alpha\wedge\m b\s$)\nobreak\s.
\s\s This is not so astonishing in view of the fact that
the partition function of the \s$SL(2,\NC)/SU(2)\s$ WZW
may be expected, by formal arguments
similar to the ones used in \cite{Wittenfact},
to be the hermitian square of a
holomorphic section of a negative power of the
determinant bundle. But there are no such sections.
There exist, however distributional solutions
of the corresponding Ward identities and the right hand
side of (\ref{HYPF}) is one of them.
\vskip 1cm

\nsection{\hspace{-.6cm}.\ \ Assembling the final formula}
\vs 0.5cm

The main results (\ref{DETer}) and (\ref{HYPF}) of the
calculations of the last two Sections permit to
reduce the formal scalar product formula (\ref{newI})
to the following finite-dimensional integral\nobreak:
\qq
\|\Psi\|^2\ \ \ \s\s\s=\ \ \ \s\s\s{\rm const}.\m\ i^{^{-N}}
\s\ {\rm det}\s
({\rm Im}\s\tau)^{-1}\s
\left({_{{\rm det}'\m(-\Delta)}\over^{{\rm area}}}
\right)^{\hs{-0.07cm}1/2}\ee^{-{ik\over 2\pi}\int_{_\Sigma}
\hs{-0.05cm}\tr\s\m A_0^{10}\wedge\m A^{01}_0}
\ \s\s\s\s\m\cr
\cdot\s\int{\rm det}\m(\m\smallint_{_\Sigma}\m\langle\kappa^r,
\m\kappa^s\rangle\s{\rm vol}\s)^{-1}\s\s\s{\rm det}'\m
(\m\de_{L^{2}_\bx}^\dagger\m
\de_{L^{2}_\bx})\ \m\s\ee^{-\s2\pi k\s(\smallint_{_{
x_0}}^{^{{\bf x}}}\hs{-0.03cm}
\bar\omega)\s(\m{\rm Im}\s\tau\m)^{-1}
(\smallint_{_{x_0}}^{^{{\bf x}}}\hs{-0.03cm}
\omega)}\ \s\s\s\s\cr
\cdot\s\ |\m\sum\limits_{j=1}^g\s(-1)^{j}\s\s
{\rm det}\m(
\smallint_{_\Sigma}\kappa^r\omega^i
\wedge\m b\s)_{\hs{-0.01cm}_{i\not=j}}\
\omega^j(x)\s\s\s\psi(\bx,\m b)\s|^{^{\wedge 2}}
\ \bigg(\hs{-0.09cm}\prod\limits_m
\m\da_{\bar z^{\alpha_m}}\da_{z^{\beta_m}}
\hs{-0.09cm}\prod\limits_\alpha
\delta(z^{\alpha})\hs{-0.03cm}\bigg)\ \s\s\m\cr
\cdot\ \s|\s\epsilon_{\alpha_1,\dots,\alpha_{{N}}}
\s\m z^{\alpha_1}\m dz^{\alpha_2}\hs{-0.03cm}\wedge\m
\dots\m\wedge dz^{\alpha_{{N}}}\s|^{^{\wedge 2}}\s
\bigg(\prod\limits_{m_1\not=m_2}
\ee^{-{4\pi\over k+2}\m G(x_{m_1},\m x_{m_2})}\bigg)\ \s\s\m\cr
\cdot\ \bigg(\prod\limits_m \ee^{-{4\pi\over k+2}
\m :G(x_m,\m x_m):}
\s\m{_1\over^i}\langle\eta^{\alpha_m},\wedge\eta^{\beta_m}
\hs{-0.05cm}\rangle(x_m\hs{-0.04cm})\bigg)\ .
\qqq
The integral is, for fixed \s$\bx\s$, \s over
the \s$(N-1)\equiv(3g-4)$-dimensional projective space
with homogeneous coordinates \s$(z^{\alpha})\s$, over the
Cartesian product of \s$M\equiv(k+1)(g-1)\s$ copies of
\s$\Sigma\s$ (variables \m\s$x_m\s$) \s and, finally,
over the projection \s$x\s$ of \s$\bx\in\tilde
\Sigma\s$ to \s$\Sigma\s$.
Let us discuss first the \s$z^{\alpha}\s$ integral.
It has the form
\qq
I_z\s\equiv\hs{-0.2cm}\int\limits_{\NP H^1(L^{-2}_\bx)}
\hs{-0.3cm}|P(z)|^2\s\bigg(\hs{-0.09cm}\prod\limits_m
\m\da_{\bar z^{\alpha_m}}\da_{z^{\beta_m}}
\hs{-0.09cm}\prod\limits_\alpha
\delta(z^{\alpha})\hs{-0.03cm}\bigg)\s
\s|\s\epsilon_{\alpha_1,\dots,\alpha_{{N}}}
\s z^{\alpha_1} dz^{\alpha_2}\hs{-0.08cm}\wedge
\dots\wedge\hs{-0.05cm}dz^{\alpha_{{N}}}\m|^{^{\wedge 2}}
\s,\ \ \ \
\qqq
where
\qq
P(z)\s=\s
\sum\limits_{j=1}^g\s(-1)^j\s\s{\rm det}\m(
\smallint_{_\Sigma}\kappa^r\omega^i
\wedge\m b\s)_{\hs{-0.01cm}_{i\not=j}}\
\omega^j(x)\s\s\s\psi(\bx,\m b)
\label{P(z)}
\qqq
is a homogeneous polynomial in \s$(z^{\alpha})\s$ of degree
\s$M\s$ (with values in \s$K|_x\s$)\m. \s The integrand
is a distributional \s$2(N-1)$-form on \s$H^1(L^{-2}_\bx)\s$
invariant under complex rescalings of \s$z\s$.
Note that formally
\qq
\int\limits_{H^1(L^{-2}_\bx)}
\hs{-0.3cm}|P(z)|^2\s\s\bigg(\hs{-0.09cm}\prod\limits_m
\m\da_{\bar z^{\alpha_m}}\da_{z^{\beta_m}}
\hs{-0.09cm}\prod\limits_\alpha
\delta(z^{\alpha})\hs{-0.09cm}\bigg)
\s|\s dz^1\hs{-0.04cm}\wedge\cdots\wedge dz^{{N}}\m|^2\s
=\s{_1\over^{(N-1)!}}\m
\s(\smallint_\NC|\lambda^{-1}d\lambda|^2)
\ I_z\s ,\ \ \
\qqq
where the divergent integral on the right hand side is
over the fibers of \m the projection
of \s\s$H^1(L^{-2}_\bx)\s\s$ onto \s\s$\NP
H^1(L^{-2}_\bx)\s$. \s Of course,
the left hand side is perfectly well defined and we shall
take it as a definition of the
right hand side. One may expect
to reabsorb this way the infinite
constant \s$(-1)^M\m\Gamma(-M)\s$
produced by the integration of the zero mode of the field
\s$\varphi\s$, see Eq.\s\s(\ref{GL}). This is more then
formal gymnastics. In Appendix E, we show
that changing the order
of integration in the above arguments by computing
the integral over \s$\NP H^1(L^{-2}_\bx)\s$ modular
degrees of freedom just after the \s$w\s$
functional integration
and the one over the scalar field
\s$\varphi\s$ only afterwards,
one obtains the same final result but no infinite constants,
apart of those of the Wick ordering, appear in the intermediate
steps. This way, it is rather the convergent integration
over the (part of) the modular
degrees of freedom then the divergent
Gupta-Trivedi-Wise-Goulian-Li trick which removes the
cumbersome Liouville-type terms from the effective
action for \s$\varphi\s$ and renders the \s$\varphi\s$ integral
calculable. It is an interesting question whether similar
arguments may be used to substantiate the Goulian-Li
trick in the gravity case.
\vskip 0.5cm

With the above interpretation of \s$I_z\s$, we obtain
the following expression for the scalar product of genus
\s$g\s$ CS states\m:
\qq
\|\Psi\|^2\ \ \ \s\s=\ \ \ \s\s{\rm const}.\m
\s\ {\rm det}\s
({\rm Im}\s\tau)^{-1}\s
\left({_{{\rm det}'\m(-\Delta)}\over^{{\rm area}}}
\right)^{\hs{-0.07cm}1/2}\ee^{-{ik\over 2\pi}\int_{_\Sigma}
\hs{-0.05cm}\tr\s\m A_0^{10}\wedge\m A^{01}_0}
\s\hs{1.1cm}\s\s\cr
\cdot\s\int{\rm det}\m(\m\smallint_{_\Sigma}\m
\langle\kappa^r_\bx,
\m\kappa^s_\bx\rangle\s{\rm vol}\s)^{-1}\s\s\s{\rm det}'\m
(\m\de_{L^{2}_\bx}^\dagger\m
\de_{L^{2}_\bx})\ \m\s\ee^{-\s2\pi k\s(\smallint_{_{
x_0}}^{^{{\bf x}}}\hs{-0.03cm}
\bar\omega)\s(\m{\rm Im}\s\tau\m)^{-1}
(\smallint_{_{x_0}}^{^{{\bf x}}}\hs{-0.03cm}
\omega)}\ \s\s\s\s\cr
\cdot\ \prod\limits_m
\m\da_{\bar z^{\alpha_m}_{\bf x}}\da_{z^{\beta_m}_{\bf x}}
|_{_{b=0}}\s\s\s
\bigg({_1\over^i}\s\m
|\m\sum\limits_{j=1}^g\s(-1)^{j}\s\s
{\rm det}\m(\smallint_{_\Sigma}\kappa^r_\bx\omega^i
\wedge\m b\s)_{\hs{-0.01cm}_{i\not=j}}\
\omega^j(x)\s\s\s\psi(\bx,\m b)\s|^{^{\wedge 2}}
\hs{-0.02cm}\bigg)\ \s\s\s\cr
\cdot\ \s
\bigg(\hs{-0.1cm}\prod\limits_{m_1\not=m_2}
\ee^{-{4\pi\over k+2}\m G(x_{m_1},\m x_{m_2})}\bigg)
\bigg(\prod\limits_m \ee^{-{4\pi\over k+2}
\m :G(x_m,\m x_m):}
\s\m{_1\over^i}\langle\eta^{\alpha_m}_\bx,
\wedge\eta^{\beta_m}_\bx
\hs{-0.05cm}\rangle(x_m\hs{-0.04cm})\bigg)\ ,
\label{final}
\qqq
with the \s$(M+1)$-fold integration over \s$\Sigma\s$
(over the projection \s$x\s$ of \s$\bx\s$ to \s$\Sigma\s$
and over the positions \s$x_m\s$ of \s$M\s$
``screening charges''). \s We have restored the \s$\bx\s$
subscript to stress the \s$\bx\s$ dependence
of various entries in the integrated function.
In fact, it is easy to see that the latter
depends only on \s$x\s$. \s In Appendix F, we submit
formula (\ref{final}) to few consistency checks
showing that \s$\|\Psi\|^2\s$ does
not depend on the choices of the bases \s$(\kappa^r_\bx)\s$
of \s$H^0(L_\bx^2)\s$ and \s$(\eta^\alpha_\bx)\s$ of
\s$H^0(L^2_\bx K)\s$, and of the choice of a hermitian structure
of \s$L_0\s$. \s We also show that upon multiplication
of the Riemannian
metric of \s$\Sigma\s$ by a function \s$\ee^\sigma\s$,
\s\s$\|\Psi\|^2\s$ picks up the factor
\qq
\exp[\m-{_i\over^{24\pi}}\m{_{3k}\over^{k+2}}
\s\smallint_{_\Sigma}\m(\m
\hf\m\da\sigma\wedge\de\sigma\m+\m\sigma R\m)\m]
\qqq
which guarantees the
right value \s$c={3k\over k+2}\s$ of the Virasoro central
charge of the theory with partition function given
by Eq.\s\s(\ref{PartF}).
\vs 0.5cm

We shall rewrite the scalar product
formula putting it into a form both more geometric
and closer to the spirit
of discussion of Sect.\s\s4. To this end consider
\qq
\chi(\bx;\m x_1,\dots,x_{k(g-1)})\s=\sum_{(\alpha_m)}\prod
\limits_{m=1}^{k(g-1)}
\left(\s\eta^{\alpha_m}_\bx(x_m)\s\s\da_{z^{\alpha_m}_{\bf x}}
\right)\s\psi(\bx,b)\ .
\label{End}
\qqq
Clearly, the relation (\ref{hoMO}) holds
so that \s$\chi\s$ is the holomorphic
\s$k(g-1),0$-form on \s$\Sigma\times
\Sigma^{k(g-1)}\s$ with values
in the h.l.b. $B_k\s$ of Eq. (\ref{B})
discussed at the end of Sect.\s\s3.
It is essentially the same object as \s$\chi\s$
introduced by Eq.\s\s(\ref{chi}) is Sect.\s\s4,
directly related to Bertram's picture \cite{Bertram}
of CS states. The precise relation between
the two \s$\chi$'s is given by Eq.\s\s(\ref{JaBert})
of Appendix C. With its use, one
obtains from Eq.\s\s(\ref{final})
a fully normalized formula for the scalar product
which uses the description of CS states
by polynomials \s$\psi'\s$ discussed in Sect.\s\s4.
\vs 0.4cm

The expression
\qq
\ee^{-\s2\pi k\s(\smallint_{_{x_0}}^{^{{\bf x}}}\hs{-0.03cm}
\bar\omega)\s(\m{\rm Im}\s\tau\m)^{-1}
(\smallint_{_{x_0}}^{^{{\bf x}}}\hs{-0.03cm}
\omega)}\ |\m\Phi(\bx)\m|^2
\label{SPro}
\qqq
with \m\s$\Phi\s$ as in Eq.\s\s(\ref{PhI}) defines an
admissible hermitian structure on \s$L^2 K((2-2g)x_0)\s$,
\s see the beginning of Sect.\s\s5: it
induces a connection with the curvature
\s$2\pi\s\omega\s({\rm Im}\s\tau)^{-1}\hs{-0.1cm}
\wedge\bar\omega\s=\s-4\pi ig\s\alpha\s$.
\s In order to find the geometric interpretation,
of the other terms in the scalar product formula
(\ref{final}), let us return to the linear map (\ref{Map})
\qq
H^0(K)\ni\nu\s\ {\smash{\mathop{\longmapsto}
\limits^{l({\bf x},b)}}}\ \s\nu\m[b]\in H^1(L_\bx^{-2}K)
\cong H^0(L_\bx^2)^*\ .
\qqq
Recall from Sect.\s\s6.3 that surjectivity of \s$l(\bx,b)\s$
assured the local regularity of the projection
from \s$\CA^{01}\s$ into the orbit space
\s$\CA^{01}/\CG^\NC\s$.
\s We may view \s\s$\wedge\hs{-0.05cm}^{^{g-1}}\hs{-0.05cm}
l(\bx,b)\s\s$ as a
holomorphic \s$1,0$-form on \s$\Sigma\s$ with values
in the bundle \s\s${\rm det}^{-1}H^0(K)\m\otimes\m
{\rm det}^{-1}R^0{pr_1}_*\CL_0^2\s\s$ with the
representation
\qq
\wedge\hs{-0.05cm}^{^{g-1}}\hs{-0.05cm}
l(\bx,b)\s\s=\s\s\sum\limits_{j=1}^g(-1)^j\s
{\rm det}\m(\smallint_{_\Sigma}\kappa^r_\bx\omega^i
\wedge\m b\s)_{\hs{-0.01cm}_{i\not=j}}\s\m
\omega^j(x)\s\s\otimes\s\m(\smash{\mathop{\wedge}\limits_i}
\omega^i)^{-1}\m(\smash{\mathop{\wedge}\limits_r}
\kappa_\bx^r)^{-1}\ ,
\qqq
compare the discussion after (\ref{RANK}) in
Sect.\s\s5. Setting
\qq
&\phi(\bx;\m x_1,\dots,x_{g-1})\s\ \equiv\ \s
\sum\limits_{(\alpha_r)}\s\prod\limits_{r=1}^{g-1}
\eta_\bx^{\alpha_r}(x_r)\s\s\da_{z_{\bf x}^{\alpha_r}}
\wedge\hs{-0.07cm}^{^{g-1}}\hs{-0.05cm}l(\bx,b)\ \ \ &\cr
&=\s\s(g-1)!\s\s\s\CS_{g-1}\bigg(\sum\limits_{j=1}^g
(-1)^j\s\m{\rm det}\m(\m\kappa_\bx^r(x_r)\m\omega^i
(x_r)\m)_{\hs{-0.01cm}_{i\not=j}}\s\m
\omega^j(x)\bigg)\m\otimes\s(\smash{\mathop{\wedge}\limits_i}
\omega^i)^{-1}\m(\smash{\mathop{\wedge}\limits_r}
\kappa_\bx^r)^{-1}\s,\ \ \ &
\qqq
with \s$\CS_{g-1}\s$ symmetrizing the variables \s$x_r\s$,
we obtain a holomorphic \s$g,0$-form
on \s$\Sigma\times\Sigma^{g-1}\s$ with values in
\qq
&&C_k\s\s\s\equiv\s\s\s
{\rm det}^{-1}H^0(K)\s\otimes\s{pr_1}_*({\rm det}^{-1}
R^0{pr_1}_*\CL_0^2)\s\s S^{g-1}(\CL_0^2)\cr
&\cong&{\rm det}^{-1}H^0(K)\s\otimes\s{pr_1}_*(
{\rm det}^{-1}R^0{pr_1}_*{\CL_0'}^2)\s\s S^{g-1}
({\CL_0'}^2)\s\s\s\equiv\s\s\s C'_k
\qqq
where the isomorphism of the h.l.b.'s
on \s$\Sigma\times\Sigma^{g-1}\s$ \s$C_k'\s$
and \s$C_k\s$
is induced by the isomorphism (\ref{notdiff}).
In the right hand realization,
\s$\phi\s$ coincides with the one introduced
by Eq.\s\s(\ref{jesz}) in Sect.\s\s5.
Now, notice that
\qq
\sum\limits_{(\alpha_m)}\prod\limits_{m=1}^M
\hs{0.2cm}\left(\s\eta^{\alpha_m}_\bx(x_m)\s\s
\da_{z^{\alpha_m}_{\bf x}}\hs{-0.1cm}\right)
\wedge\hs{-0.1cm}^{^{g-1}}
\hs{-0.05cm}l(\bx,b)\s\s\s\psi(\bx,b)\hs{5cm}\cr
\s=\s({{_M}\atop^{g-1}}\hs{-0.06cm})\s\s\s
\CS_M\left(\s\phi(\bx;\m x_1,\dots,
x_{g-1})\s\s\m\chi(\bx;\m x_g,\dots,x_M\s)\right)\s.
\qqq
Clearly, the right hand side is a holomorphic
\s$(1+M),0$-form on \s$\Sigma\times\Sigma^{M}\s$ with values
in the h.l.b.
\qq
&{\rm det}^{-1}H^0(K)\s\otimes\s{pr_1}^*(\m({\rm det}^{-1}
R^0{pr_1}_*\CL_0^2)(L^2
K((2-2g)x_0)\m)^k\s\s S^M(\CL_0^2)\ &\cr
&\equiv\s\s{\rm det}^{-1}H^0(K)\s\otimes\s{pr_1}^*
(\m({\rm det}^{-1}R^0{pr_1}_*{\CL_0'}^2)(L^2K)\m)^k\s\s
S^M({\CL_0'}^2)\s.&
\label{8.18}
\qqq
It is easy to see that
\qq
{\rm det}\s(\m{\rm Im}\s
\tau\hs{-0.03cm})^{-1}\s\s\ee^{\hs{-0.03cm}
-\s2\pi k\s(\smallint_{_{x_0}}^{^{{\bf x}}}\hs{-0.05cm}
\bar\omega)\s{1\over{\rm Im}\s\tau}\s
(\smallint_{_{x_0}}^{^{{\bf x}}}\hs{-0.05cm}
\omega)}\s\s
\bigg|\sum_{(\hs{0.03cm}\alpha_m\hs{0.05cm})}\hs{-0.04cm}
\prod\limits_{m=1}^{M}\hs{-0.15cm}\left(\hs{-0.02cm}
\eta^{\alpha_m}_\bx
(\hs{-0.02cm}x_m\hs{-0.03cm})
\s\m\da_{z^{\alpha_m}_{\bf x}}\hs{-0.09cm}\right)
\hs{-0.05cm}\sum\limits_{j=1}^g\m(-1
)^{j}\s\m{\rm det}\m(\smallint_{_\Sigma}
\hs{-0.03cm}\kappa^r_\bx\omega^i\hs{-0.04cm}
\wedge b)_{\hs{-0.01cm}_{i\not=j}}\s\m\cr
\cdot\ \omega^j(x)\ \m\psi(\bx,b)\m\bigg|^{\wedge 2}\s\s
=\s\s\s\m({_{M}\atop^{g-1}})^{\hs{-0.02cm}2}\s\s\s
|\m\CS_M(\m\phi(\bx;\m x_1,\dots,
x_{g-1})\s\s\m\chi(\bx;\m x_g,\dots,x_M\s)\m)\m
|^{^{\wedge 2}},\ \ \hs{0.8cm}
\qqq
where, if we interpret \s$\CS_M(\phi\s\chi)\s$ as a
\s$(1+M),0$-form on \s$\Sigma\times\Sigma^M\s$ with values
in the bundle on the left hand side of (\ref{8.18}),
we should use on the latter the hermitian metric
induced by the Riemannian
metric of \s$\Sigma\s$, the hermitian structure
of \s$L_0\s$  and an admissible hermitian structure
of \s$L^2K((2-2g)x_0)\s$. \s It will be then
simpler to work only with
the admissible hermitian metrics on all occurring
line bundles, including the holomorphic tangent bundle
whose hermitian structure is given by the Riemannian metric.
With such choices, we may rewrite
\qq
\|\Psi\|^2\ \ \ \s\s=\ \ \s\s{\rm const}.\s\s\s i^{^{-M-1}}
\int\s\s{\rm det}'\m
(\m\de_{L(-x)^{2}}^\dagger\m
\de_{L(-x)^2})\ \
|\m\CS_M(\m\phi(x;\m x_1,\dots,x_{g-1})\s\hs{1.5cm}
\s\cr\s\s\s
\cdot\ \chi(x;\m x_g,\dots,x_M)\m)\m|^{^{\wedge 2}}
\hs{-0.05cm}\prod\limits_{m_1\not=m_2}\hs{-0.3cm}
\ee^{-{4\pi\over k+2}\m G(x_{m_1},\m x_{m_2})}\s
\prod\limits_m\hs{-0.02cm}\ee^{-{4\pi\over k+2}
\m :G(x_m,\m x_m):}\m,\hs{0.6cm}
\label{final2}
\qqq
including the prefactors into the normalization of the
hermitian metric and replacing \s$\bx\s$ by \s$x\s$
in accordance
with the interpretation of the \s$\bx$-dependence as
giving rise to geometric objects on \s$\Sigma\s$.
The Green function \s$G(x,y)\s$ in (\ref{final2})
should be orthogonal to the \s$2$-form \s$\alpha
\equiv{i\over 2g}\s
\omega\s({\rm Im}\s\tau)^{-1}\hs{-0.1cm}
\wedge\bar\omega\s$.
\s We have rewritten \s${\rm det}'(\m\de_{L_{\bf x}^2}^\dagger
\de_{L_{\bf x}^2})\s$ as \s${\rm det}'(\m\de_{L(-x)^2}^\dagger
\de_{L(-x)^2})\s$ using the fact that the latter determinant
is independent of the normalization of the hermitian
structure on \s$L(-x)\s$ so that it takes the same value
for any admissible metric on \s$L(-x)\s$.
When specified to the case of Arakelov metric on \s$\Sigma\s$,
\s this is exactly the expression (\ref{final0}) of Sect.\s\s5,
if we reinterpret \s$\CS_M(\m\phi\s\s\chi\m)\s$
according to the right hand side of (\ref{8.18})
and use the relation between the hermitian structures
induced by isomorphism (\ref{notdiff}), see the
discussion in Appendix B.
\vs 0.5cm

Similarly as for the lower genus case, see \cite{Quadr},
the natural conjecture is that the integral on the right
hand side of Eq.\s\s(\ref{final2}) converges
if and only if the function \s$\psi\s$
defines a globally non-singular CS state \s$\Psi\s$.
The singularities under the integral in Eq.\s\s(\ref{final2})
come from the product
\qq
\prod\limits_{m_1\not=m_2}\hs{-0.2cm}
\ee^{-{4\pi\over k+2}\m G(x_{m_1},\m x_{m_2})}\s\s\s\s\sim\s\s
\prod\limits_{m_1\not=m_2}\hs{-0.2cm}d(x_{m_1},
\m x_{m_2})^{^{-{2\over k+2}}}\ .
\qqq
The power counting when \s$Q+1\s$ of \s$x_{m}\s$ converge
shows that
\qq
\CS_M(\m\phi\m\s\chi\m)(x;\m y,y+y_1,\dots,y+y_Q,x_{Q+2},\dots,x_M)
\label{conV}
\qqq
has to have the vanishing Taylor expansion at zero in
\s$y_1,\dots,y_Q\s$
up to order \s$\leq\s Q({{Q+1}\over{k+2}}-1)\s$.
This gives also a set of sufficient conditions for
the convergence of the integral in (\ref{final2}).
Notice that for \s$g=2\s$ when \s$M=k+1\s$ these conditions
are always satisfied. For \s$g>2\s$, \s taking
\s$Q=k+1\s$, we infer that
if the integral converges then
\s$\CS_M(\phi\s\chi)(x;\m x_1,\dots,x_M)\s$ has to vanish
whenever \s$k+2\s$ of \s$x_m$'s coincide.
Let us see that this
condition is, indeed, satisfied for \s$\phi\s$ corresponding
to CS states. As we have explained at the end
of Sect.\s\s4, such states give sections
\s$\chi(x;\m x_g,\dots,x_M)\s$ which vanish
whenever \s$k+1\s$ of \s$x_m$'s \s coincide.
On the other hand,
\qq
\sum\limits_{j=1}^g(\hs{-0.05cm}-1
\hs{-0.05cm})^j\s\m{\rm det}(
\m\kappa^r_x(x_{r})\s\omega^i(x_{r})
\m)_{\hs{-0.01cm}_{i\not=j}}\s\m\omega^j(x)\s\s=\s\s-\s\s
{\rm det}\left(\matrix{\kappa_x^1
(x_1)\s\omega^i(x_1)\cr\cdot\cr
\cdot\cr\kappa_x^{g-1}(x_{g-1})\s\omega^i(x_{g-1})\cr
\omega^i(x)}\right)
\qqq
vanishes whenever two of \s$x_r$'s coincide. Hence
\s$\CS_M(\phi\s\chi)(x;\m x_1,\dots,x_M)\s$
vanishes whenever \s$k+2\s$ of
\s$x_m$'s \s are equal. It is clear that a complete
analysis of the convergence of the integral in
Eq.\s\s(\ref{final2}) and of the related ``fusion
rule conditions'' should be based on the geometry
studied in \cite{Bertram}\cite{Thad} and we shall postpone
it to a future work.

\vskip 2cm

\nappendix{A}
\vs 0.5cm

\no Let us show that, for \s$c\s$
a non-vanishing function on \s$\Sigma\s$,
\s for \s$g_c\s=\left(\matrix{_{c^{-1}}&_0\cr^0&^c}\right)\s$
and for \s$h_c=U\m g_c U^{-1}\s$,
\qq
\exp[\s S(h_{c},\s A^{10}_0+A^{01}_{{\bf x},0})]\s=
\s\ee^{{_1\over^{2\pi i}}\int_{_\Sigma} c^{-1}\da c\wedge
\m(c^{-1}\de c+2a_{{\bf x}})}\s\ \nu\m(c)\ ,
\label{topr}
\qqq
where \s$\nu\s$ is given by Eq.\s\s(\ref{corrt}).
Recall, that \s$U\s$ is a smooth isomorphism
of rank \m$2\m$ vector bundles with trivial determinant,
\s\s$U:\m L_0\oplus L_0^{-1}\m\rightarrow\m
\Sigma\times\NC^2\s$. \s\s
The gauge field \s$A^{10}_0+A^{01}_{{\bf x},0}\s$
represents the image under \s$U\s$ of the diagonal connection
\s$\nabla+\de-\sigma^3 a_{{\bf x}}\s$. First note that,
by the gauge invariance of the WZW action, \s$S(h_{c},
\s A^{10}_0+A^{01}_{{\bf x},0})\s$ is independent
of the choice of \s$U\s$. Moreover, since under an infinitesimal
change of the field \s$h\s$,
\qq
\delta\m S(h,\s A)\s=\s{_i\over^{2\pi}}\hs{-0.04cm}\int_{_\Sigma}
\hs{-0.04cm}
\tr\s\s h^{-1}\delta h\ \s F(A^{10}+
\s{}^{h^{-1}}\hs{-0.16cm}A^{01})
\label{var}
\qqq
which transforms covariantly, one infers that
under small changes of the function \s$c\s$,
\qq
\delta\m S(h_c,\s A^{10}_0+A^{01}_{\bx,0})\s=
\s{_i\over^\pi}\int_{_\Sigma}\hs{-0.04cm} c^{-1}\delta c\s\m(\m
\da\m(c^{-1}\de c)\m+\m\da a_{\bf x}\m+\m F_0\m)\ ,
\qqq
which is the variation of the right hand side of
Eq.\s\s(\ref{topr}).
We may then assume that \s$c=1\s$ on a small disc \s$D\s$.
Using a trivialization of the bundle \s$L_0\s$ over \s$D\s$
and over \s$\Sigma\setminus D\s$ with the transition
function \s$f\s$ defined around the boundary of \s$D\s$,
we may take the isomorphism \s$U\s$ equal to the identity
on \s$\Sigma\setminus D\s$ and interpolating smoothly
\s$\left(\matrix{_{f^{-1}}&_0\cr ^0&^f}\right)\s$ inside \s$D\s$.
\s In any case, \s$h_c\s=\left(\matrix{_{c^{-1}}&_0\cr^0&^c}
\right)=\s g_c\s\s$
everywhere. Let \s$a_0\s$ be the \s$1$-form representing on
\s$\Sigma\setminus D\s$ the metric connection of \s$L_0\s$.
\s It follows easily, that
\qq
S(h_{c},\s A^{10}_0+A^{01}_{{\bf x},0})\s=
\s S(g_c)\s-\s{_i\over^{\pi}}\int_{_{\Sigma\setminus D}}
\hs{-0.15cm}
c^{-1}dc\wedge(a_0+a_{{\bf x}})
\qqq
from which Eq.\s\s(\ref{topr}) follows by integration by parts
on the cut surface.
\vs 0.5cm

Let us identify the flat bundle corresponding to the
character \s$\Pi_1\ni p\s\mapsto\s\nu(c_p)\in S^1\s$
of the fundamental group \s$\Pi_1\s$ of \s$\Sigma\s$.
First note that \s$\nu(c_p)\s$ is independent of the
choice of the metric on the \s h.l.b. $L_0\s$.
\s Suppose that \s$L_0=L(-x_0)\s$ has divisor
\s$D\equiv\sum\limits_{m=1}^{Q+g}y_m-\sum
\limits_{n=0}^{Q}x_n\s$ so that \s$L_0\cong\CO(D)\s$.
Let us choose a hermitian metric on \s$\CO(D)\s$
so that \s\s$|\m 1(x)\m|^2\s=\s\exp[\m4\pi\m(\sum
\limits_{m}G(x,y_m)-\sum\limits_{n}
G(x,x_n)\m)\m]\s\s$ where \s$1\m\s$ is the
canonical section of \s$\CO(D)\s$ with zeros at
\s$y_m\s$ and poles
at \s$x_n\s$ and \s$G(x,y)\s$ is a Green
function of the Laplacian on \s$\Sigma\s$. \s$1\s$
trivializes \s$\CO(D)\s$ on
\s$\Sigma\setminus\{y_m,x_n\}\s$ and the \s$1,0$-form
\qq
a\s\equiv\s 4\pi\m\da\s(\sum
\limits_{m=1}^{Q+g}G(\m\cdot\m,y_m)-\sum\limits_{n=0}^{Q}
G(\m\cdot\m,x_n)\m)
\label{A.5}
\qqq
represents there the metric connection of \s$\CO(D)\cong
L_0\s$. \s In particular, its curvature
\s$F_0\s$ is equal to \s$\de\da a\s$. \s
Cutting \s$\epsilon$-balls around the points
\s$y_m\s$ and \s$x_n\s$ (region \s$B_\epsilon\s$)
\m and integrating by parts, we obtain
\qq
&\nu(c_p)\s\equiv\s\ee^{\s{i\over\pi}
\smallint_{_\Sigma}F_0\s{\rm ln}\m c_p}\s\s\prod
\limits_{j=1}^g\bigg(W_{a_j}^{-{_i\over^\pi}
\smallint_{_{b_j}}c_p^{-1}dc_p}\s W_{b_j}^{{_i\over^\pi}
\smallint_{_{a_j}}c_p^{-1}dc_p}\bigg)\s&\cr
&=\s\exp[\m{_i\over^\pi}\s\m{\displaystyle{
\lim_{\epsilon\rightarrow 0}}}
\smallint_{_{\Sigma\setminus B_\epsilon}}(\de a)\s{\rm ln}\m c_p]
\s\s\prod\limits_{j=1}^g\bigg(W_{a_j}^{-{_i\over^\pi}
\smallint_{_{b_j}}c_p^{-1}dc_p}\s W_{b_j}^{{_i\over^\pi}
\smallint_{_{a_j}}c_p^{-1}dc_p}\bigg)&\cr\cr
&=\s\exp[\m{_i\over^\pi}\s\m{\displaystyle{
\lim_{\epsilon\rightarrow 0}}}
\smallint_{_{\Sigma\setminus B_\epsilon}}(\s a\s d\s{\rm ln}\m
c_p\m-\m\smallint_{_{\da B_\epsilon}}a\s{\rm ln}\m c_p\s)\m]
\ .&
\qqq
Since \s$\de G(x,y)={1\over{4\pi\m(x-y)}}\s+\s$ a smooth
function, \s the boundary term contributes
\qq
\prod\limits_{m=1}^{Q+g} c_p(y_m)^2\prod\limits_{n=0}^Q
c_p(x_n)^{-2}
\label{charact}
\qqq
whereas the volume term \s\s${\displaystyle{
\lim_{\epsilon\rightarrow 0}}}
\smallint_{_{\Sigma\setminus B_\epsilon}}a\s\m d\s{\rm ln}\m
c_p\s\s$
may be shown to vanish by using Eq.\s\s(\ref{A.5})
and integrating once more by parts (\s$c_p\s$ is harmonic).
It is easy to see that the flat bundle corresponding
to the character (\ref{charact})
of \s$\Pi_1\s$ is equivalent to the trivial
bundle with the \s$\de$-operator
$$\de\s-2\pi\s(\sum\limits_m
\smallint_{x_0}^{y_m}\omega\m-\m\sum\limits_n
\smallint_{x_0}^{x_n}\omega\s)\s({\rm Im}\s\tau)^{-1}
\bar\omega$$
and is isomorphic to $\CO(D-(g-1)x_0)^2\cong L(-gx_0)^2\s$.
\vs 1cm

\nappendix{B}
\vs 0.8cm

\no Consider a function \s$f_{x_0}\s$ on \s$\tilde\Sigma
\times\Sigma\s$ given by
\qq
f_{x_0}({\bf x},y)\s\s\equiv\s\s
\ee^{\s2\pi\m i\m(\int_{_{x_0}}^{^{\bf x}}\hs{-0.1cm}\omega)\s
({\rm Im}\s\tau)^{-1}\m(\int_{_{x_0}}^{^{y}}\hs{-0.1cm}
{\rm Im}\s\omega)\m}\
\s\m{\vartheta(\m -a+\smallint_{_{x_0}}^{^{\bf x}}
\hs{-0.1cm}\omega\s|\s\tau\m)
\s\s\s\vartheta(\m a+\smallint_{_{x_0}}^{^{y}}\hs{-0.1cm}\omega
\s|\s\tau\m)\over
\vartheta(\m a+\smallint_{_{\bf x}}^{^{y}}\hs{-0.07cm}\omega
\s|\s\tau\m)}\ ,
\label{B.1}
\qqq
where \s$a\s$ is an odd characteristic. One has
\qq
f_{x_0}(p{\bf x},y)\s=\s
c_{p}(y)^{-1}\s\s f_{x_0}({\bf x},y)\ ,\hs{1cm}
(\m\de\s+\s\pi(\smallint_{_{x_0}}^{^{\bf x}}
\hs{-0.1cm}\omega)\s
({\rm Im}\s\tau)^{-1}\m\bar\omega(y)\m)\s\s\m
f_{x_0}({\bf x},y)\s=\s0\ .\nonumber
\qqq
Besides, \s$f_{x_0}\s$ has first order zeros
at \s$x=x_0\s$ and \s$y=x_0\s$
and a first order pole at \s$x=y\s$. \s It follows
that multiplication by \s$f_{x_0}\s$ establishes
an isomorphism between the h.l.b.'s \s$\CL'_0\s$ and
\s${pr_1}^*(\CO(-x_0))\s\CL_0\s$
over \s$\Sigma\times\Sigma\s$.
\vs 0.4cm

Note that the hermitian structure on the h.l.b. $\CL_0\s$
coming from an admissible hermitian metric on the bundle
\s$L_0\s$ (see the beginning of Sect.\s\s5 for
the definition of admissibility)
induces the connection with curvature
$$\pi\s\m{pr_1}^*(\omega)\s(\m{\rm Im}\s\tau\m)^{-1}
{pr_2}^*(\bar\omega)-\pi\s\m{pr_1}^*(\bar\omega)
\s(\m{\rm Im}\s\tau\m)^{-1}{pr_2}^*(\omega)+(g-1)\s
{pr_2}^*\alpha\ .$$
Taking also an admissible hermitian structure on
the bundle \s$\CO(-x_0)\s$ we obtain a hermitian
metric on the h.l.b. ${pr_1}^*(\CO(-x_0))\CL^0\s$
corresponding to the curvature form
\qq
-{pr_1}^*\alpha+\pi\s\m{pr_1}^*(\omega)\s(\m{\rm Im}\s\tau\m)^{-1}
{pr_2}^*(\bar\omega)-\pi\s\m{pr_1}^*(\bar\omega)
\s(\m{\rm Im}\s\tau\m)^{-1}{pr_2}^*(\omega)+(g-1)
{pr_2}^*\alpha\s,\ \ \
\qqq
the same as the curvature induced by the hermitian
structure of \s$\CL_0'\s$ described around
Eq.\s\s(\ref{G1}) in Sect.\s\s5. Hence multiplication
by \s$f_{x_0}\s$ must carry one hermitian structure
into the other one, up to a constant factor.
\vs 1cm

\nappendix{C}
\vs 0.5cm

Let us discuss in more detail the relation
between the two descriptions of the CS states:
the one discussed in Sect.\s\s3 using functions
\s$\psi(\bx,b)\s$ with \s$\bx\in\tilde\Sigma\s$,
\s$b\in\wedge^{01}(L_0^{-2})\s$
and the one of \cite{Bertram},
discussed in Sect.\s\s4, employing polynomials
\s$\psi'(b')\s$, \s$b'\in\wedge^{01}(L^{-2})\s$.
Let us fix \s$\bx\s$ with \s$x\not=x_0\s$
and \s$b\s$. Viewing
\s$b\s$ as an element of \s$\wedge^{01}(L_\bx^{-2})\s$,
\s we may define a form
\s$b''\in\wedge^{01}(L(-x)^{-2})\s$
by setting
\qq
b''\s\equiv\s f_{x_0}(\bx,\m\cdot\m)^2\m\s b
\label{C.1}
\qqq
since the multiplication by \s$f_{x_0}(\bx,\m\cdot\m)\s$
given by Eq.\s\s(\ref{B.1}) establishes an
isomorphism between \s$L(-x)\s$
and \s$L_{\bx}\s$. \s We shall choose \s$\kappa\in
H^0(L^2(-x))\setminus H^0(L(-x)^2)\s$ s.\s t.
\qq
\smallint_{_\Sigma}\hs{-0.01cm}\kappa\m\nu\wedge b''\s=\s0
\hs{1cm}{\rm for}\hs{1cm}\nu\in H^0(K(-x))\ .
\label{guaran}
\qqq
Except on a subset of \s$b''\m$'s of codimension at least
\s$2\s$ in \s$H^1(L(-x)^{-2})\s$, such \s$\kappa\s$
exists and is unique up to normalization
(compare the discussion around (\ref{Map}) in Sect.\s\s6.3).
Eq.\s\s(\ref{guaran}) guarantees that there exists
a function \s$f_1\in\Gamma(\CO(x))\s$ s.\s t.
\qq
\de\m f_1\s=\s b\s\kappa\ .
\qqq
Let \s$f_t\equiv 1+tf_1\in\Gamma(\CO(x))\s$ for
\s$t\in\NC\s$. \s Since, by our assumptions,
\s$\kappa(x)\not=0\s$ as an element of \s$L^2(-x)\s$
and, for small \s$t\s$, \s$f_t\s$ may have zeros only close
to \s$x\s$, it follows that, for such \s$t\s$, \s the map
\qq
L^{-1}\ni l\s\s\longmapsto\s\s(lf_t,\m l\kappa)\in
L(-x)^{-1}\oplus L(-x)
\qqq
is an embedding which, moreover, is holomorphic
if we modify the \s$\de\s$ operator of \s$L(-x)^{-1}\oplus
L(-x)\s$ by replacing it by
\s\s$\de\m+(\matrix{_0&_{t\m b''}\cr
^0&^0})\s$. \s Choose now \s$\zeta\in\Gamma(\CO(-x))\s$
and \s$s\in\Gamma(L^{-2}(x))\s$ s.\s t.
\qq
\zeta-s\kappa=1\ .
\label{=1}
\qqq
Note that \s$s(x)\s$ has to be a non-vanishing element
of \s$L^{-2}(x)\s$ because otherwise Eq.\s\s(\ref{=1})
could not be satisfied as \s$\zeta\s$, viewed as
a function on \s$\Sigma\s$, \s vanishes at \s$x\s$.
Now, We shall perturb \s$\zeta\s$ and \s$s\s$ by
taking \s$\zeta_t\in\Gamma(\CO(x))\s$ and \s$s_t\in
\Gamma(L^{-2}(x))\s$ s.\s t.
\qq
\zeta_t f_t-s_t\kappa=1
\label{=12}
\qqq
and \s$\zeta_t=\zeta+t\zeta_1+o(t)\s$, \s$s_t=s+ts_1+o(t)\s$
are analytic in (small) \s$t\s$. \s This may be easily
achieved by solving Eq.\s\s(\ref{=12}) for \s$\zeta_t\s$
with \s$s_t=s\s$ outside a small ball \s$B_\epsilon(x)\s$ around
\s$x\s$ and for \s$s_t\s$ with \s$\zeta_t=\zeta\s$
on \s$B_{2\epsilon}(x)\s$ and by interpolating between the two
solutions in \s$B_{2\epsilon}(x)\setminus B_\epsilon(x)\s$.
Consider now the smooth isomorphism
\qq
V_t:L^{-1}\oplus L\ \longrightarrow\ L(-x)^{-1}\oplus
L(-x)\ ,\hs{1.3cm}V_t\s\s=\s\left(\matrix{_{f_t}&_{s_t}\cr
^\kappa&^{\zeta_t}}\right)
\qqq
depending analytically on (small) \s$t\s$. \s A straightforward
computation shows that
\qq
V_t^{-1}\left(\matrix{_0&_{t\m {b''}}\cr^0&^0}\right)V_t
\s\s+\s\m V_t^{-1}\de\m V_t\s\s=\s\s
\left(\matrix{_0&_{t\zeta_t^2b''+\zeta_t\de s_t-s_t\de\zeta_t}
\cr^0&^0}\right)\ .
\label{speci}
\qqq
Notice that \s\s$t\zeta_t^2b''+\zeta_t\de s_t-s_t\de\zeta_t
\s\equiv\s b'_t=b_0'+tb'_1+o(t)\s\in\s\wedge^{01}(L^{-2})\s$.
In particular,
\qq
b_0'&=&\m\zeta\de s-s\m\de \zeta\s=\s(1+s\kappa)\de s
-s\m\de(s\kappa)\s=\s\de s\ ,\label{C.9}\\
b_1'&=&\zeta^2b''+\zeta\m\de s_1+\zeta_1\de s-s\m\de
\zeta_1-s_1\de\zeta\s=\s b''-\de(f_1s)+\de s_1\ ,
\label{C.10}
\qqq
where we have used the relations \s$\zeta=1+s\kappa\s$
and \s$\zeta_1=-\zeta f_1+s_1\kappa\s$ following
from Eqs.\s\s(\ref{=1}) and (\ref{=12}).
Let us define two gauge fields
\qq
A^{01}_{\bx,tb}&=& U\left(\matrix{_0&_{tb}
\cr^0&^0}\right)U^{-1}
\s+\s U\m\de U^{-1}\ ,\cr
{A'}^{01}_{b'_t}&=& U'\left(\matrix{_0&_{b'_t}\cr^0&^0}
\right){U'}^{-1}\s+\s U'\m\de{U'}^{-1}\ ,
\qqq
where \s\s$U:L_0^{-1}\oplus L_0\longrightarrow\Sigma\times
\NC^2\s\s$ and \s\s$U':L^{-1}\oplus L\longrightarrow
\Sigma\times\NC^2\s\s$ are smooth isometric isomorphisms,
see Sects.\s\s3 and 4.
$A^{01}_{\bx,tb}\s$ and \s${A'}^{01}_{b'_t}\s$ are
gauge related:
\qq
{A'}^{01}_{b'_t}\s=\s h_t^{-1}A^{01}_{\bx,tb}\m h_t\s+
\s h_t^{-1}\de h_t\ ,
\qqq
where
\qq
h_t\s\s=\s\s U\left(\matrix{_{f_{x_0}
({\bf x},\m\cdot\m)^{-1}}&
_0\cr^0&^{f_{x_0}({\bf x},\m\cdot\m)}})
\right) V_t\s\s{U'}^{-1}\ .
\qqq
Expressing the same CS state \s$\Psi\s$ in two
descriptions corresponding to Eq.\s\s(\ref{vnewCS})
and Eq.\s\s(\ref{newCS}) and comparing them using
the gauge invariance of \s$\Psi\s$, \s we obtain
the relation
\qq
\psi'(b'_t)\s\s=\s\s\exp[\m k\m S(h_t,\m A^{01}_{\bx,tb})\s
+\s{_{ik}\over^{2\pi}}
\smallint_{_\Sigma}\hs{-0.04cm}{\rm tr}
\m\s\s({A'}^{10}_0\wedge{A'}^{01}_{b'_t}-\m A^{01}_0
\wedge A^{01}_{\bx,tb})\m]\ \s\psi(\bx,tb)\ .
\label{C.13}
\qqq
Now, due to the homogeneity of \s$\psi(\bx,\m\cdot\m)\s$,
\qq
\psi(\bx,tb)\s=\s t^{^{k(g-1)}}\s\psi(\bx,b)\ .
\label{C.14}
\qqq
On the other side, it is easy to see with the use of
Eq.\s\s(\ref{C.9}) that,
for \s$\eta'\in H^0(L^2 K)\s$,
\qq
\smallint_{_\Sigma}\hs{-0.03cm}\eta'\wedge b_0'\s=
\s\smallint_{_\Sigma}\hs{-0.03cm}\eta'\wedge\de s\s=\s
{\displaystyle{\lim_{\epsilon\rightarrow 0}}}
\s\s\smallint_{_{\Sigma\setminus B_{\epsilon}(x)}}
\hs{-0.25cm}\eta'\wedge\de s\s=\s
{\displaystyle{\lim_{\epsilon\rightarrow 0}}}
\smallint_{_{\da B_{\epsilon}(x)}}\hs{-0.18cm}\eta\m s\s=\s
2\pi i\s{_{\eta'(x)}\over^{d(s^{-1})(x)/dx}}\ ,\hs{0.5cm}
\qqq
where \s$s^{-1}\s$ is differentiated as a section
of \s$L^2\s$ vanishing at \s$x\s$. \s
Hence the class of \s$b'_0\s$ in \s$\NP H^1(L^2)\s$
coincides with the image of \s$x\in\Sigma\s$
under the embedding (\ref{Emb})
of \s$\Sigma\s$ into \s$\NP H^1(L^2)\s$.
Specifying Eq.\s\s(\ref{speci}) to \s$t=0\s$, we infer that
that the corresponding rank \s$2\s$ holomorphic bundle
is isomorphic by \s$V_0\s$ with the split bundle
\s$L(-x)^{-1}\oplus L(-x)\s$. \s
Using the integral presentation (\ref{present})
of \s$\psi'\s$ and Theorem 2a of \cite{Bertram},
see Sect.\s\s4, we obtain the relation
\qq
\psi'(b'_t)\s\s=\s\s{_{(2\pi i)^k}\over^{k!}}\s\s
t^{^{k(g-1)}}\s\s({_{d(s^{-1})(x)}\over^{dx}})^{^{-k}}
\ \int_{_{\Sigma^{k(g-1)}}}\hs{-0.2cm}
\chi'({\smash{\mathop{x,\dots,x}\limits_{k\ {\rm times}}}};
\m x_{k+1},\dots,x_{kg})\hs{2.4cm}\ \s\cr
\cdot\ b'_1(x_{k+1})\s\cdots\s
b_1'(x_{kg})\s\s\s\s+\s\s\s o(t^{^{k(g-1)}})\s.\hs{0.8cm}
\label{+HO}
\qqq
Besides, using Eq.\s\s(\ref{C.10}), we may replace
\s$b_1'\s$ by \s\s$b''=f_{x_0}(\bx,\m\cdot\m)^2\s b\s\s$
on the right hand side.
Hence the relations (\ref{C.13}), (\ref{C.14}) and
(\ref{+HO}) imply that
\qq
\chi'({\smash{\mathop{x,\dots,x}\limits_{k\ {\rm times}}}};
\m x_{k+1},\dots,x_{kg})\s&=&\s
\CU_{x_0}(\bx)\s\ \chi(\bx;\m x_{k+1},\dots,
x_{kg})\cr
&\cdot&f_{x_0}(\bx,x_{k+1})^{-2}\cdots
\s f_{x_0}(\bx,x_{kg})^{-2}\ ,
\label{JaBert}
\qqq
where
\qq
\CU_{x_0}(\bx)\s\s\equiv\s\s{_{k!}\over^{(2\pi i)^k}}
\s\s({_{d(s^{-1})(x)}\over^{dx}})^{^{k}}\s\m
\exp[\m k S(h_0, A^{01}_{\bx,0})
+{_{ik}\over^{2\pi}}
\smallint_{_\Sigma}\hs{-0.03cm}{\rm tr}
\s\s({A'}^{10}_0\hs{-0.1cm}\wedge\hs{-0.05cm}
{A'}^{01}_{b'_0}\hs{-0.07cm}
-\hs{-0.03cm}A^{01}_0\hs{-0.1cm}
\wedge\hs{-0.05cm}A^{01}_{\bx,0})\m]\s.\hs{0.8cm}
\label{C.19}
\qqq
$\CU_{x_0}(\bx)$ takes values in \s$(L^2K)^k\s$.
\s It must be independent of the choice of \s$\kappa,
\ \zeta\s$ and \s$s\s$ since the other terms in
Eq.\s\s(\ref{JaBert}) are.
It gives an explicit realization of the isomorphism
between functions \s$\Phi(\bx)\s$ on \s$\tilde\Sigma\s$
transforming by Eq.\s\s(\ref{PhI}) and behaving
like \s$(x-x_0)^{^{-2k(g-1)}}\s$ around \s$x_0\s$
and sections of \s$(L^2K)^k\s$, \s see Appendix A.
Such isomorphism is unique up to normalization
and Eq.\s\s(\ref{C.19}) fixes this normalization
completely (in a way dependent on the isometric
isomorphisms \s$U\s$ and \s$U'\s$)\m.
\s Eq.\s\s(\ref{JaBert})
establishes the precise relation between functions
\s$\psi\s$ used to represent the CS states in this paper
and polynomials \s$\psi'\s$ introduced in
Sect.\s\s4 and corresponding to the description
of \cite{Bertram}\cite{Thad}.
\vs 1cm

\nappendix{D}
\vs 0.5cm

We shall prove here the formula (\ref{determi})
for the zeta-function regularized determinant
of the operator \s$\bar D_n^\dagger\m\bar D_n\s$.
Let us consider the determinant line bundle \s$\CF\s$
of the \s$\de$-family \s\s$(\s\de\m+\m[A^{01}_{\bx,b},
\s\cdot\s\m]\m\equiv\m\bar D_n\m)\s\s$ of operators
acting in the trivial bundle \s$\Sigma\times{sl(2,\NC)}\s$.
It is a holomorphic line bundle over the space of pairs
\s$(\bx,b)\s$ with the fibers $${\rm det}\m
(\m{\rm ker}(\bar D_n)\m)^{-1}\s\m{\rm det}\m(\m{\rm coker}
(\bar D_n)\m)\ .$$
Generically, \s${\rm ker}(\bar D_n)=0\s$ and the dual space
to the \s${\rm coker}(\bar D_n)\s$ is spanned by the
\s$sl(2,\NC)$-valued \s$1,0$-forms
\s\s$\omega^\alpha(\bx,b)\equiv U\s(\matrix{_{-\mu^\alpha}
&_{\lambda^\alpha}\cr^{\eta^\alpha}&^{\mu^\alpha}})\s U^{-1}\s\s$
constructed in Sect.\s\s5.3., see formulae (\ref{5.33})
and (\ref{5.34}). The complex gauge transformations
\s$h_{c,v}\equiv U\m g_{c,v}\m U^{-1}\s$, \s with \s$g_{c,v}\s$
as in Eq.\s\s(\ref{2.13}), \s$c\s$ a non-zero constant
or \s$c=c_p\s$, \s act on \s$\CF\s$. Division by their
action gives the \s$4^{\rm th}\s$ power \s (\s$4\m=\m 2\s\s\times$
the dual Coxeter number of \s$SU(2)\s$) \s of the h.l.b.
${\rm DET}\s$ over the compact space \s$\NP W_0\s$ discussed
in Sect.\s\s3. The formula
\qq
\|1\otimes\smash{\mathop{\wedge}\limits_\alpha}
\omega^\alpha(\bx,b)\s\|^2\s=\s\m
{\rm det}\s(\bar D_n^\dagger\m\bar D_n)
\ \s{\rm det}\m(\m\Omega(1,n))^{-1}
\qqq
defines Quillen's hermitian metric \cite{Quillen}
on \s$\CF\s$. \s Its curvature is easily calculable (from the
Riemann-Roch-Grothendick Theorem, see e.\s g. \cite{Bost})
to be
\qq
{_{2}\over^{\pi i}}\int_{_\Sigma}\tr\s\s\s
(\delta A^{01}_{\bx,b})^\dagger\wedge\m\delta A^{01}_{\bx,b}\s=\s
{_{2}\over^{\pi i}}\int_{_\Sigma}(\m 2\m{\overline{\delta a_\bx}}
\wedge\m\delta a_\bx\s+\s\langle\delta b,\m\wedge\m\delta
b\rangle\m)\ .
\qqq
The change of the Quillen metric on \s$\CF\s$ under
the complex gauge transformations \s$h_{c,v}\s$
may be inferred from the chiral anomaly formula (\ref{ins2})
with \s\s$S(h_{c,v}h_{c,v}^\dagger\m,\m A(n))\s$
given by Eqs.\s\s(\ref{param}) and (\ref{toprove}).
It is then easy to see that the modified metric
\qq
{\|\s\cdot\s\|'}^2\s=\s\|\s\cdot\s\|^2\s\s\ee^{\s{2i\over\pi}
\int_{_\Sigma}\hs{-0.03cm}\langle\m b\m,\m\wedge b\m\rangle}
\qqq
is invariant under the \s$h_{c,v}\s$ transformations
and descends to \s${\rm DET}^4\s$. \s Its curvature is
\s\s${4\over\pi i}\smallint_{_\Sigma}\hs{-0.05cm}
{\overline{\delta a_\bx}}\wedge\m\delta a_\bx\s$.
On the other hand, the right hand side of
Eq.\s\s(\ref{QuiL}) multiplied by \s\s$\exp[{2i\over\pi}
\int_{_\Sigma}\hs{-0.03cm}\langle\m b\m,\m\wedge b\m\rangle]\s\s$
also defines a hermitian structure on the h.l.b. ${\rm DET}^4\s$
and the Riemann-Roch-Grothendick Theorem shows that
the curvatures agree. Hence, the two metrics are proportional
with the proportionality constant which might {\it a priory}
depend on the metric (i.\s e. also on the complex
structure) of \s$\Sigma\s$.
\vs 0.4cm

In order to see that Eq.\s\s(\ref{determi}) for \s${\rm det}\m(
\bar D_n^\dagger\bar D_n)\s$ represents
properly also the dependence on the metric of \s$\Sigma\s$,
it is enough to show that it produces the right
behavior of the determinant in the limit when \s$b\s$ is
replaced by \s$tb\s$ and \s$t\rightarrow 0\s$. This limit
may be studied by the \s$2^{\m{\rm nd}}\s$ order perturbation
theory. For \s$t=0\s$, \s$\bar D_n^\dagger\bar D_n\s$ has
the \s$g$-dimensional kernel \s$U\m(H^0(L_\bx^2)\m\sigma^-\m
+\m\NC\m\sigma^3)\m U^{-1}\s$. \s For \s$t\not=0\s$,
\s the operator is modified by a relatively compact
perturbation \cite{Kato}. If the rank of the matrix
\s$(\smallint_{_\Sigma}\hs{-0.05cm}\kappa^r\omega^i
\wedge\m b\m)\s$ is \s$g-1\s$, \s all the zero eigenvalues
of \s$\bar D_n^\dagger\bar D_n\s$ move up and their
product is easily calculated to be
\qq
R\m t^{2g}\m\equiv\m 2^{^{-g}}\m t^{2g}\s{\rm area}^{-1}\s
{\rm det}\m(K_0)^{-1}\s|z^1|^2\m\s(H_0^{-1})_{_{11}}\s\m
\m{\rm det}\s\m(\m(\smallint_{_\Sigma}\hs{-0.05cm}
{\overline{\kappa^r\omega^i\wedge\m b}})\s({\rm Im}\s\tau)_{ij}
\m(\smallint_{_{\Sigma}}\hs{-0.05cm}\kappa^s
\omega^j\wedge\m b)\m)
\ \ \ \
\qqq
in the leading nontrivial order (we have assumed that
\s$z^\alpha(b)=0\s$ for \s$\alpha>1\s$)\m. \s It follows
that, when \s$t\rightarrow 0\s$
(and with the zeta-functions regularized determinants),
\qq
{\rm det}\m(\bar D_n^\dagger\bar D_n)\s=\s R\s\s t^{2g}\s\s\m
{\rm det}'\m(\bar D_n^\dagger\bar D_n)|_{_{t=0}}\s\s+\s
o\m(t^{2g})\ .
\label{Goodb}
\qqq
On the other hand,
\qq
{\displaystyle{\lim_{t\rightarrow 0}}}\s\s\m{\rm det}
\m(\m\Omega(1,\bx,tb)\m)\s=\s{_2\over^i}\m(\smallint_{_\Sigma}
\hs{-0.05cm}\bar\mu_1\wedge\m\mu_1\m)\s\s\s
{\rm det}_{_{\alpha,\beta\not=1}}(H_0^{\alpha\beta})
\qqq
in the notation of Eq.\s\s(\ref{5.33}). Using also
the relation (\ref{5.37}), we obtain
\qq
{\displaystyle{\lim_{t\rightarrow 0}}}\s\m{\rm det}
\m(\m\Omega(1,\bx,tb)\m)\s\s=\s\s 4\s\m|\m{\rm det}
\m(\smallint_{_\Sigma}\hs{-0.05cm}\kappa^r\omega^i
\wedge\m b\m)_{_{i<g}}|^2\hs{6cm}\cr
\cdot\ \sum\limits_{i',\m j'}\s
(-1)^{^{i'+j'}}\s\m{\rm det}\m(\smallint_{_\Sigma}
\hs{-0.05cm}{\overline{\kappa^r\omega^i
\wedge\m b}}\m)_{_{i\not=i'}}\s\s({\rm Im}\s\tau)_{i'
\hs{-0.05cm}j'}
\s\s{\rm det}(\smallint_{_\Sigma}\hs{-0.05cm}
\kappa^s\omega^j\wedge\m b\m)_{_{j\not=j'}}\s\s\m
{\rm det}\m(H_0)\s\s(H_0^{-1})_{_{11}}\ \s\cr
=\s 4\s\m{\rm det}\m({\rm Im}\s\tau)
\s\s\m{\rm det}\m(H_0)\m\s\s
(H_0^{-1})_{_{11}}\s\s\m|\m{\rm det}\m(\smallint_{_\Sigma}
\hs{-0.05cm}\kappa^r\omega^i\wedge\m b\m)_{_{i<g}}|^{-2}
\hs{4cm}\cr
\cdot\s\ {\rm det}\m(\m(\smallint_{_\Sigma}
\hs{-0.05cm}{\overline{\kappa^r\omega^i
\wedge\m b}})\s\s({\rm Im}\s\tau)^{^{-1}}_{\hs{0.07cm}ij}\s
(\smallint_{_\Sigma}\hs{-0.05cm}\kappa^s\omega^j
\wedge\m b)\m)\ ,\ \
\label{MaTra}
\qqq
where the last equality is a consequence of the identity
\s\m${\rm det}\m(\sum_{_j}\hs{-0.07cm}
{\overline{A^{rj}}}\hs{0.03cm} A^{sj}\m)
=\sum_{_j}\hs{-0.07cm}|\m
{\rm det}\m(A^{ri})_{_{i\not=j}}|^2\s\m$
for \s$(A^{rj})\s$ a \s$(g-1)\times g\s$ matrix
which may be easily verified by taking
\s$(A^{rj})\s$ with first \s$(g-1)\s$ columns forming
a unit matrix. It follows now from Eq.\s\s(\ref{llast}) that
\qq
{\displaystyle{\lim_{t\rightarrow 0}}}\s\s t^{-2g}
\left(\int\hs{-0.07cm}\delta(BV)\s\s\ee^{-\|CV\|^2}
\s\m DV\hs{-0.1cm}\right)^{\hs{-0.1cm}-1}
=\s 4\s|z^1|^2\s\s(H_0^{-1})_{_{11}}
\hs{4cm}\cr
\cdot\s\s\s{\rm det}\m(\m(\smallint_{_\Sigma}
\hs{-0.05cm}{\overline{\kappa^r\omega^i
\wedge\m b}})\s\s({\rm Im}\s\tau)^{^{-1}}_{\hs{0.07cm}ij}\s
(\smallint_{_\Sigma}\hs{-0.05cm}\kappa^s\omega^j
\wedge\m b)\m)
\qqq
and, consequently, the right hand side of
Eq.\s\s(\ref{determi}) behaves when \s$t\rightarrow 0\s$
in accordance with (\ref{Goodb}).
This ends the proof of formula (\ref{determi}).
\vs 1cm

\nappendix{E}
\vs 0.5cm

\no Let us denote
\qq
\CP(\varphi)&\equiv&{\rm det}\m(H_\varphi)^{-1}
\s|z^1|^{^2}\hs{-0.1cm}\int\limits_{\NC^{N-1}}\hs{-0.07cm}
|P(z)|^{^2}\s\s\m\ee^{-{k+4\over 2\pi}
\s\bar z^\alpha(H^{-1}_\varphi)_{_{\beta\alpha}} z^\beta}\s
\prod\limits_{\alpha=2}^{N}d^2z^\alpha\ \s\cr
&&\hs{-1.5cm}=\s({_2\over^{k+4}})^{^{N}}\s|z^1|^{^2}
\hs{-0.1cm}\int\limits_{\NC^{N-1}}\hs{-0.1cm}|P(z)|^{^2}\s
\bigg(\hs{-0.08cm}\int\hs{-0.06cm}
\ee^{-{2\pi\over k+4}\s\m\bar c_\alpha\s
(H_\varphi)^{\alpha\beta}
\s c_\beta\m\s+\s\m i\s c_\alpha z^{\alpha}\s
+\s\s i\s\bar c_\alpha\bar z^{\alpha}}
\prod\limits_\alpha d^2c_\alpha\hs{-0.1cm}\bigg)
\hs{-0.04cm}\prod\limits_{\alpha=2}^{N}d^2z^\alpha.
\hs{0.8cm}
\label{nowy}
\qqq
where \s$P(z)\s$ is a homogeneous polynomial in variables
\s$(z^\alpha)\s$ of degree \s$M\equiv(k+1)(g-1)\m$:
\qq
P(z)\s=\s\sum\limits_{(\alpha_m)}\m{_1\over{M!}}\bigg(
\prod\limits_{m=1}^M\da_{z^{\alpha_m}}P(z)\bigg)
\prod\limits_{m=1}^M z^{\alpha_m}\s\equiv\s
\s\sum\limits_{(\alpha_m)}P_{_{(\alpha_m)}}\s
\prod\limits_{m=1}^M z^{\alpha_m}\ .
\qqq
$\CP(\varphi)\s\s$ with \s$P(z)\s$ given by Eq.\s\s(\ref{P(z)})
is the \s$z$-integral to be computed after the \s$w$-integration
in Sect.\s\s 7 if we postpone the \s$\varphi\s$ integral till
after the one over \s$z^\alpha\m$'s\m, \s
see Eq.\s\s(\ref{winteg}).
The integrals in (\ref{nowy}) clearly converge.
We shall show that the zero-mode integral
\qq
\int\limits_{\NR}\ee^{-2aM}\s\CP(\varphi+a)\ da\s=\s
\hf\m\pi^{^{M-1}}\m({_2\over^{k+4}})^{^{M+N}}
\s\sum\limits_{(\alpha_m),(\beta_m)}
{\overline{P_{_{(\alpha_m)}}}}\s\s P_{_{(\beta_m)}}\m\prod
\limits_{m=1}^M H_\varphi^{\alpha_m\beta_m}\ .
\label{D.3}
\qqq
Consequently, integrating in our calculation of the right hand
side of (\ref{newI}) first over \s$w\s$ then over
\s$z^\alpha,\s\s\alpha>1\s$ and at the end over \s$\varphi\s$
one obtains the expression (\ref{final}) without encountering
other infinities then the standard ones removed by the
zeta-function regularization of the determinants and
the Wick ordering of the \s$\varphi$-field exponentials.
In order to proof formula (\ref{D.3}), let us rewrite
\qq
&{\displaystyle{\int\limits_{\NR}}}
\ee^{-2aM}\s\CP(\varphi+a)\ da\s=\s
{_1\over^{2\pi}}{\displaystyle{\int\limits_\NC}}|t|^{^{2(M-1)}}
\s\s\CP(\varphi
-\ln\m|t|)\s\s\s d^2t&\cr
&=\s{_1\over^{2\pi}}\s\s{\rm det}\m(H_\varphi)^{-1}\s|z^1|^{^2}
\hs{-0.1cm}{\displaystyle{\int\limits_{\NC^{N}}}}\hs{-0.01cm}
|t|^{^{2(M+N-1)}}
\m\s|P(z)|^{^2}\s\s\s\ee^{-{k+4\over 2\pi}\s|t|^{^2}
\s\bar z^\alpha(H^{-1}_\varphi)_{_{\beta\alpha}}\hs{-0.03cm}
z^\beta}\s
\prod\limits_{\alpha=2}^{N}d^2z^\alpha\ .&
\qqq
The integral clearly converges. By the change of variables
\qq
\zeta^1=t\m z^1\s,\ \ \zeta^2=t\m z^2\s,\ \dots\ ,\ \
\zeta^{N}=t\m z^{N}\ ,
\qqq
one obtains
\qq
{\displaystyle{\int\limits_{\NR}}}
\ee^{-2aM}\s\CP(\varphi+a)\ da\s=\s
{_1\over^{2\pi}}\s\s{\rm det}\m(H_\varphi)^{-1}
{\displaystyle{\int\limits_{\NC^{N}}}}
|P(\zeta)|^{^2}
\s\s\s\ee^{-{k+4\over 2\pi}\s
\bar\zeta^\alpha(H^{-1}_\varphi)_{_{\beta\alpha}}\hs{-0.03cm}
\zeta^\beta}\s
\prod\limits_{\alpha=1}^{N}d^2\zeta^\alpha\s.\
\qqq
Now Eq.\s\s(\ref{D.3}) follows by simple Gaussian integration.
\vs 1cm

\nappendix{F}
\vs 0.5cm

We shall check the consistency of the formula
(\ref{final}) for the scalar product of the
CS states. First of all, the integrand on the right hand
side is independent of the choice of the
bases \s$(\kappa^r_\bx)\s$
of \s$H^0(L_\bx^2)\s$ and \s$(\eta^\alpha_\bx)\s$ of
\s$H^0(L_\bx^2 K)\s$. \s Indeed, a change of
\s$(\kappa^r_\bx)\s$ in \s\s$|\s\sum\limits_{j=1}^g\s(-1)^{j}\s\s
{\rm det}\m(\smallint_{_\Sigma}\kappa^r_\bx\omega^i
\wedge\m b\s)_{\hs{-0.01cm}_{i\not=j}}\
\omega^j(x)\s\s|^{\wedge 2}\s\s$ is compensated by
that in \s\s${\rm det}\m(\m\smallint_{_\Sigma}
\m\langle\kappa^r_\bx,
\m\kappa^s_\bx\rangle\s{\rm vol}\s)^{-1}\s\s$
and \s\s$\eta^\alpha_\bx\m\da_{z^{\alpha}_{\bf x}}\s\s$ is
independent of the choice of \s$(\eta^\alpha_\bx)\s$.
Another consistency check is the independence
of the integrand under change of the hermitian structure
of \s$L_0\s$. \s Recall, that the the Green function
\s$G(x,y)\s$ of the Laplacian was chosen so that
\s$\smallint_{_\Sigma}\hs{-0.05cm}
G(\m\cdot\s,\m y)\s((k+2)F_0(y)+\hf R(y))=0\s$,
\s where \s$F_0\s$ is the curvature form of \s$L_0\s$.
\s The multiplication of the hermitian metric of \s$L_0\s$
by a positive function
\s$\ee^{\varphi/(k+2)}\s$ leads to the replacement
\s$F_0\m\mapsto\m F_0+{\de\da\varphi\over k+2}\s$ and
\qq
G(x,y)\s\ \ &\mapsto\ \ &G(x,y)\s+
\s{_1\over^{4\pi M}}\s(\varphi(x)+\varphi(y))\cr
&&-\s{_{i}\over^{8\pi^2M^2}}\s
\smallint_{_\Sigma}\m(\m\da\varphi\wedge
\de\varphi\s+\s\varphi(2(k+2)F_0+R)\m)\ .
\qqq
Since \s$\langle\eta^{\alpha_m},\wedge\eta^{\beta_m}
\hs{-0.05cm}\rangle(x_m\hs{-0.04cm})\s$ picks up
the factor \s$\ee^{2\varphi(x_m)/(k+2)}\s$, the last line
of Eq.\s\s(\ref{final}) is multiplied by
$$\exp[\m{{i}\over{2\pi(k+2)}}\s
\smallint_{_\Sigma}\m(\m\da\varphi\wedge
\de\varphi\s+\s\varphi(2(k+2)F_0+R)\m)\m]\s.$$
By the chiral anomaly formula,
\s\s${\rm det}\m(\m\smallint_{_\Sigma}\m\langle\kappa^r,
\m\kappa^s\rangle\s{\rm vol}\s)^{-1}\s\s\s{\rm det}'\m
(\m\de_{L^{2}_\bx}^\dagger\m
\de_{L^{2}_\bx})\s\s$ changes by the factor
$$\exp[-{i\over{\pi(k+2)^2}}\smallint_{_\Sigma}\m
(\m\da\varphi\wedge\de\varphi\m+\m(k+2)\varphi(2F_0+\hf R)
\m)\m]\s.$$
Finally, we shall show that
\s\s$\exp[-{ik\over 2\pi}\int_{_\Sigma}
\hs{-0.05cm}\tr\s\m A_0^{10}\wedge\m A^{01}_0]\
|\psi(\bx,b)|^2\s\s$, which depends on the metric
of \s$L_0\s$ through the smooth isometric isomorphism
\s\s$U:\m L_0^{-1}\oplus L_0\m\rightarrow\m
\Sigma\times\NC^2\s$, \s yields upon the multiplication
of the metric by \s\s$\ee^{\varphi/(k+2)}\s\s$ the factor
$$\exp[\m-{ik\over2\pi(k+2)^2}\smallint_{_\Sigma}\m(\m
\da\varphi\wedge\de\varphi\m+\m2(k+2)\varphi F_0\m)\m]\s$$
cancelling the previous ones.
First note that, by Eq.\s\s(\ref{5.20}),
\qq
\ee^{-{ik\over 2\pi}\int_{_\Sigma}
\hs{-0.05cm}\tr\s\m A_0^{10}\m\wedge\m A^{01}_0}\
|\psi(\bx,b)|^2\s
=\s c(\bx)\ |\s\Psi(A^{01}_{\bx,b})\m|^2\
\ee^{-{ik\over 2\pi}\int_{_\Sigma}
\hs{-0.05cm}\tr\s\m(A_{{\bf x},b}^{10})^\dagger\wedge
\m A^{10}_{{\bf x},b}\s+\s{ik\over 2\pi}\int_{_\Sigma}
\hs{-0.05cm}\langle b,\wedge b\rangle}
\label{S}
\qqq
with \s$c(\bx)\s$ independent of the choice of the metric
on \s$L_0\s$. \s As follows from the transformation
properties of \s$\Psi\s$ and of the WZW action \s$S\s$,
\qq
|\s\Psi({}^{h^{\hs{-0.06cm}-1}}\hs{-0.19cm}A^{01})\m|^2\
\ee^{-{ik\over 2\pi}\int_{_\Sigma}
\hs{-0.05cm}\tr\s\s({}^{h^{\hs{-0.06cm}-1}}\hs{-0.16cm}
A^{10})^\dagger\wedge
\m{}^{h^{\hs{-0.06cm}-1}}\hs{-0.16cm}A^{10}}\hs{3.5cm}\cr
\s=\s\ee^{\s k\m S(hh^\dagger,\m-(A^{01})^\dagger
+\s A^{01})}\ \s|\s\Psi(A^{01})\m|^2
\ \s\ee^{-{ik\over 2\pi}\int_{_\Sigma}
\hs{-0.05cm}\tr\s\s(A^{10})^\dagger\wedge
\m A^{10}}\ .\label{S1}
\qqq
In particular, the expression does not change if \s$h\s$
takes values in the compact group. It follows that, for
a fixed metric on \s$L_0\s$, \s\s$\exp[-{ik\over 2\pi}
\int_{_\Sigma}
\hs{-0.05cm}\tr\s\m A_0^{10}\m\wedge\m A^{01}_0]\
|\psi(\bx,b)|^2\s\s$ is independent of the choice
of the isometric isomorphism \s$U:\m L_0^{-1}\oplus
L_0\s\rightarrow\s\Sigma\times\NC^2\s$ since
\s$A^{01}_{\bx,b}\s$ for two different choices are
related by an \s$SU(2)$-valued gauge transformation.
\vs 0.4cm

After the multiplication of the hermitian structure
on \s$L_0\s$ by \s$\ee^\varphi\s$, we may take
\s$U\m\ee^{-\varphi\sigma^3/2}\s$ as the new
isometric isomorphism from \s$L_0^{-1}\oplus L_0\s$
to \s$\Sigma\oplus\NC^2\s$. As a result of the change
of \s$U\s$, \s$A^{01}_{\bx,b}\s$ changes to
\s${}^{h^{\hs{-0.05cm}-1}}\hs{-0.2cm}A^{01}_{\bx,b}\s$
with \s$h\m=\m U\m\ee^{\varphi\sigma^3/2}\m U^{-1}\s$.
By virtue of Eqs.\s\s(\ref{S}) and (\ref{S1}),
\s\s$\ee^{-{ik\over 2\pi}\int_{_\Sigma}
\hs{-0.05cm}\tr\s\m A_0^{10}\m\wedge\m A^{01}_0}\s
|\psi(\bx,b)|^2\s\s$ is then multiplied by
\qq
\ee^{\s k\m S(hh^\dagger,\m-(A^{01}_{{\bf x},b})^\dagger
+\m A^{01}_{{\bf x},b})\s-\s{ik\over 2\pi}\int_{_\Sigma}
\hs{-0.05cm}(\ee^{-2\varphi}-1)\langle b,\wedge b\rangle}\ .
\label{FaC}
\qqq
We are left with the calculation of \s\s$
S(hh^\dagger,\m-(A^{01}_{{\bf x},b})^\dagger
+\m A^{01}_{{\bf x},b})\s$. \s By Eq.\s\s(\ref{var}),
under the infinitesimal change of \s$\varphi\s$
\qq
\delta\m S(hh^\dagger,\m-(A^{01}_{\bx,b})^\dagger+
\m A^{01}_{\bx,b})\s=\s{_i\over^{2\pi}}\int_{_\Sigma}
\hs{-0.05cm}\tr\s\s\m U\m\delta\varphi\s\sigma^3\m U^{-1}
\s F(-(A^{01}_{\bx,b})^\dagger+
\m {}^{(\hs{-0.02cm}hh^\dagger\hs{-0.03cm})^{\hs{-0.05cm}-1}}
\hs{-0.2cm}A^{01}_{\bx,b})\cr
=\s{_i\over^{2\pi}}\int_{_\Sigma}
\hs{-0.05cm}\tr\s\s\m \delta\varphi\s\sigma^3\s
{\rm curv}\left(\nabla+\de+\left(\matrix{\bar a_\bx-a_\bx
+\de\varphi&\ee^{-2\varphi}\m b\cr
b^\dagger&-\bar a_\bx+a_\bx-\de\varphi}\right)\right)
\hs{0.5cm}\cr\cr
=\s{_i\over^{2\pi}}\int_{_\Sigma}
\hs{-0.05cm}\tr\s\s\m \delta\varphi\s\sigma^3\s
\left(\matrix{-F_0+\da\de\varphi-\ee^{-2\varphi}
\langle b,\wedge b\rangle&
\nabla(\ee^{-2\varphi}b)+2\m\ee^{-2\varphi}\m
\bar a_\bx\wedge\m b\cr
(\de+2a_\bx-\de\varphi)\m b^\dagger&F_0-\da\de\varphi+
\ee^{-2\varphi}\langle b,\wedge b\rangle}\right)\cr\cr
=\s\delta\bigg(\hs{-0.07cm}-{_i\over^{2\pi}}
\smallint_{_\Sigma}\m(\m\da\varphi
\wedge\de\varphi+2\varphi F_0-
(\ee^{-2\varphi}\hs{-0.05cm}-1)\langle b,\wedge
b\rangle)\bigg)
\qqq
so that the factor (\ref{FaC}) becomes
\s\s$\ee^{-{ik\over 2\pi}\smallint_{_\Sigma}\m(\m\da\varphi
\wedge\de\varphi+2\varphi F_0\m)}\s,\s\s$ as required.
\vs 0.4cm

As for the change of \s$\psi(\bx,b)\s$ itself, a straightforward
calculation shows that if
\s$h:\m\Sigma\rightarrow SL(2,\NC)\s$
then
\qq
\ee^{\s{ik\over 2\pi}\int_{_\Sigma}
\hs{-0.05cm}\tr\s\s\m{}^{h^{\dagger}}\hs{-0.16cm}
A_0^{10}\m\wedge{}^{h^{\hs{-0.05cm}-1}}
\hs{-0.18cm}A^{01}_{{\bf x},b}}\s\m\psi({}^{h^{\hs{-0.05cm}-1}}
\hs{-0.2cm}A^{01}_{{\bf x},b})\s
=\s\ee^{\s k\m S(hh^\dagger\hs{-0.05cm},\m A^{10}_0+
\m A^{01}_{{\bf x},b})\m-\m k\m S(hh^\dagger)\m+\m k\m S(h)
\m-{ik\over2\pi}\int_{_\Sigma}
\hs{-0.05cm}\tr\s\s\m A^{10}_0\wedge
(h^{\dagger})^{-1}\de h^{\dagger}}\ \s\cr
\cdot\s\ \ee^{\s{ik\over2\pi}\int_{_\Sigma}
\hs{-0.05cm}\tr\s\s\m A^{10}_0\wedge\m A^{01}_{{\bf x},b}}\
\Psi(A^{01}_{{\bf x},b})\ .\hs{1.3cm}
\qqq
It follows that if \s\s$U\m\mapsto\m U'=h^{-1}U\s$ with
\s$SU(2)$-valued \s$h\s$ then \s$\psi(\bx,b)\s$
changes only by a constant factor \s\s$\ee^{\s-k\m S(h)
\m-{ik\over2\pi}\int_{_\Sigma}
\hs{-0.05cm}\tr\s\s\m A^{10}_0
\wedge(h^{\dagger})^{-1}\de h^{\dagger}}\s$.
\s On the other hand, if the metric of \s$L_0\s$
is multiplied by \s$\ee^\varphi\s$ and \s$U\m\mapsto\m U'=
U\m\ee^{-\varphi\sigma^3/2}\m\equiv\m h^{-1} U\s$, \s then
\s$\psi(\bx,b)\s$ picks up the factor
\qq
&\ee^{\m k\m S(hh^{\dagger},\m A^{10}_0
+\m A^{01}_{{\bf x},b})\m-\m k\m
S(hh^\dagger)\m+\m k\m S(h)
\m-{ik\over2\pi}\int_{_\Sigma}
\hs{-0.05cm}\tr\s\s\m A^{10}_0
\wedge(h^\dagger)^{-1}\de h^\dagger}&\cr
&=\s\ee^{\m k\m S(hh^{\dagger},\m A_0)\m-\m k\m
S(hh^\dagger)\m+\m k\m S(h)
\m-{ik\over2\pi}\int_{_\Sigma}
\hs{-0.05cm}\tr\s\s\m A^{10}_0
\wedge(h^{\dagger})^{-1}\de h^{\dagger}}\ ,&
\qqq
where the last inequality may be checked by differentiating
\s$S(hh^{\dagger},\m A^{10}_0
+\m A^{01}_{{\bf x},b})\s$
with respect to \s$\varphi\s$ with
the use of Eq.\s\s(\ref{var}). Again
\s$\psi(\bx,b)\s$ is multiplied by a constant independent
of \s$\bx\s$ and \s$b\s$.
\vskip 0.5cm

In the next check, let us find the dependence
of the right hand side of Eq.\s\s(\ref{final})
on the conformal factor of the Riemannian metric
of \s$\Sigma\s$. If the metric is multiplied by
a positive function \s$\ee^{\sigma}\s$, then
\qq
G(x,y)\s\ \ &\mapsto\ \ &G(x,y)\s+
\s{_1\over^{8\pi M}}\s(\sigma(x)+\sigma(y))\cr
&&-\s{_{i}\over^{32\pi^2M^2}}\s
\smallint_{_\Sigma}\m(\m\da\sigma\wedge
\de\sigma\s+\s2\sigma(2(k+2)F_0+R)\m)\ ,\cr
:G(x,x):\s\s&\mapsto\ \ &:G(x,x):\s-
\s{_{M-1}\over^{4\pi M}}\s(\sigma(x)+\sigma(y))\cr
&&-\s{_{i}\over^{32\pi^2M^2}}\s
\smallint_{_\Sigma}\m(\m\da\sigma\wedge
\de\sigma\s+\s2\sigma(2(k+2)F_0+R)\m)\ ,
\qqq
and the last line of Eq.\s\s(\ref{final}) is multiplied
by
$$\exp[\m{_i\over^{4\pi(k+2)}}\smallint_{_\Sigma}\m
(\m\hf\m\da\sigma\wedge
\de\sigma\m+\m\sigma\m(2(k+2)F_0+R)\m)\m]\ .$$
In virtue of the conformal anomaly formula,
\s\s${\rm det}\m(\m\smallint_{_\Sigma}\m\langle\kappa^r,
\m\kappa^s\rangle\s{\rm vol}\s)^{-1}\s\s\s{\rm det}'\m
(\m\de_{L^{2}_\bx}^\dagger\m
\de_{L^{2}_\bx})\s\s$ and
\s\s$\left({_{{\rm det}'\m(-\Delta)}\over^{{\rm area}}}
\right)^{\hs{-0.07cm}1/2}\s\s$
change, respectively, by the factors
$$\exp[\m-{_i\over^{12\pi}}\smallint_{_\Sigma}\m(\m
\hf\m\da\sigma\wedge\de\sigma\m+\m\sigma
\m(6F_0+R)\m)\m]\hs{0.6cm}
\s{\rm and}\hs{0.6cm}\exp[\m-{_i\over^{24\pi}}
\smallint_{_\Sigma}\m(\m
\hf\m\da\sigma\wedge\de\sigma\m+\m\sigma R\m)\m]\s.\ \ \ $$
Altogether, Eq.\s\s(\ref{final}) picks the factor
\qq
\exp[\m-{_i\over^{24\pi}}\m{_{3k}\over^{k+2}}
\s\smallint_{_\Sigma}\m(\m
\hf\m\da\sigma\wedge\de\sigma\m+\m\sigma R\m)\m]
\qqq
when the Riemannian metric is multiplied by
\s$\ee^{\m\sigma}\s$. \s This guarantees the
right value \s$c={3k\over k+2}\s$ of the Virasoro central
charge of the theory with partition function given
by the formula (\ref{PartF}).
\vs 0.5cm

Another easy check of formula (\ref{final}) shows that
its right hand side does not depend on the choice
of \s$x_0\s$ used to fix the bundle \s$L_0=L(-x)\s$.
We leave it to the reader. A more involved problem
which we have not addressed is the independence of
the scalar product expression of the choice of
the h.l.b. $L\s$ of degree \s$g\s$.
\vs 2.5 cm


\begin{thebibliography}{bib}


\bibitem{Bertram}
A. Bertram: {\it Moduli of Rank-2 Vector Bundles,
Theta Divisors and the Geometry of Curves in
Projective Space}. J. Diff. Geom. {\bf 35},
429-469 (1992)

\bibitem{WittenJones}
E. Witten: {\it Quantum Field Theory and Jones Polynomial},
Commun. Math. Phys. {\bf 121}, 351-399 (1989)

\bibitem{WitBos}
E. Witten: {\it Non-Abelian Bosonization in Two Dimensions}.
Commun. Math. Phys. {\bf{92}}, 455-472 (1984)

\bibitem{Sesh}
C. S. Seshadri: {\it Space of Unitary Vector Bundles on a Compact
Riemann Surface}. Ann. of Math. {\bf 85}, 303-336 (1967)

\bibitem{NaraRama}
M. S. Narasimhan, S. Ramanan: {\it Moduli of Vector
Bundles on a Compact Riemann Surface}. Ann. Math. {\bf 89},
19-51 (1969)

\bibitem{Verl}
E. Verlinde:
{\it Fusion rules and modular transformations
in 2-d conformal field theory}.
Nucl. Phys. {\bf B 300}, 360-376 (1988)

\bibitem{Bernard}
D. Bernard: {\it On the Wess-Zumino-Witten Models on
Riemann Surfaces}.
Nucl. Phys. {\bf B 309}, 145-174 (1988)

\bibitem{Hitchin}
N. Hitchin: {\it Flat Connections and Geometric Quantization}.
Commun. Math. Phys. {\bf 131}, 347-380 (1990)

\bibitem{Karpacz}
K. Gaw\c{e}dzki: {\it Constructive Conformal Field Theory}.
In: Functional integration, geometry and strings, eds. Z. Haba,
J. Sobczyk, Birkh\"{a}user, Basel, Boston, Berlin 1989,
pp. 277-302

\bibitem{Axel}
S. Axelrod, S. Della Pietra, E. Witten:
{\it Geometric Quantization of Chern-Simons Gauge Theory}.
J. Differential Geometry {\bf 33}, 787-902 (1991)

\bibitem{KZ}
V. Knizhnik, A. B. Zamolodchikov:
{\it Current Algebra and Wess-Zumino Model
in Two Dimensions}. Nucl. Phys. {\bf B 247}, 83-103 (1984)

\bibitem{1}
K. Gaw\c{e}dzki, A. Kupiainen: {\it SU(2) Chern-Simons Theory
at Genus Zero}. Commun. Math. Phys. {\bf 135}, 531-546 (1991)

\bibitem{Quadr}
K. Gaw\c{e}dzki: {\it Quadrature of Conformal Field Theories}.
Nucl. Phys. {\bf B 328}, 733-752 (1989)

\bibitem{FalGaw0}
F. Falceto, K. Gaw\c{e}dzki, A. Kupiainen: {\it Scalar
Product of Current Blocks in WZW Theory}. Phys. Lett. {\bf B260},
101-108 (1991)

\bibitem{Flume}
P. Christe, R. Flume: {\it The Four-Point Correlations
of All Primary Operators of the d=2 Conformally
Invariant \s$SU(2)\s$ $\sigma$-Model with Wess-Zumino
Term}. Nucl. Phys. {\bf B 282}, 466-494 (1987)

\bibitem{FadZam}
A. B. Zamolodchikov, V. A. Fateev: {\it Operator Algebra
and Correlation Functions of the Two-Dimensional
Wess-Zumino $SU(2)\times SU(2)$ Chiral Model}.
Yad. Fiz. {\bf 43}, 1031-1044 (1986)

\bibitem{VarSchecht}
V. V. Schechtman, A. N. Varchenko: {\it Hypergeometric
Solutions of Knizhnik-Zamolodchikov Equations}.
Lett. Math. Phys. {\bf 20}, 279-283 (1990)

\bibitem{Resh}
N. Reshetikhin: seminar in Strasbourg, May 93

\bibitem{Varch}
A. Varchenko: {\it Critical Points of the Product
of Powers of Linear Functions and Families
of Bases of Singular Vectors}. Chapel Hill preprint 1993

\bibitem{Babu}
H. M. Babujian, R. Flume: {\it Off-Shell Bethe Ansatz
Equation for Gaudin Magnets and Solutions of
Knizhnik-Zamolodchikov Equations}. Preprint hep-th/9310110

\bibitem{FeiFrResh}
B. Feigin, E. Frenkel, N. Reshetikhin: {\it Gaudin
Model, Bethe Ansatz and Correlation Functions at the
Critical Level}. Preprint hep-th/9402022

\bibitem{Thad}
M. Thaddeus: {\it Stable Pairs, Linear Systems
and the Verlinde Formula}, Berkeley preprint 1993

\bibitem{JaffeQ}
A. Jaffe, F. Quinn: {\it ``Theoretical Mathematics'':
towards a Cultural Synthesis of Mathematics and
Theoretical Physics}. B.A.M.S. {\bf 29}, 1-13 (1993)

\bibitem{Ja}
K. Gaw\c{e}dzki: {\it $SU(2)\s$ WZNW Model at Higher Genera from
Gauge Field Functional Integral}. Preprint IHES/P/93/66,
hep-th 9312051

\bibitem{Feigin}
B. Feigin: lectures at the Conference ``Vector Bundles on Curves''.
Bombay Dec. 13-24, 1993

\bibitem{Haba}
Z. Haba: {\it Correlation Functions of \s$\sigma\s$
Fields with Values in a Hyperbolic Space}.
Int. J. Mod. Phys. {\bf A 4}, 267-286 (1989)

\bibitem{GK}
K. Gaw\c{e}dzki, A. Kupiainen, {\it Coset Construction
from Functional Integrals}: Nucl. Phys. {\bf B 320},
625-668 (1989)

\bibitem{GTW}
A. Gupta, S. P. Trivedi, M. B. Wise: {\it Random Surfaces
in Conformal Gauge}. Nucl. Phys. {\bf B 340}, 475-490 (1990)

\bibitem{GL}
M. Goulian, M. Li: {\it Correlation Functions
in Liouville Theory}. Phys. Rev. Lett. {\bf 66}, 2051-2055
(1991)

\bibitem{Wittenfact}
E. Witten: {\it On Holomorphic Factorization of WZW and
Coset Models}. Commun. Math. Phys. {\bf 144}, 189-212 (1992)

\bibitem{AGBMNV}
L. Alvarez-Gaum\'{e}, J.-B. Bost, G. Moore, P. Nelson,
C. Vafa: {\it Bosonization on Higher Genus Riemann Surfaces}.
Commun. Math. Phys. {\bf 112}, 503-552 (1987)

\bibitem{Arakel}
S. J. Arakelov: {\it Intersection Theory of Divisors on
an Arithmetic Surface}. Math. USSR Izv.
{\bf 8}, 1167-1180 (1974)

\bibitem{Quillen}
D. Quillen: {\it Determinants of Cauchy-Riemann
Operators over a Riemann
Surface}. Funct. Anal. Appl. {\bf 19}, 31-34 (1985)

\bibitem{Bost}
J.-B. Bost: {\it Fibr\'{e}s D\'{e}terminants, D\'{e}terminants
R\'{e}gularis\'{e}s et Mesures sur les Espaces de Modules
des Courbes Complexes}. Ast\'{e}risque {\bf 152-153},
113-149 (1987)

\bibitem{Kato}
T. Kato: Perturbation theory for linear operators.
Berlin, Heidelberg, New York: Springer 1966


\end{thebibliography}
\end{document}